\documentclass[a4paper,twocolumn,11pt,accepted=2019-01-28]{quantumarticle}
\pdfoutput=1
\usepackage{graphicx}
\usepackage{braket}
\usepackage{amsmath}
\usepackage{amsfonts}
\usepackage{amssymb}
\usepackage{bbm}
\usepackage{amsthm}
\usepackage{pgf}
\usepackage[backend=bibtex,sorting=none,firstinits=true,style=phys,maxbibnames=5,biblabel=brackets,chaptertitle=false,pageranges=false,doi=false]{biblatex}
\newtheorem*{theorem}{Main Result}

\usepackage{hyperref}

\begin{document}

\title{Computational universality of symmetry-protected topologically ordered cluster phases on 2D Archimedean lattices}
\author{Austin K. Daniel}\email{AustinDaniel@unm.edu}
\author{Rafael N. Alexander}
\author{Akimasa Miyake}
\address{Center for Quantum Information and Control, 
Department of Physics and Astronomy,
University of New Mexico, Albuquerque, NM 87131, USA}

\date{December 11, 2019}

\begin{abstract}
What kinds of symmetry-protected topologically ordered (SPTO) ground states can be used for universal measurement-based quantum computation in a similar fashion to the 2D cluster state? 2D SPTO states are classified not only by global on-site symmetries but also by subsystem symmetries, which are fine-grained symmetries dependent on the lattice geometry. Recently, all states within so-called SPTO cluster phases on the square and hexagonal lattices have been shown to be universal, based on the presence of subsystem symmetries and associated structures of quantum cellular automata. Motivated by this observation, we analyze the computational capability of SPTO cluster phases on all vertex-translative 2D Archimedean lattices. There are four subsystem symmetries here called ribbon, cone, fractal, and 1-form symmetries, and the former three are fundamentally in one-to-one correspondence with three classes of Clifford quantum cellular automata. We conclude that nine out of the eleven Archimedean lattices support universal cluster phases protected by one of the former three symmetries, while the remaining lattices possess 1-form symmetries and have a different capability related to error correction.
\end{abstract}

\maketitle

\section{Introduction}

\begin{figure}
\includegraphics[width=\linewidth]{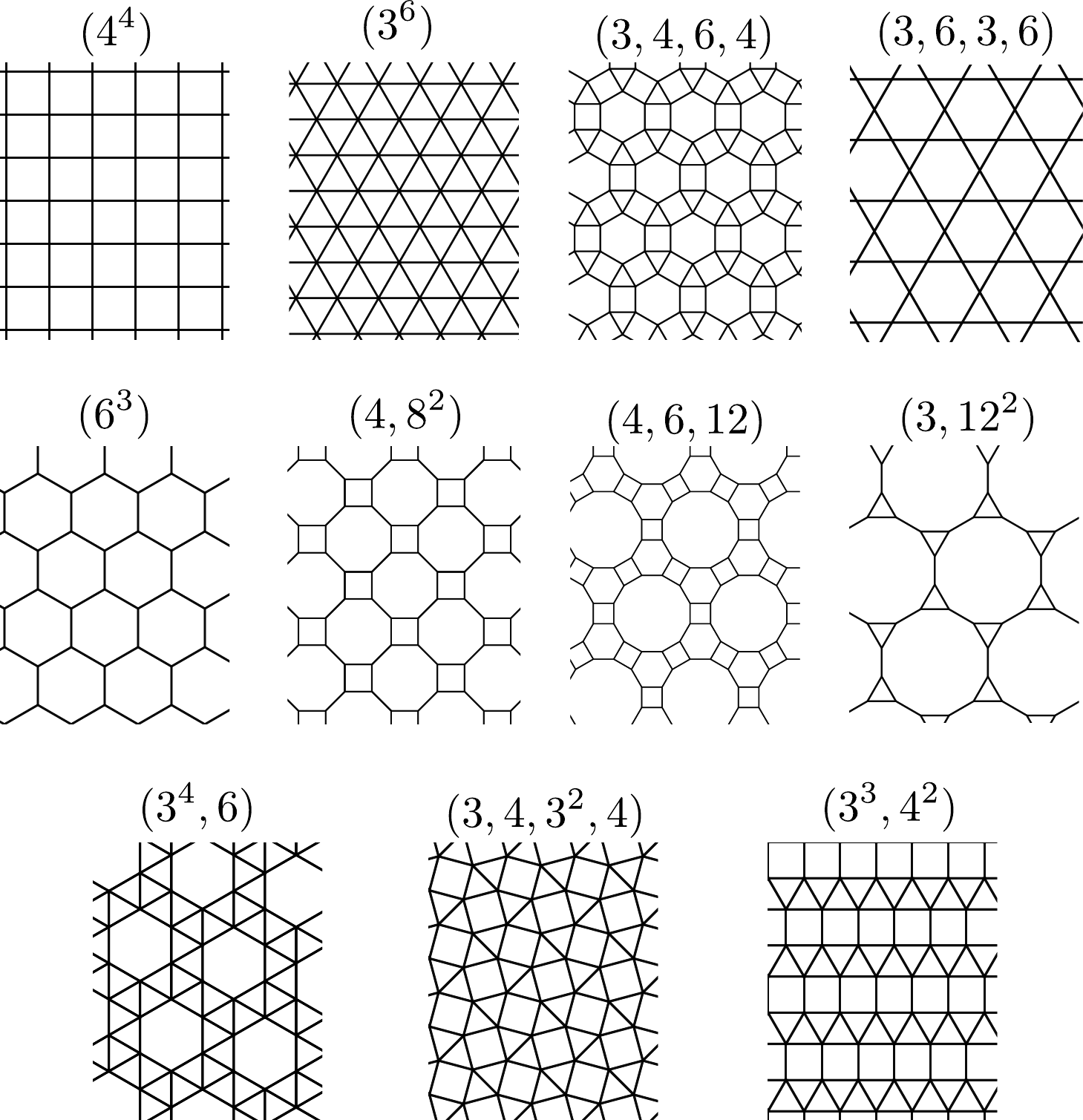}
\caption{The eleven Archimedean lattices.  Each lattice is vertex translative, meaning that the same pattern of shapes meet about each vertex.  They are then labeled according to this rule. For example, at each vertex in the $(4,8^2)$ lattice a square and two octagons meet.}
\label{Archlat}
\end{figure}

Geometry plays an important role in both quantum information and many-body physics. Quantum states can inherit symmetries from the way their composite parts are arranged geometrically, which can in turn result in novel physical properties. In measurement-based quantum computation (MBQC) \cite{Raussendorf2001}, many-body entanglement is converted to quantum computation via local measurements and classical communication, and the importance of geometry is twofold. First, it directly dictates the computational utility of entangled resources states known as graph states \cite{Briegel2001, hein2006entanglement}. Second, the geometry of the entangled state can give rise to symmetries, which are known to play key roles, directly or indirectly, in constructing and characterizing many-body entangled states that are universal for MBQC~\cite{verstraete2004valence, nest2006universal, nest2007fundamentals, gross2007novel, gross2007measurement, doherty2009identifying, miyake2011quantum, wei2011affleck, wei2012two, wei2013quantum, wei2014hybrid, wei2015universal, 
miller2016hierarchy, wei2017universal, miller2018latent, gachechiladze2019changing}. Recent progress reveals that some of these states possess topological orders under a symmetry restriction, known as symmetry-protected topological orders (SPTO), which have been of recent interest in condensed matter physics and the modern classification of quantum phases of matter \cite{gu2009tensor, chen2011classification, schuch2011classifying, pollmann2012symmetry, chen2013symmetry}.

It has been observed in Ref.~\cite{miyake2010quantum} that all ground states of a certain 1D SPTO phase, known as the \emph{Haldane phase} \cite{affleck1987rigorous, affleck1988valence}, have an equivalent computational capacity provided that the symmetries remain unbroken. Furthermore, any ground state residing in a 1D SPTO phase protected by a finite Abelian symmetry group has been shown to act as a 1D MBQC resource~\cite{else2012symmetry, else2012symmetry2, miller2015resource, Stephen2017, Raussendorf2017}. 
Recently, these results have been extended to 2D resource states lying in a quasi-1D SPTO phase, protected by so-called \emph{subsystem symmetries}, giving rise to quantum phases of matter in which every state is universal resource for MBQC.
Remarkably, the computational power of such phases is a direct consequence of the symmetries they possess~\cite{Stephen2017}. In particular, the first example of a computationally universal phase, known as the \emph{2D cluster phase}, was constructed from the rigid line-like symmetries of the square lattice cluster state in Ref.~\cite{Raussendorf2018}, followed by the fractal symmetries of the hexagonal lattice in Ref.~\cite{Devakul2018}. A recent paper~\cite{Stephen2019} has constructed tensor network states with underlying Clifford quantum cellular automaton (QCA) in their virtual space, so that they have subsystem symmetries and support computationally universal subsystem SPTO phases.


In this Article, we will take a ``lattice-first'' approach, constructing 2D cluster phases from the subsystem symmetries common to all the ground states on a given 2D lattice and identifying the structure of QCA that underlies its tensor network description. It is known that for graph states, computational power depends strongly on its lattice or graph ~\cite{nest2006universal, nest2007fundamentals, browne2007generalized, mhalla2011graph}. By performing an in-depth characterization of each of the eleven Archimedean lattices (shown in Fig.~\ref{Archlat}), we analyze the roles the lattice plays in the resource quality of the corresponding subsystem SPTO phases.
Besides being of independent geometric interest (c.f.~\cite{Richter2004})---they are the only vertex-translative lattices in 2D---they contain lattices more exotic than those studied previously, thus offering an important testbed for our method for constructing cluster phases, which complements the methods of Ref.~\cite{Stephen2019}.  Our lattice-first approach yields several new insights, such as a counterexample case to the conjecture of Ref.~\cite{Stephen2019} that cluster phases with glider QCA should be constructed using \emph{line-like} symmetries, as well as examples of lattices with \emph{one-form} symmetries, which represent foliated error correcting codes and have underlying non-unitary QCA.

Following the background materials in Sec.~\ref{sec:background}, we provide a general procedure to identify relevant subsystem symmetries and related QCA structures of the graph state for the construction of the surrounding cluster phase in Secs.~\ref{Determining_the_QCA}, \ref{Determining_Symmetry}, \ref{Computational_Universality}. We show that nine of the eleven lattices support a universal cluster phase, corresponding to either QCA with cone or fractal symmetries described in Sec.~\ref{Cone_Symmetries} and Sec.~\ref{Fractal_Symmetries}, respectively. The other two cases support \emph{one-form} symmetries, which prevent them from forming cluster phases as described in Sec.~\ref{1form}. These results emphasize an important correspondence between the fundamental subsystem symmetries and the types of QCA, which we summarize in Table~\ref{SStable}. It is curious that none of the eleven Archimedean lattices support a \emph{periodic} QCA structure. To address this, we note in Sec.~\ref{Periodic} that when any lattice is partially decorated, it can support cluster phases with an underlying periodic QCA structure, thus providing a wealth of new examples. In Sec.~\ref{Foliation}, we study how global properties of the lattice---the location of input and output qubits on the lattice, and also how the lattice is embedded on the torus---can affect the computational properties.

\section{Preliminaries}\label{sec:background}

\subsection{Graph states}\label{Cluster_States_Review}
We begin with some definitions and notation. The Pauli operators are denoted as
\begin{eqnarray}
X&=& |0\rangle\langle 1| + |1\rangle\langle 0| \\
Y&=&-i|0\rangle\langle 1| + i|1\rangle\langle 0| \\
Z&=&|0\rangle\langle0| - |1\rangle\langle1|.
\end{eqnarray}

Let the state $|m^{(\sigma)}\rangle$ denote the $(-1)^m$ eigenstate of the Pauli operator $\sigma$ for $m\in\{0,1\}$ and $\sigma\in\{X,Y,Z\}$.  If the superscript is omitted, the state is implied to be a $Z$ eigenstate.  The $n$-qubit Clifford group is the normalizer of the $n$-qubit Pauli group. It is generated by the Hadamard, Phase, and $CZ$ gates
\begin{align}
H =&  |0^{(x)}\rangle\!\bra{0} + |1^{(x)}\rangle\!\bra{1}\\
S =& \ket{0}\! \bra{0} + i \ket{1}\! \bra{1}\\
CZ =& \ket{0}\!\bra{0}\otimes \mathbbm{1} + \ket{1}\!\bra{1}\otimes Z.
\end{align}

Traditionally, MBQC consists of preparing a graph state first, and then implementing quantum processing via a sequence of adaptive single-qubit measurements~\cite{Raussendorf2001}. Each graph state, $\ket{\psi_\mathcal{G}}$, is specified by a graph $\mathcal{G}(V,E)$, where $V$ and $E$ are the vertex and edge sets, respectively. Each vertex represents a qubit initialized in the state $|0^{(x)}\rangle$. Edges represent the action of a controlled-$Z$ ($CZ$) gate
\noindent between two adjacent qubits. Thus, 
\begin{equation}
\ket{\psi_\mathcal{G}}= \prod_{(j,k)\in E} CZ_{jk} |0^{(x)}\rangle^{\otimes |V|}.
\end{equation}

Equivalently, the graph state can be uniquely defined in terms of its stabilizer group, i.e., as the unique +1 eigenstate of the set 
\begin{equation}\label{Stabilizer_Relation}
\left\{S_v = X_v \bigotimes_{l\in\mathcal{N}(v)}Z_l~ \bigg|~ \forall v\in V\right\},
\end{equation}
where $l\in \mathcal{N}(v)$ if and only if $(l,v)\in E$.  For an extended review of graph states see Ref.~\cite{hein2006entanglement}.

The usefulness of a given graph state depends on the graph $\mathcal{G}$. For example, when $\mathcal{G}$ is a simple one-dimensional path graph 
with open boundary conditions, we can encode a single \emph{logical} qubit at the edge and perform $SU(2)$ rotations and logical measurements via an adaptive sequence of single-site measurements in the
\begin{equation}
X_\theta = \cos(\theta)X + \sin(\theta)Y
\end{equation}
and $Z$ bases, respectively.

Universal MBQC requires graphs of dimension higher than 2, and for the remainder of this article $\mathcal{G}$ is assumed to be one of the eleven Archimedean lattices (see Fig.~\ref{Archlat}) embedded on a cylinder with circumference $n$, or equivalently a torus with a single cut along the minor circumference. In principle, universal MBQC on such a graph state can be implemented by using $Z$ basis measurements to delete specific vertices in the graph, thereby carving out isolated regions of 1D wires (useful for single-qubit gates) and also leaving some transverse connectivity (useful for entangling gates)~\cite{Raussendorf2001}. However, this method does not generalize conveniently to arbitrary members of the surrounding SPTO phase since measurements away from the $X$ basis violate the relevant symmetries.  Fortunately MBQC can be performed in manner that minimizes symmetry violating operations~\cite{Raussendorf2017, Stephen2017}. We review and make use of this method in Sec.~\ref{Computational_Universality}.

\subsection{Subsystem symmetries}\label{Sub_Syms}

\begin{figure*}
\centering
\includegraphics[width=0.9\linewidth]{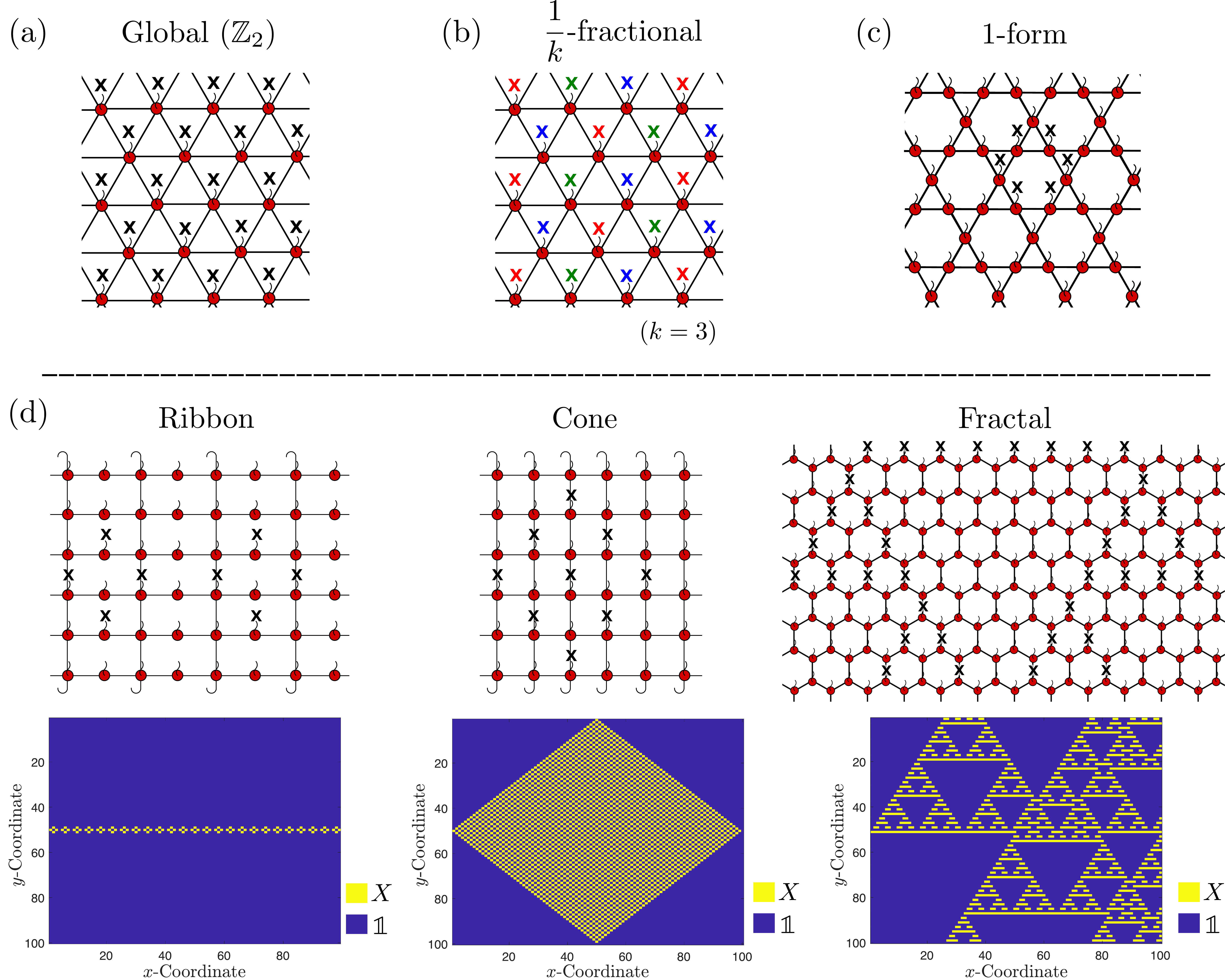}
\caption{Examples of the common types of symmetry. The tensor network notation uses the conventions described in Appendix.~\ref{TNNappendix}. \textbf{(a)} The simplest symmetry is a global $\mathbb{Z}_2$ symmetry, which involves applying $X$ to every physical degree of freedom. \textbf{(b)} The triangular lattice or $(3^6)$ lattice is $3$-colorable and thus has $\frac{1}{3}$-fractional symmetry.  The action of this symmetry corresponds to group $\mathbb{Z}_2^3$.  The three generators are depicted in red, blue, and green.  \textbf{(c)} The kagome or $(3,6,3,6)$ lattice has 1-form symmetry.  Each symmetry generator acts on the spins surrounding a hexagonal plaquette.  The resulting symmetry group is thus extensively large.  \textbf{(d)}  Examples of the three subsystem symmetries studied in this paper. In each lattice, periodic boundary conditions are imposed along the top and bottom, giving each subsystem symmetry a periodic structure.  The ribbon symmetry is depicted for two periods and the cone and fractal symmetry are depicted for a single period (above).  The three become physically distinct for larger system sizes (below).  Ribbon symmetries act in a localized narrow band of constant width whereas cone and fractal symmetries have support that spreads over the lattice in each direction.  Furthermore, cone symmetries fill out the inside of their causal cone with a regular structure whereas fractal symmetries are supported within the causal cone on a fractal subset of sites.  Each subsystem symmetry corresponds to a distinct Clifford QCA residing in the virtual level of the tensor network representation as described in Sec.~\ref{QCAandMBQC}.  We leverage this property and the Gottesman-Knill theorem to generate these plots numerically.}
\label{Symmetries_Figure}
\end{figure*} 

Now we discuss the symmetries of the 2D graph states described above.  Recall that a graph state is uniquely specified by its stabilizer group.  We wish to identify a whole family of states that have similar computational properties, so the full stabilizer group is too restrictive.  It is fruitful to instead consider subgroups of the stabilizer group, henceforth referred to as \emph{symmetries} of the graph state.  In particular, the symmetries considered here will consist only of tensor products of $X$ and $\mathbbm{1}$ operators.

Graph states can have many different symmetries, as illustrated in Fig.~\ref{Symmetries_Figure}. The simplest being a global $\mathbb{Z}_2$ symmetry, where the nontrivial element of the $\mathbb{Z}_2$ subgroup of the stabilizer group arises from taking the product of all stabilizers on all sites. Later, we consider four kinds of \emph{subsystem symmetries}, because they only have non-trivial support over a subset of qubits on the graph.  The first is the $\frac{1}{k}$-\emph{fractional symmetry} \cite{miller2018latent}, which is defined over any graph $\mathcal{G}(V,E)$ with chromatic number $k$. The $\frac{1}{k}$-fractional symmetry acts on the state as a product of $X$ operators on vertices of a common color.  This symmetry then forms a $\mathbb{Z}_2^{k}$ subgroup of the full stabilizer group.  The next examples are the \emph{ribbon}, \emph{cone}, and \emph{fractal symmetries}. Elements of these distinct symmetry groups are formed by taking minimal products of stabilizers so as to cancel all $Z$ operators.  They all form $\mathbb{Z}_2^{2n}$ subgroups of the stabilizer group.  The final symmetry is the \emph{1-form symmetry}, which has support on a compact manifold of co-dimension 1.  For 2D graph states this corresponds to symmetries whose generators are locally acting loops of $X$ operators.  Again, such loops are formed by taking products of stabilizers centered at each site on the loop.

\subsection{Finding symmetries}\label{QCAandMBQC}

Determining the existence and structure of ribbon, cone, and fractal subsystem symmetries for a given graph state by multiplying stabilizers can be challenging. Here we follow a more convenient method introduced in Ref.~\cite{Stephen2019} that leverages the tensor network representation of the graph states.  Indeed, the connection between tensor network representations and MBQC resources has long been studied \cite{verstraete2004valence, gross2007novel, gross2007measurement}.  For a brief introduction to tensor network notation see Appendix~\ref{TNNappendix}.  In this tensor network representation the structure of the virtual space is described by a Clifford quantum cellular automaton (CQCA). Translationally invariant CQCA have been classified~\cite{schlingemann2008structure,gutschow2010time}, allowing a connection to be drawn between each class and these three subsystem symmetries.  Consequently, the computational power of the 2D graph state can be attributed to an underlying CQCA structure.
In this section we review CQCA, their classification, and how they can be used in conjunction with tensor networks to determine subsystem symmetries.

Generally, cellular automata define a local update rule on state vectors. In the context of the Heisenberg picture evolution of Pauli operators via a local translationally-invariant Clifford circuit $T$, CQCA specify a transfer matrix $\mathcal{T}$ that acts on the binary vector representation $\boldsymbol{\xi} = \boldsymbol{\xi}^{(x)}\oplus\boldsymbol{\xi}^{(z)} \in\mathbb{F}_2^{2n}$ of Pauli operators, i.e., 
\begin{align}
P(\boldsymbol{\xi}) \mapsto T P(\boldsymbol{\xi}) T^{\dagger}  = P(\mathcal{T}\boldsymbol{\xi})
\end{align}
where $\oplus$ is the direct sum and
\begin{align}\label{Pauliop}
P(\boldsymbol{\xi}) = \otimes_{k=0}^{n-1} X_k^{\xi^{(x)}_k}Z_k^{\xi^{(z)}_k}.
\end{align}
The dimension of $\boldsymbol{\xi}$ is $2n$, and so this evolution can be simulated efficiently (a.k.a., the Gottesman-Knill Theorem~\cite{gottesman1998heisenberg}). Note that $T^p=\mathbbm{1}$ for some integer $p$, which we refer to as the \emph{period} of the CQCA.

Due to translational invariance, $T$ admits a compact representation in terms of Laurent polynomials~\cite{schlingemann2008structure, Stephen2019}. A Pauli operator $P(\boldsymbol{\xi})$ can be written as a two dimensional vector $\tilde{\boldsymbol{\xi}}$ whose entries are polynomials in a variable $\eta$ with degree $n-1$ and coefficients $\xi$ in $\mathbb{F}_2$, i.e.,
\begin{equation}
\tilde{\boldsymbol{\xi}} = 
\begin{pmatrix}
\sum_{k=0}^{n-1} \xi_{k}^{(x)} \eta^k \\
\sum_{k=0}^{n-1} \xi_{k}^{(z)} \eta^k
\end{pmatrix},
\end{equation}
\noindent where the first (second) entry describes $X$ ($Z$) support of the Pauli. $T$ can similarly be represented by a matrix of polynomials of the same form
\begin{equation}
\widetilde{\mathcal{T}} = 
\begin{pmatrix}
\sum_{k=0}^{n-1} t^{(xx)}_k \eta^k & \sum_{k=0}^{n-1} t^{(xz)}_k \eta^k \\
\sum_{k=0}^{n-1} t^{(zx)}_k \eta^k & \sum_{k=0}^{n-1} t^{(zz)}_k \eta^k
\end{pmatrix}.
\end{equation}

CQCA have been classified in Ref.~\cite{gutschow2010time} according to the trace of $\widetilde{\mathcal{T}}$ into three distinct classes based on how $p$ scales with the system size $n$. These are periodic, glider, or fractal.
\begin{equation}
\textrm{Tr}\left(\widetilde{\mathcal{T}}\right) = 
\begin{cases}
0,1; &\textrm{Periodic,} \rightarrow p=\Omega(1) \\
\eta^c + \eta^{-c}; &\textrm{Glider,}\rightarrow p=\Omega(n)\\
\textrm{otherwise}; &\textrm{Fractal,} \rightarrow p ~\textrm{irregular}
\end{cases}
\end{equation}
While the above classification of CQCAs was made with perfect translational invariance in space and time, we will give a more general method for determining the underlying CQCA structure of a given graph state in Sec.~\ref{Determining_the_QCA}. This will often give more general CQCAs that are invariant under translation by $\Delta~(\tau)$ steps in the space (time) direction.  We can appeal to the same classification described above by blocking the CQCA appropriately in the space and time directions.
 
Finally, we will clarify that there is a one to one correspondence between the ribbon, cone, or fractal subsystem symmetry of a graph state and the class of the CQCA structure underlying the virtual space of its tensor network representation. While this will be discussed in a more general context in Sec.~\ref{Determining_Symmetry}. for now we discuss this correspondence for a particular example, the $(4^4)$ or square lattice graph state.  The tensor network representation of the $(4^4)$ graph state, denoted as $|(4^4)\rangle$, can be written as \cite{Raussendorf2018}
\begin{equation}
\includegraphics[width=.7\linewidth]{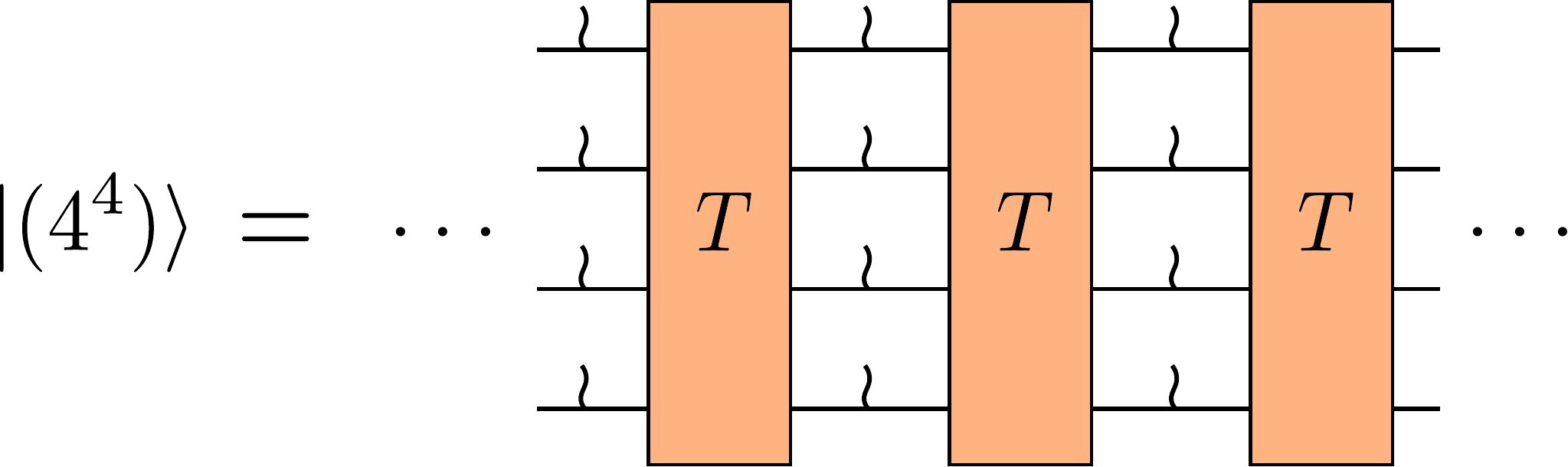},
\end{equation}
where each wavy line represents a physical degree of freedom and
\begin{equation}
T = \bigotimes_{j=0}^{n-1}H_j \prod_{j=0}^{n-1} CZ_{j,j+1}.
\end{equation}
The operator $T$ generates a CQCA residing in the glider class.  Now consider evolving a single site Pauli operator through the virtual level of the tensor network. For each column of copy tensors, $X$ operators commute through and leave behind an $X$ operator on the physical level.  Following this, the Pauli operator is updated according to the CQCA transfer matrix $T$.  After propagating this operator though the tensor network a cone symmetry, depicted in Fig.~\ref{Symmetries_Figure} is left behind on the physical degrees of freedom.  Since evolution under a CQCA can be simulated efficiently on a classical computer, subsystem symmetries can be determined in an efficient manner. In the remainder of this paper, all unitary QCA are CQCA.

\subsection{Phases of symmetry-protected topological order}\label{SPTOreview}

In light of the correspondence of Sec.~\ref{QCAandMBQC}, it is natural to ask: can \emph{any} state with such subsystem symmetries have a CQCA structure and be considered a universal resource for MBQC? Families of symmetry respecting states can naturally be discussed in terms of symmetry-protected topological order (SPTO). SPTO is a property of many-body ground states wherein the entanglement is robust to symmetry respecting perturbations \cite{chen2010local}.  Furthermore, the low-energy spectrum of the corresponding Hamiltonian is dependent on the topology of the system.  Namely, in the presence of open boundaries, the system exhibits ground state degeneracy corresponding to edge modes, whereas for periodic boundaries the ground state is unique and symmetric \cite{chen2011two, chen2013symmetry}. Such systems have been conjectured to be good candidates for MBQC resources \cite{miyake2010quantum, else2012symmetry, else2012symmetry2}.

Subsystem symmetries can protect non-trivial SPTO \cite{you2018subsystem, devakul2018classification}.  A scheme for MBQC with 1D-SPT phases that leverages the symmetry to do universal MBQC at arbitrary points in the phase was proposed in Refs.~\cite{miller2015resource, Stephen2017, Raussendorf2017}. Note, however, that this approach cannot be immediately applied to 2D SPTO because strict single-site locality of measurements is required for MBQC.
However, by considering additional lattice symmetries in 2D, the authors of Refs.~\cite{Raussendorf2018} were able to describe a 2D cluster phase, extending the 1D results to quasi-1D systems with subsystem symmetries, such as those discussed in Sec.~\ref{QCAandMBQC}, and giving rise to 2D resources phases that are universal for MBQC \cite{Raussendorf2018,Stephen2019}.  A self-contained review of MBQC protocols with quasi-1D SPT phases is given in Appendix~\ref{MBQCSSPT}.
It turns out that a cluster phase is an SPTO phase where the correspondence between subsystem symmetries and CQCA structures holds at every point. Remarkably, one can recast the MBQC scheme entirely in terms of symmetries, allowing the CQCA to be leveraged to achieve entangling gates.  For this reason, the computational power is uniform throughout the cluster phase. 

Moving between states in an SPTO phase corresponds to
 applying some constant-depth quantum circuit consisting of layers of symmetry-respecting unitary gates with disjoint support $U_{\phi}$~\cite{Hastings2005, chen2010local}. Thus, an arbitrary point in the phase $|\phi\rangle$ can be thought of as  $U_{\phi}$ applied to some reference state taken to be the graph state $|\psi_C\rangle$. One can write a tensor network for $|\phi\rangle$ by first taking a tensor network description of the fixed point $|\psi_C\rangle$, defined by tensors,
 \begin{equation}
\includegraphics[height = 15pt]{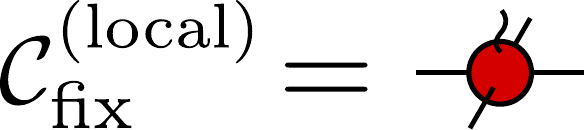}, 
\end{equation}
and then apply the unitary $U_\phi$, which can always be expressed as a matrix product unitary (MPU) due to its local nature (cf. \cite{cirac2017matrix, burak2018matrix, williamson2016matrix}).  Exploiting this fact $U_\phi$ can be written as the MPU, we describe graphically, for the case of a square lattice,  
 \begin{equation}
\includegraphics[height = 33pt]{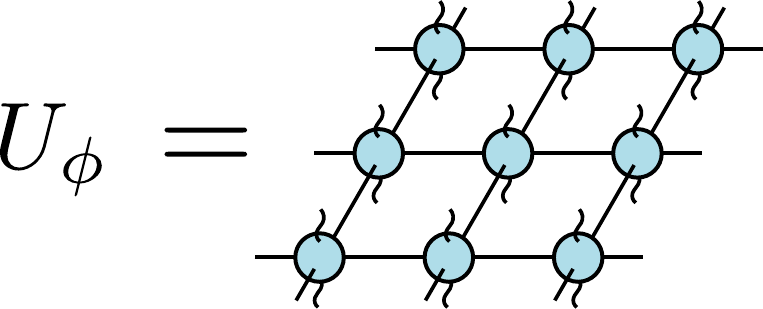},
\end{equation}
with local tensors,
 \begin{equation}
\includegraphics[height = 15pt]{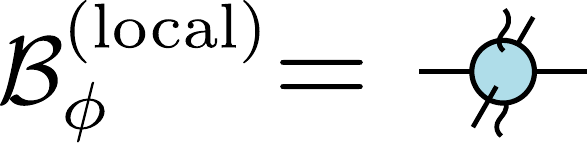}.
\end{equation}
These are commonly referred to as ``junk tensors" \cite{else2012symmetry} as they increase the bond dimension of the tensor network and are dependent on the microscopic details of the point in the phase.  We can then write a tensor network description of $|\phi\rangle$ as
 \begin{equation}
\includegraphics[height = 45pt]{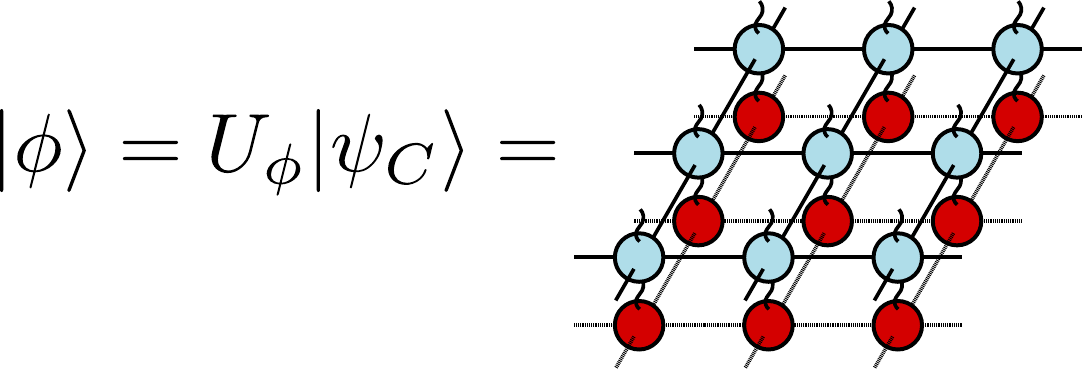} \label{Construct_Tensor}.
\end{equation}
Thus, the new tensors describing $|\phi\rangle$ are
\begin{equation}\label{Phase_Tensor}
\includegraphics[height = 20pt]{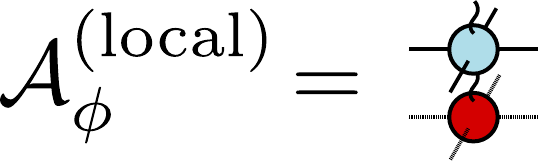}.
\end{equation}

The bottom layer of tensors generates the CQCA, as seen in Sec~\ref{QCAandMBQC}.  To enforce that $|\phi\rangle$ belongs to a cluster phase, it is sufficient to require that the MPU commutes with local $X$ operators, i.e.,
\begin{equation}
\includegraphics[height = 28pt]{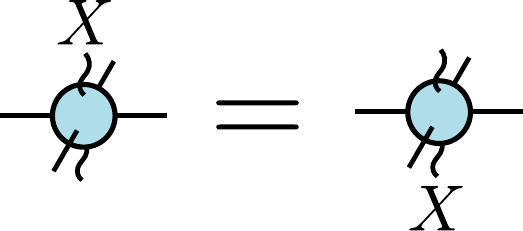}\label{Bsym}.
\end{equation}
Let $\boldsymbol{\xi}=\boldsymbol{\xi}^{(x)}\oplus\boldsymbol{\xi}^{(z)}\in\mathbb{F}_2^{2|V|}$ be used to denote an arbitrary Pauli operator $P(\boldsymbol{\xi})$, as defined in Eq.~(\ref{Pauliop}), with support on the lattice.  Furthermore, let $X(\boldsymbol{\xi}^{(x)})$ and $Z(\boldsymbol{\xi}^{(z)})$ denote the $X$ and $Z$ part of $P(\boldsymbol{\xi})$ so that $P(\boldsymbol{\xi})=X(\boldsymbol{\xi}^{(x)})Z(\boldsymbol{\xi}^{(z)})$. We can then expand $U_\phi$ in terms of Pauli operators,
\begin{equation}
U_\phi = \sum_{\boldsymbol{\xi}} c_{\boldsymbol{\xi}} X\left(\boldsymbol{\xi}^{(x)}\right)Z\left(\boldsymbol{\xi}^{(z)}\right). \label{Udef}
\end{equation} 
For Eq.~(\ref{Bsym}) to hold, it is sufficient to require that all $Z$ operators in the above expression must be of the form
\begin{equation}
\label{ZProductConstraint}
 Z\left(\boldsymbol{\xi}^{(z)}\right)= \prod_{v\in V'}\bigotimes_{j\in\mathcal{N}(v)} Z_j,
\end{equation}
\begin{figure}
\centering
\includegraphics[width=0.4\linewidth]{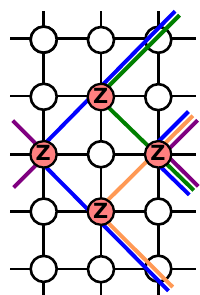}
\caption{A product of $Z$ operators that commutes with all cone symmetries of the $(4^4)$ lattice.  Each circle represents a site in the lattice with the physical index suppressed.  Distinct cone symmetries correspond to lines of a particular color that intersect the sites where the particular symmetry supports an $X$ operator.  Each cone symmetry intersects an even number of $Z$ operators.  Notice this product of $Z$ operators has support on the neighbors of the central site and thus is stabilizer equivalent to a single $X$ operator at that site.}
\label{Commuting_Z}
\end{figure}
where $V^{\prime}$ is a local, bounded-size subset of vertices.  An example of such an operator for the $(4^4)$ lattice is shown in Fig.~\ref{Commuting_Z} where $V'$ is the central qubit in the figure. 

To see that Eq.~(\ref{ZProductConstraint}) implies Eq.~(\ref{Bsym}), recall that our reference state is the graph state, and thus, we may use the stabilizer relation Eq.~(\ref{Stabilizer_Relation}) to write
\begin{equation}
Z \left(\boldsymbol{\xi}^{(z)}\right)|\psi_C\rangle = \prod_{v\in V'} X_v |\psi_C\rangle.
\end{equation}
Therefore, all $Z$ operators in Eq. (\ref{Udef}) can be replaced by $X$ operators.  Hence, Eq. (\ref{Udef}) can be recast as
\begin{equation}
U_\phi = \sum_{\boldsymbol{\zeta}} c_{\boldsymbol{\zeta}} X(\boldsymbol{\zeta}), \label{Usimp}
\end{equation}
where $\boldsymbol{\zeta}\in\mathbb{F}_2^{|V|}$ is the length $|V|$ binary vector ${\boldsymbol{\zeta} = \boldsymbol{\xi}^{(x)} + \boldsymbol{\xi}_{V'}~(\textrm{mod}~ 2)}$ where $\boldsymbol{\xi}_{V'}$ is binary vector with nonzero entries corresponding to vertices in the subset $V'$.  $U_\phi$ has the MPU decomposition
\begin{equation}
\mathcal{B}_{k,\phi} = \sum_{\boldsymbol{\zeta}} c_{\boldsymbol{\zeta}}^{\frac{1}{|V|}} X^{\zeta_k}_k \otimes|\boldsymbol{\zeta}\rangle^{\otimes 4},
\end{equation}
where $\mathcal{B}_{k,\phi}$ is the MPU tensor at the $k^{\textrm{th}}$ site. Hence,  {Eq.~(\ref{Bsym})} holds.

One key takeaway is that, in order for an SPTO phase constructed around a graph state to be a cluster phase, it is sufficient to show that the only products of $Z$ operators in Eq.~(\ref{Udef}) that commute with all subsystem symmetries are of the form of Eq.~(\ref{ZProductConstraint})---see for example Fig.~\ref{Commuting_Z}.

\section{Cluster phases on Archimedean lattices}\label{Main_Sec}

\begin{table*}
\centering
\begin{tabular}{ | | c | c | c | c | c | | }
\hline
Real space & Real space & Virtual space & Computational& Lattices  \\
symmetry & symmetry group & QCA structure & phase & \\
\hline
\hline
$\frac{1}{k}$Fractional & $\mathbb{Z}_2^{k}$ & - & - & All \\
Ribbon & $\mathbb{Z}_2^{2n}$ & Periodic & Yes & Partially decorated \\
Cone & $\mathbb{Z}_2^{2n}$ & Glider & Yes & $(4^4)$, $(3^6)$, $(3,4,6,4)$ \\
Fractal & $\mathbb{Z}_2^{2n}$ & Fractal & Yes & $(6^3)$, $(4,8^2)$, $(4,6,12)$, \\
 & & & & $(3^4,6)$, $(3,4,3^2,4)$, $(3^3,4^2)$ \\
1 - Form & $\mathbb{Z}_2^{\mathcal{O}(nN)}$ & No & No & $(3,6,3,6)$, $(3,12^2)$\\
\hline
\end{tabular}
\caption{\label{SStable} Classification of Archimedean cluster phases according to subsystem symmetry, QCA structure, and computational capability.  To each lattice, we identify the subsystem symmetry its corresponding fixed-point tensor (i.e., graph state) possesses. The three fundamental symmetries giving rise to a universal cluster phase (ribbon, cone, and fractal as defined in Sec.~\ref{Sub_Syms}) always correspond to Clifford QCA (with periodic, glider, and fractal structure, respectively) in the virtual space of the tensor network representation (see Sec.~\ref{QCAandMBQC}).  Notice, the two lattices supporting one-form symmetries do not support a computational phase (see Sec.~\ref{1form}).  Furthermore, none of the Archimedean lattices have ribbon symmetries, however, any of the nine lattices with cone or fractal symmetry can be partially decorated to give a computationally universal phase protected by the ribbon symmetries as described in Sec.~\ref{Periodic}.}
\end{table*}

Our approach is to find the underlying QCA structure and subsystem symmetries for lattices that can be appropriately partitioned into quantum wires for MBQC. We use this procedure to systematically study subsystem SPTO states $|\phi\rangle$'s on the Archimedean lattices. Note, however, since they share the bottom layer of tensors in Eq.~(\ref{Construct_Tensor}) determined by a corresponding graph state, most of the following analysis can be made as if we handled graph states.  We find that nine of these lattices support an underlying QCA structure, two of which were previously studied in Refs.~\cite{Raussendorf2018, Devakul2018}.  We use the subsystem symmetries in each of the nine cases to define a cluster phase and prove universality for MBQC.  Our results are summarized as follows together with Table~\ref{SStable}.
\begin{theorem}
Let $|\phi\rangle$ be any SPTO state in a 2D cluster phase constructed on one of the vertex-translative Archimedean lattices, excluding $(3,6,3,6)$ and $(3,12^2)$, and protected by its fundamental subsystem (i.e., cone or fractal) symmetry. All states $|\phi\rangle$'s in the same phase share an underlying (i.e., glider or fractal) QCA structure respectively, so that they are uniformly universal for MBQC, namely universal quantum computation is feasible under a common protocol of measurements, regardless of microscopic specification of $|\phi\rangle$.
\end{theorem}

As shown in Table \ref{SStable}, the different lattices have different types of symmetries.  We describe the features of lattices with cone symmetries in Sec.~\ref{Cone_Symmetries} by focusing on the $(3,4,6,4)$ lattice.  This is a particularly illuminating example because it reveals the fundamental importance of cone symmetries in defining the cluster phase in comparison to the emphasis on line symmetries in Ref.~\cite{Stephen2019}).  In Sec.~\ref{Fractal_Symmetries} we describe features of lattices with fractal symmetries by studying the $(4,8^2)$ lattice.  Note that none of the Archimedean lattices have an underlying periodic QCA structure. In Sec.~\ref{Periodic}, we describe how to convert lattices possessing either a glider or fractal QCA structure into partially decorated lattices that have a periodic QCA structure.

\subsection{Determining the QCA}\label{Determining_the_QCA}

\begin{figure*}
\includegraphics[width=\linewidth]{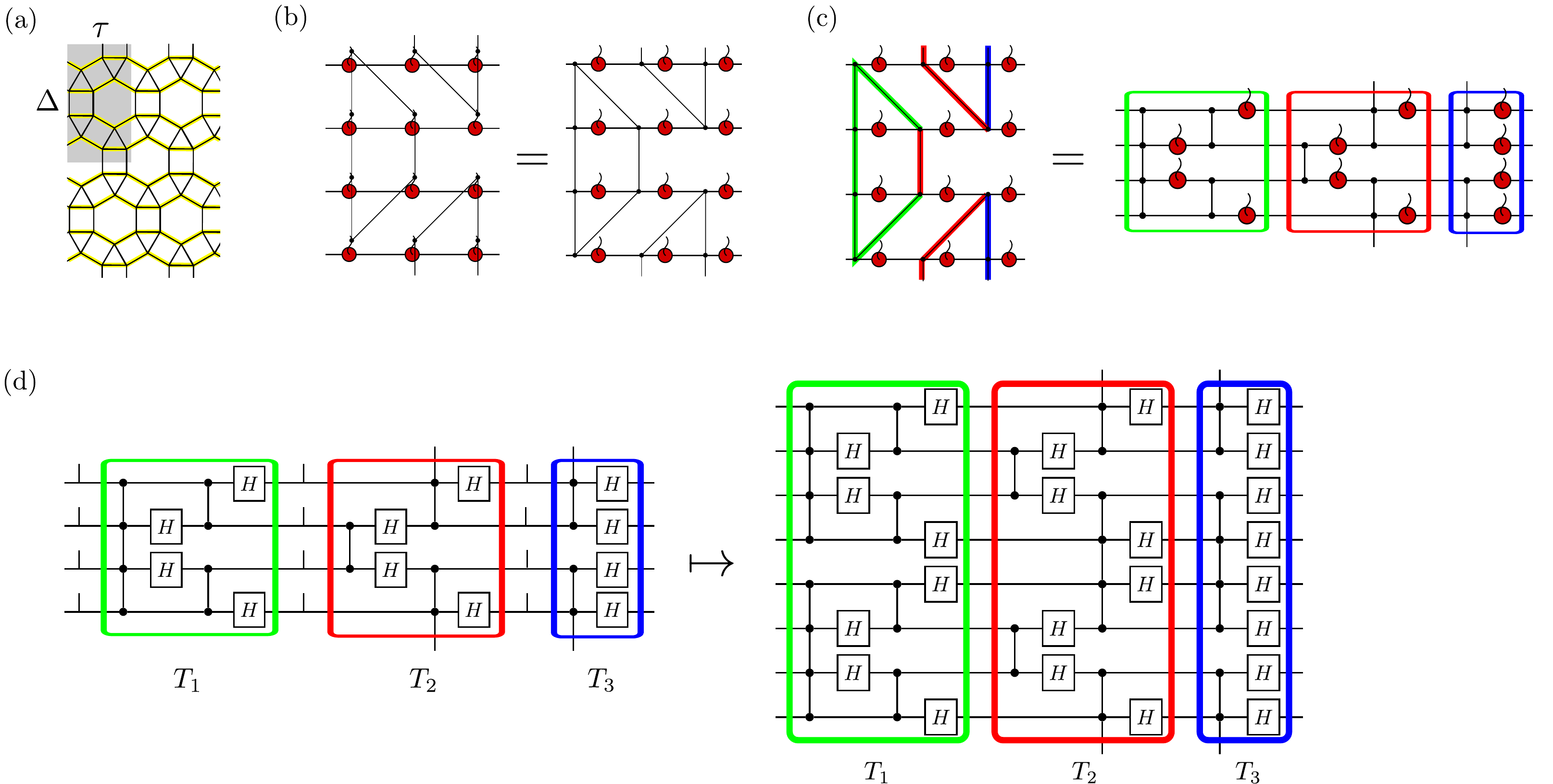}
\caption{An outline of the procedure to construct the underlying QCA of a 2D graph state, focusing on the $(3,4,6,4)$ lattice as an example. \textbf{(a)} Given a lattice, first identify the $\tau\times\Delta$ sized translationally invariant structure, shaded in grey.  Next, partition the lattice into disjoint sets of wires, outlined in yellow. \textbf{(b)} Deform the lattice so that each vertex lies on a square grid.  Each wire may then be replaced by the MPS for a 1D graph state.  All remaining edges become $CZ$ gates coupling the neighboring wires.  These $CZ$ gates can then be pushed down to the virtual level by virtue of Eq.~(\ref{CZ_Symmetry}).  \textbf{(c)} To enforce a proper causal structure, sort the components of the tensor network into common time slices.  $CZ$ gates belonging to a common time slice are shaded in green, red, and blue, respectively.  This results in blocks of tensors that are connected only by wires.  \textbf{(d)} Decompose each MPS tensor into a copy tensor and Hadamard gate using Eq. (\ref{1D_MPS_Tensor}).  Moving all copy tensors to the front of their respective time slice and contracting them with $|0^{(x)}\rangle$ states gives the transfer matrix for the QCA, $T = T_3T_2T_1$.  For all circuit diagrams drawn, $CZ$ gates implemented in a single step are drawn so they overlap.}
\label{torussquare}
\end{figure*}

We first discuss our general method for determining the underlying QCA structure for a given lattice.  The idea is to describe the 2D graph state using several coupled 1D graph states written in MPS form.  The resulting tensor network can then be converted into a quantum circuit describing the QCA.  

Assume the lattice is embedded on a cylinder.  When we do MBQC with a resource state on this lattice, the length and circumference of the cylinder will represent the time and space directions of a (1+1) dimensional quantum circuit, respectively.
Notice, each lattice is invariant under translation by $\Delta$ and $\tau$ sites in the space and time directions, respectively.  For example, for the square lattice $\Delta = \tau = 1$ whereas for the $(3,4,6,4)$ lattice $\tau=3$ and $\Delta=4$ (see Fig.~\ref{Archlat}).  In order to ensure that the periodic boundary conditions in the spatial direction are consistent, the number of sites around the circumference must be $n = j\Delta$ for some $j\in \mathbb{N}$.  Furthermore, denote the length of the cylinder by $N$ where $N>>n$.  The upshot of translational invariance is that the analysis can be reduced to considering a single $\tau \times \Delta$ sized patch of the lattice.  For the $(3,4,6,4)$ lattice, this patch is shown in Fig.~\ref{torussquare}~(a).

For this procedure to succeed in giving a unitary QCA structure on the $n$ encoded qubits at the edge, it is necessary for the lattice to have a partitioning into $n$ induced path graphs---1D linear subgraph that contains all edges connecting its vertices in the original graph---along the time direction such that every qubit in the lattice lies in some partition.
Edges in each path graph make up distinct \emph{wires} and all remaining edges correspond to logical $CZ$ gates between neighboring wires.  These are represented in Fig.~\ref{torussquare} (a) by the yellow shaded edges.  Given such a partitioning, we can deform the lattice so as to straighten out the wires and align the vertices on a square grid as shown in Fig.~\ref{torussquare} (b).  Importantly, the $(3,6,3,6)$ and $(3,12^2)$ lattices fail to meet this condition and consequently have no unitary QCA structure.  We will revisit these two examples in Sec.~\ref{1form}.

We can then describe the remaining nine lattices as $n$ disjoint 1D graph states coupled by logical $CZ$ gates.
By rewriting each 1D graph state in terms of its MPS representation, we obtain a tensor network description of the state, shown in Fig.~\ref{torussquare}~(b).  These MPS tensors are defined as
\begin{equation}
\includegraphics[scale=0.65]{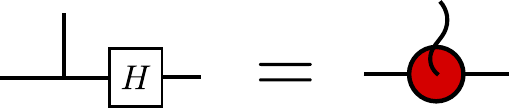},
\label{1D_MPS_Tensor}
\end{equation}
where the appropriate tensor network notational definitions are given in Appendix \ref{TNNappendix}.  
The logical $CZ$ gates coupling the wires can be pushed down to the virtual degrees of freedom via the identity
\begin{equation}
\includegraphics[scale=0.65]{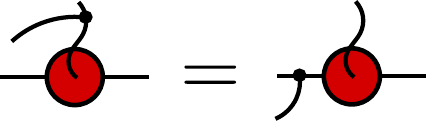},
\label{CZ_Symmetry}
\end{equation}
where the dangling wire represents half of a $CZ$ gate.  This procedure is visually depicted in Fig.~\ref{torussquare}~(b). 

To make the temporal structure of the effective circuit description apparent, we will place each node of the tensor network on a square grid (where the wires correspond to horizontal edges).  Next, we partition the network into common time slices that contain one node on each wire and some additional logical $CZ$ gates.  We can arrange all $CZ$ gates such that neighboring time slices are only connected by wires as shown in Fig.~\ref{torussquare}~(c).

Finally, we may use Eq.~(\ref{1D_MPS_Tensor}) to decompose each node into a copy tensor and Hadamard.  Each time slice can then be turned into a Clifford circuit by moving each copy tensor to the front of the time slice and contracting each with a $|0^{(x)}\rangle$ state as shown in Fig.~\ref{torussquare}~(d).  Since a QCA is time translationally invariant by definition, we should compose $\tau$ many of the Clifford circuits, given by unitaries $T_1,...,T_\tau$, to get the time translationally invariant transfer operator for the QCA, $T=T_\tau \cdot\cdot\cdot T_1$.

We note that the above procedure does not guarantee a unitary QCA structure.  Namely, the Clifford circuits, $T_j$, may not have a valid causal ordering.  This property is dependent on the initial embedding of the lattice on the cylinder, and is explored more in Sec.~\ref{Foliation}.

\subsection{Determining the symmetry}\label{Determining_Symmetry}

\begin{figure}
\includegraphics[width=\linewidth]{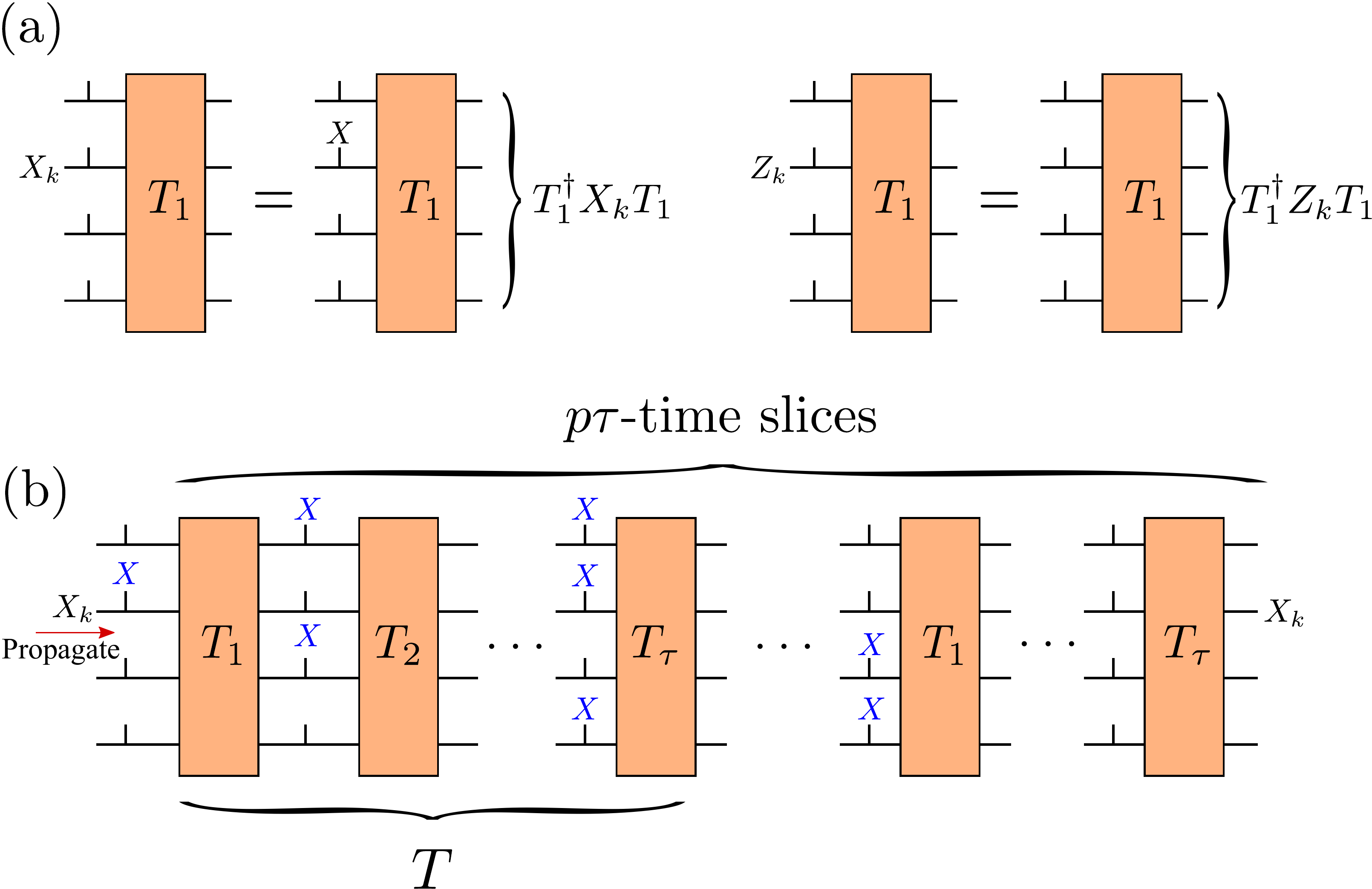}
\caption{Determining subsystem symmetries via the QCA.  \textbf{(a)} A site dependent update rule for evolving Pauli operators through the virtual space of the tensor network.  Propagating an $X$ through any virtual degree of freedom always leaves behind a physical $X$ operator on the corresponding physical site whereas $Z$ operators leave behind no physical operators.  \textbf{(b)} Propagating each of the $2n$ generators of the Pauli group through $p\tau$ time slices on the virtual level maps each back to itself, leaving behind a subsystem symmetry on the physical degrees of freedom shown in blue (color available online).  Treating the $p\tau$ time slices as one large block of sites in a quasi-1D system, each subsystem symmetry generated gives an onsite representation of $\mathbb{Z}_2^{2n}$.}
\label{Det_Sym}
\end{figure}

The subsystem symmetries can be determined by the commutation relations of Pauli operators with the tensors in each time slice.
One can see from the structure of each 1D MPS tensor shown in Eq.~(\ref{1D_MPS_Tensor}) that commuting an $X$ on the $k^\textrm{th}$ virtual wire through the collection of tensors at a given time slice results in an $X$ operator appearing on the $k^{\textrm{th}}$ physical index as shown in Fig.~\ref{Det_Sym}~(a).  Using this fact we may write down an explicit expression for the symmetry in terms of the QCA.

For $\alpha,\beta\in\mathbb{Z}$ and $\gamma\in\mathbb{Z}_\tau$ such that $\alpha = \beta \tau + \gamma$, let us define an accumulated transfer matrix, 
\begin{equation}
T^{[\alpha]} = T_\gamma \cdot\cdot\cdot T_2 T_1 \left( T \right)^\beta.
\end{equation}
This is the unitary accumulated after evolving through the QCA by $\alpha$ elementary time steps.  Let the same notation hold for the binary representation of these Clifford unitaries $\mathcal{T}_1,...,\mathcal{T}_\tau$.  Furthermore, let $\mathbf{e}_l\in\mathbb{F}_2^{2n}$ be the binary representation of a generator of the Pauli group acting on a virtual degree of freedom.  Namely,
\begin{equation}
P(\mathbf{e}_l) =
\begin{cases}
X_l ~&\textrm{if}~1\leq l\leq n\\
Z_{l-n} ~&\textrm{if}~n+1 \leq l\leq 2n
\end{cases}.
\end{equation}
Now suppose the coordinates of a physical site on the lattice embedded on the square grid are parameterized as $(x,y)$ where $y$ increases from top to bottom along the grid. If we evolve the $l^\textrm{th}$ such single site Pauli operator through $x-1$ elementary time steps,
then the components of the updated vector $\mathcal{T}^{[x-1]}\mathbf{e}_l$ specify the support of the evolved Pauli operator.  Namely, if the $y^{\textrm{th}}$ component of this vector is 1, then the $y^{\textrm{th}}$ virtual wire in the tensor network will support an $X$ operator.  Consequentially, this operator will be pushed up to the $(x,y)$ physical degree of freedom as described in Fig.~\ref{Det_Sym}~(a).  In summary, the $y^{\textrm{th}}$ component of the vector $\mathcal{T}^{[x-1]}\mathbf{e}_l$ specifies whether or not the $l^{\textrm{th}}$ symmetry generator has non-trivial support on site $(x,y)$.

Iterating this procedure generates the subsystem symmetry as shown in Fig.~\ref{Det_Sym} (b).  Therefore, we may express $l^\textrm{th}$ symmetry generator as
\begin{equation}
S_l = \bigotimes_{x,y} X_{x,y}^{(\mathcal{T}^{[x-1]}\mathbf{e}_l)_{y}} ,
\end{equation}
where the superscript simply denotes raising to the power of the binary variable ${(\mathcal{T}^{[x-1]}\mathbf{e}_l)_{y}}$.

\subsection{Computational universality}\label{Computational_Universality}

\begin{figure}[h]
\includegraphics[width=\linewidth]{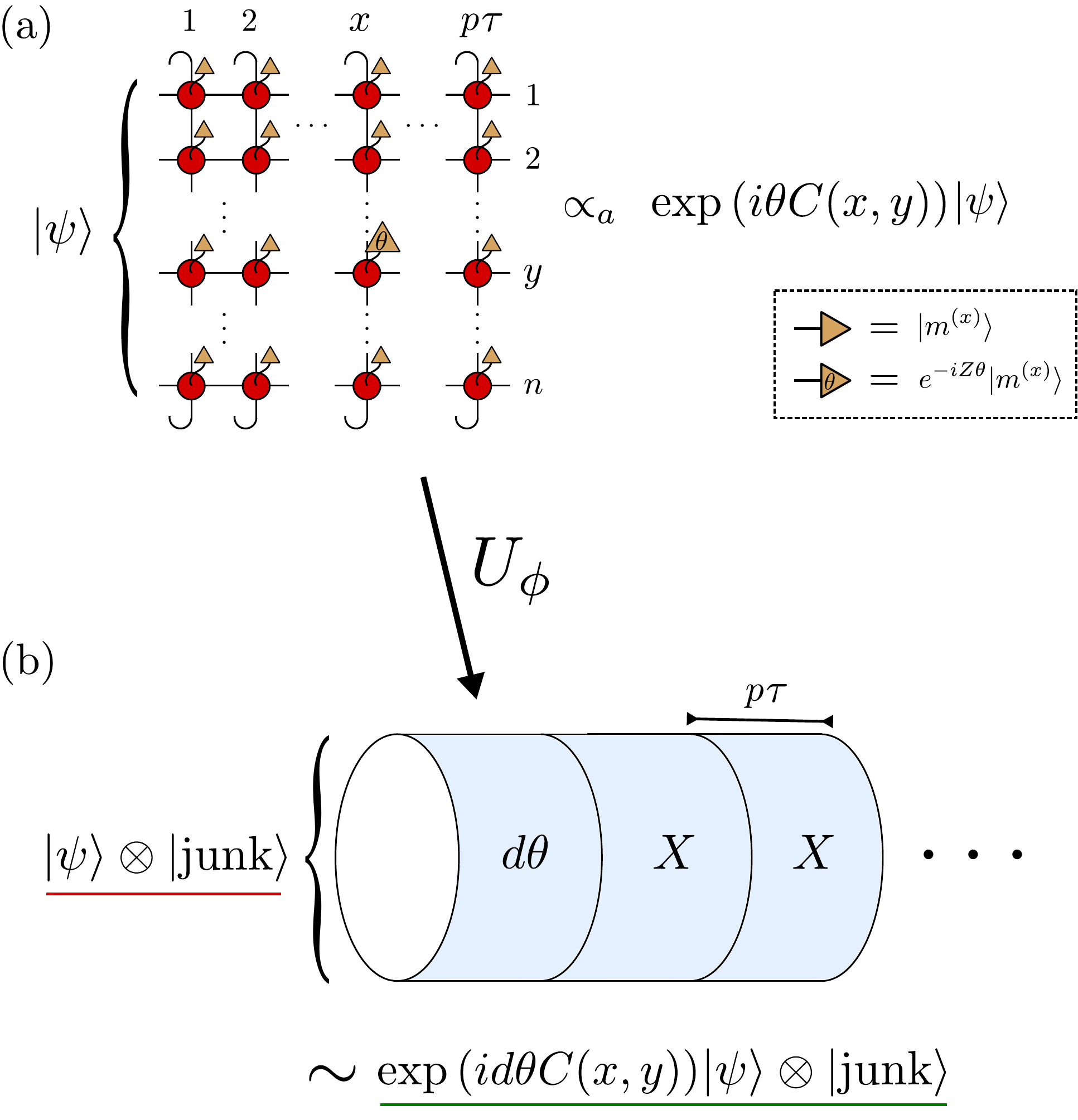}
\caption{MBQC scheme at the fixed point versus MBQC at arbitrary points in the cluster phase. \textbf{(a)} MBQC with the quasi-1D fixed point state can be achieved by measuring all qubits in the $X$ basis except for the qubit at site $(x,y)$, which is measured in the $X_\theta$ basis.  This implements a unitary generated by the MPS tensor component $C(x,y)$ by an angle $\theta$. \textbf{(b)} At an arbitrary point in the quasi-1D SPTO phase, the same procedure inevitably couples the logical and junk subsystems.  To avoid this, the input (red) undergoes a small rotation by an angle $d\theta$, followed by a segment of ``oblivious wire", in which many blocks are measured in the $X$ basis and the outcome is discarded after undoing the measurement byproduct operators.  To linear order in $d\theta$, this results in a rotated state at the output (green) while keeping the logical and junk subsystems decoupled.}
\label{Computation}
\end{figure}

A key component of our main result is that any state in the cluster phase constructed about each of the nine Archimedean lattices with a QCA structure is universal for MBQC.  To prove this we determine the universal gate set available in each case.  Once we have this, we may appeal to the techniques of Ref.~\cite{Stephen2017} for the remaining details of a computational protocol.  For completeness, these techniques are reviewed in the context of quasi-1D SPTO phases in Appendix~\ref{MBQCSSPT}.

First we introduce relevant notation.  Recall that the state consists of $n$ wires and the period of the subsystem symmetry and QCA is denoted as $p$.  Let $|\mathbf{j}\rangle$ with ${\mathbf{j}\in\mathbb{Z}_2^{np\tau}}$ represent the state of the $np\tau$ qubits in one QCA period of the tensor network.  We shall index the elements of the vector by the $x$ and $y$ coordinates of each qubit in this block, assigning the state $|0^{(x)}\rangle, |1^{(x)}\rangle$ to the qubit at site $(x,y)$ whenever the corresponding component ${\mathbf{j}_{(x,y)}=0,1}$, respectively.  Finally, let $\mathbf{e}_{(x,y)}$ denote the unit vector with all entries being 0 except that associated with site $(x,y)$.

The available gate set is determined by the fixed point tensors making up the quasi-1D MPS description of the SPTO state $|\phi\rangle = U_{\phi}|\psi_C\rangle$. The quasi-1D MPS description is obtained by contracting a $n\times p\tau$ sized block of the tensors $\mathcal{A}_\phi^{(\textrm{local})}$ defined in Eq.~(\ref{Phase_Tensor}) around a cylinder. The resulting local tensors, denoted as $\mathcal{A}_\phi$, take the form of MPS tensors,
\begin{equation}
\label{Cluster_Phase_Tensor}
\mathcal{A}_\phi = \sum_{\mathbf{j}\in\mathbb{Z}_2^{np\tau}} C(\mathbf{j}) \otimes B(\mathbf{j})~|\mathbf{j}\rangle,
\end{equation}
where $C(\mathbf{j})$ are the logical tensors coming from the graph state fixed point and $B(\mathbf{j})$ are the junk tensors coming from the symmetric constant-depth unitary $U_\phi$. In Ref.~\cite{else2012symmetry}, it was shown that the fixed point tensors can be uniquely determined by the onsite representation of the symmetry and corresponding edge operators in the projective representation of the symmetry.  The structure of the fixed point tensors for each lattice is explicitly derived in Appendix~\ref{Proofofphase}.  

To determine the gate set, we need only consider the tensors $C(x,y) := C(\mathbf{e}_{(x,y)})$.  The gate set native to the cluster phase is,
\begin{equation}
\left\{ U_{x,y}(\theta) = \textrm{exp}\left( i\theta C(x,y) \right) |~\forall(x,y) \right\}.
\end{equation}
To implement such gates physically, we measure the qubit at site $(x,y)$ in the $X_{\theta}$-basis and all others in the $np\tau$ sized block in the $X$-basis. This will, up to adaptive corrections of measurement byproduct operators, implement the desired gate.  An illustration of this is shown in Fig.~\ref{Computation}.  At arbitrary points in the cluster phase, the edge state is made up of a logical and junk subsystem.  In order to avoid losing logical information to the junk system, the qubit at site $(x,y)$ is measured in the $X_{d\theta}$-basis for small $d\theta$.  To build up to a substantial angle $\theta$, we repeat this many times, interleaving each iteration with a large number of blocks measured entirely in the $X$-basis.  In this way we have to break the symmetry gradually.  The protocol is discussed in more detail in Appendix \ref{MBQCSSPT}.

\subsection{Lattices with cone symmetries}\label{Cone_Symmetries}

\begin{table*}
\centering
\includegraphics[width=0.9\linewidth]{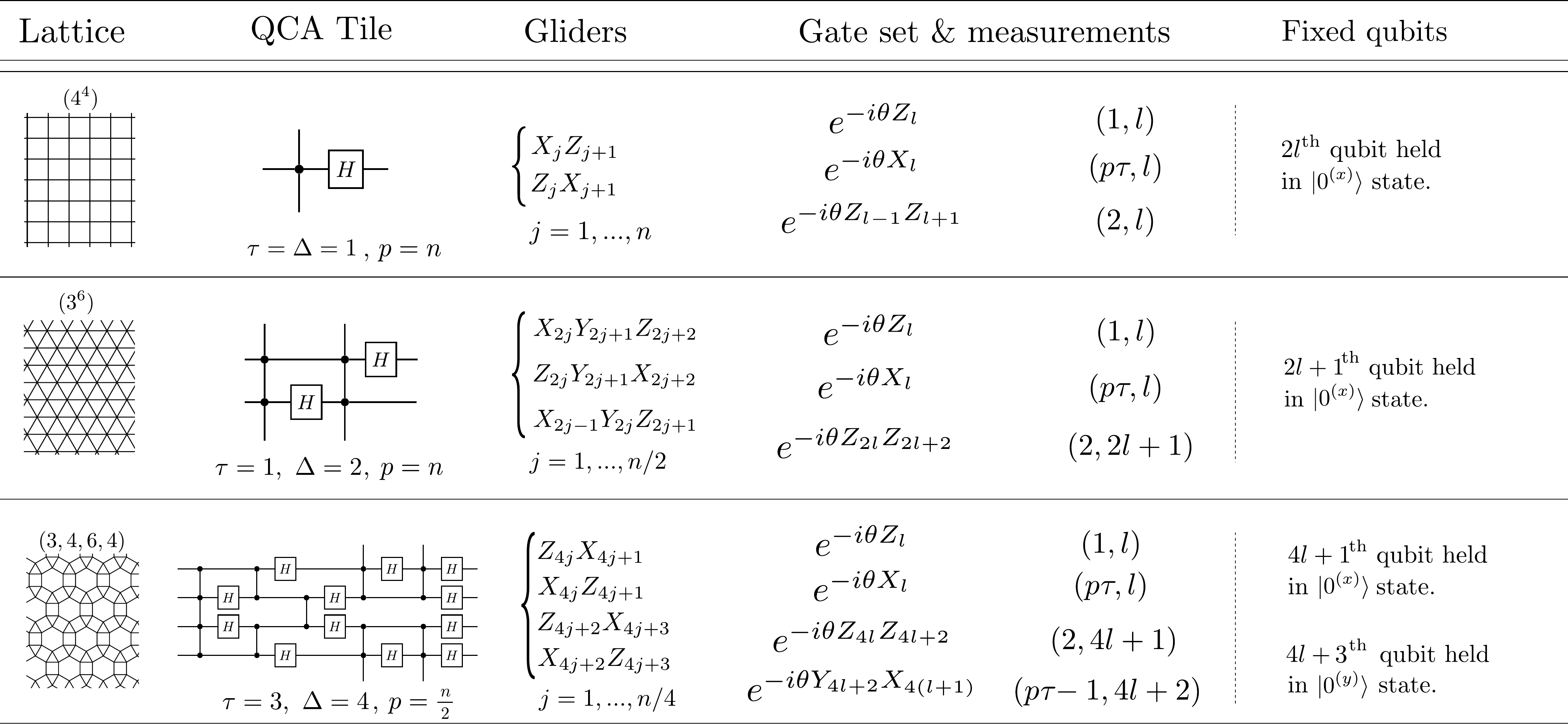}
\caption{The three lattices supporting glider QCA are shown.  By tiling space with each lattice's QCA tile, the QCA for that lattice is obtained.  Also listed are the gliders and gate sets used to prove computational universality of each lattice.  To implement each gate, every qubit in a $n\times p\tau$ sized block must be measured in the $X$ basis except for one qubit located at site $(x,y)$, which is measured in the $X_{d\theta}$ basis. Furthermore, to achieve the gate set we must restrict some qubits to be always fixed in a specific state to get the appropriate two body interaction desired.}
\label{Cone_Sym_Table}
\end{table*}

In this section, we discuss the computational capability of cluster phases constructed around Archimedean lattices with physical cone symmetries and underlying glider QCA.  These are the $(4^4)$, $(3^6)$, and $(3,4,6,4)$ lattices.  
Furthermore, we emphasize the fundamental role of cone symmetries in constructing the phase (c.f. the use of line symmetries in Ref. \cite{Stephen2019}).  The resulting properties of each lattice are summarized in Table~\ref{Cone_Sym_Table}.

 \begin{figure}
 \includegraphics[width=\linewidth]{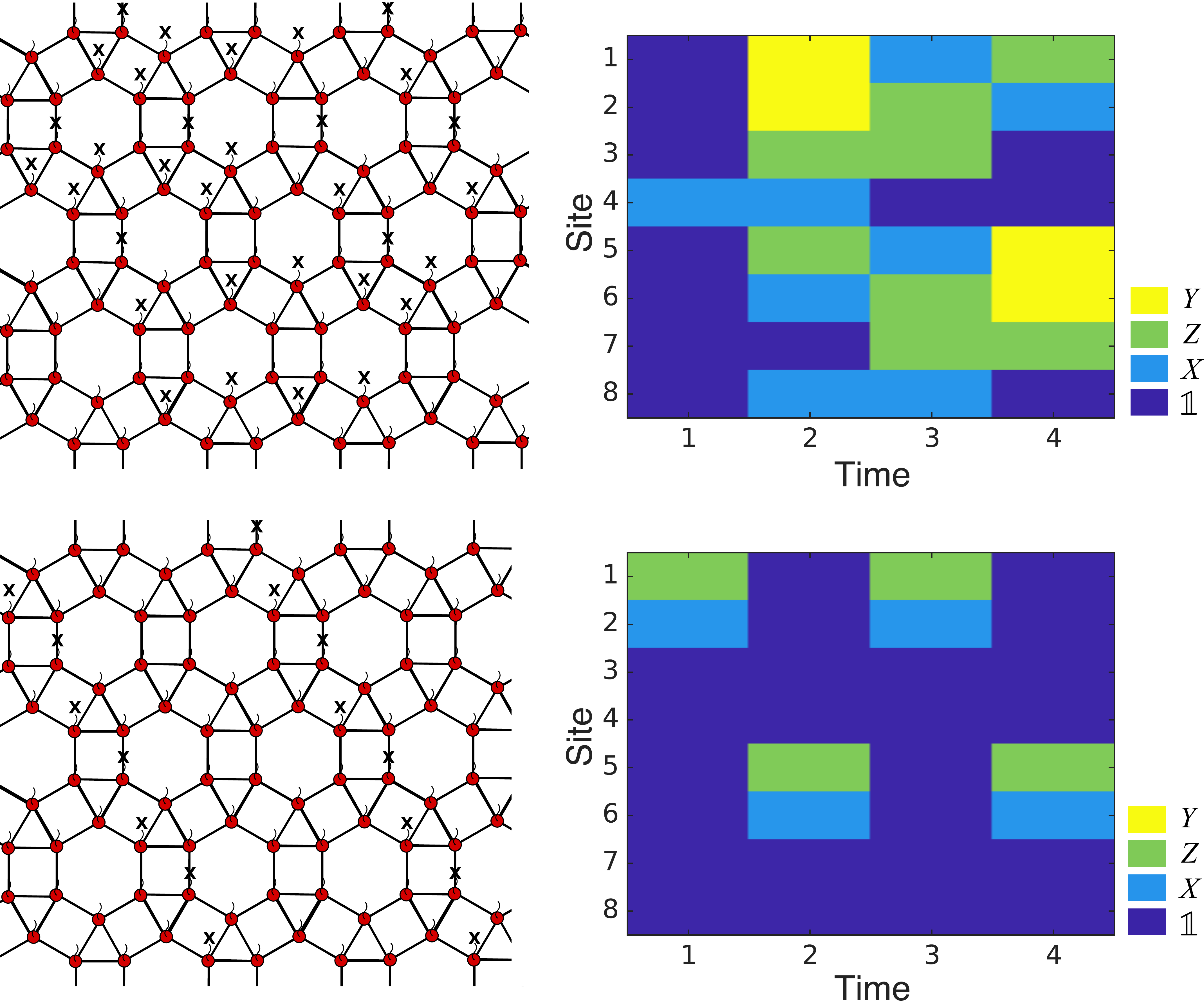}
 \caption{Top: An example of a cone symmetry and the corresponding QCA evolution for the $(3,4,6,4)$ lattice. Bottom: An example line symmetry and corresponding glider in the QCA.}
 \label{3_4_6_4_conesym}
 \end{figure}
 
 \begin{figure*}
 \centering
\includegraphics[width = 0.8\linewidth]{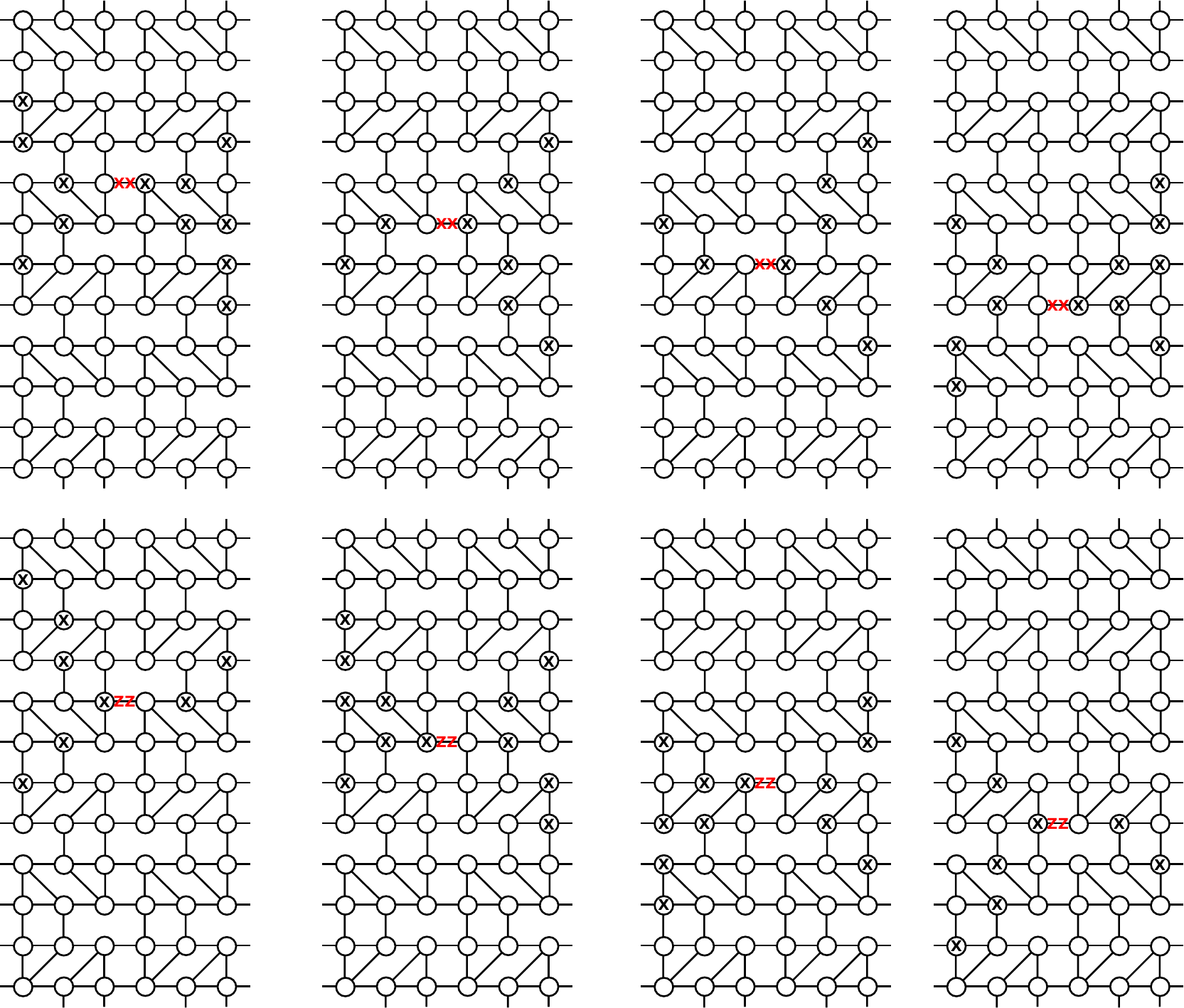}
\caption{Generators for the group of cone symmetries of the (3,4,6,4) lattice.  Shown is a tensor network description of the (3,4,6,4) lattice with the physical indices suppressed.  The $X$ operators in each white circle act on the physical degrees of freedom and give the generators for the cone symmetry group of the $(3,4,6,4)$ lattice up to vertical translation by $\Delta$ sites.  Shown at a particular internal edge in the middle of the lattice are two red Pauli operators in the virtual space of the tensor network. The symmetry on the left (right) half of the lattice can be generated by propagating the left (right) red Pauli in the left (right) direction using the rules of Fig.~\ref{Det_Sym}.  Notice, however, the red Pauli's cancel, indicating the resulting black $X$ operators represent a symmetry of the state.}
\label{3464}
\end{figure*}

\begin{figure*}
\centering
\includegraphics[width=.8\linewidth]{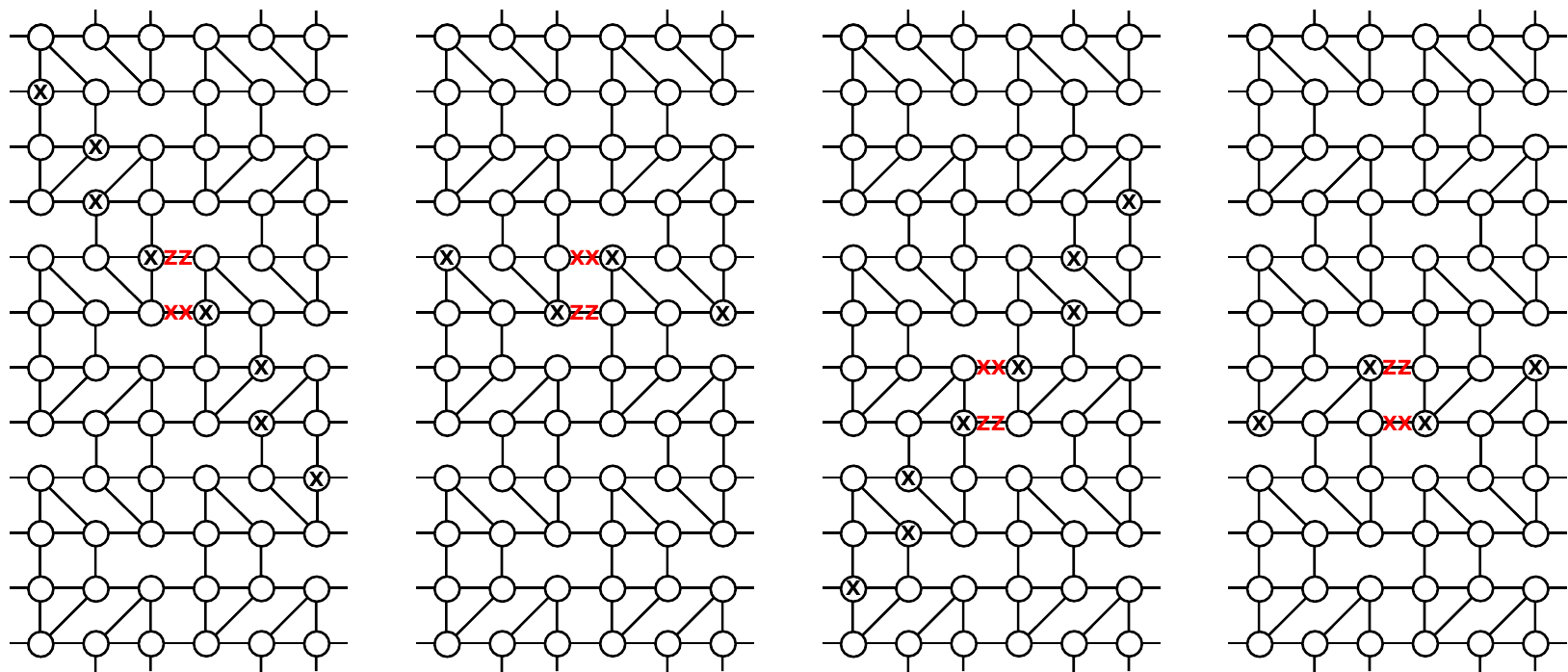}
\caption{Generators for the group of line symmetries of the $(3,4,6,4)$ lattice. All conventions are the same as in Fig.~\ref{3464}.  These physical symmetries on the left (right) half of the lattice are generated by commuting pairs of Pauli operators to the left (right).  Notice each vertical pair of Pauli's is a glider of the corresponding QCA.}
\label{3464lines}
\end{figure*}

The defining property of glider QCA is the existence of \emph{gliders}, which are operators whose support is simply shifted by $\Delta$ sites under the action of $T$.  On the physical space gliders correspond to subsystem symmetries called \emph{line symmetries}, which are composed of 1D strings of X operators .  In Sec.~4.2 of Ref. \cite{Stephen2019}, the line symmetries, which are a subgroup of the group of cone symmetries, were conjectured to protect cluster phases with underlying glider QCA at the fixed point.  We will show the $(3,4,6,4)$ lattice is a counterexample to this conjecture for the following reason; the line symmetry group is too small.  The implications of this are twofold.  First, the line symmetry group forms a $\mathbb{Z}_2^n$ subgroup of the cone symmetry group and thus has a much larger commutant that restricts the construction of a cluster phase based on these symmetries.  Furthermore, the support of each line symmetry is disjoint and so the set of logical tensors $C(x,y)$ cannot generate entangling gates when exponentiated.  Thus, the available gate set throughout the SPTO phase is not universal.  For comparison, we will also discuss in parallel the line symmetries for the $(4^4)$ and $(3^6)$ lattices, which turn out to be sufficient for defining a computationally universal cluster phase in those cases.
 
 Let us first understand the subgroup structure of the line symmetries. Using the techniques of Sec.~\ref{Determining_Symmetry}, we can determine the $2n$ generators of the group of cone symmetries for the $(3,4,6,4)$ lattice. See Fig.~\ref{3_4_6_4_conesym} for an illustration.  The injectivity of the map from virtual to physical space ensures that the group of cone symmetries is isomorphic to the Pauli group on $n$ qubits modulo phases, i.e.,
 \begin{equation}
 \mathcal{S}_{\textrm{cone}} \cong \left\langle\{ X_j, Z_j\}_{j=1}^n\right\rangle / U_4 \cong \left(\mathbb{Z}_2\times\mathbb{Z}_2\right)^{n},
 \label{conegroup}
 \end{equation}
where $U_4 = \{ 1,i,-1,-i \}$ denotes the fourth roots of unity. The relation between generators of the Pauli group and generators of the group of cone symmetries is depicted in Fig.~\ref{3464}.  On the other hand, the gliders are constructed from the following subset of Pauli operators
\begin{equation}
\label{Gamma}
\Gamma = \{Z_{4l}X_{4l+1}, X_{4l}Z_{4l+1}, Z_{4l+2}X_{4l+3}, X_{4l+2}Z_{4l+3}\}_{l=1}^{n/4}.
\end{equation}
The group of line symmetries is then isomorphic to
\begin{equation}
\mathcal{S}_{\textrm{line}} \cong \left\langle\Gamma \right\rangle/U_4 \cong \mathbb{Z}_2^n.
\end{equation}
The line symmetries are a subgroup of the cone symmetries because the generators $\Gamma$ form a subset of the generators of the cone symmetries in Eq.~(\ref{conegroup}).  The gliders for the $(4^4)$ and $(3^6)$ lattices are given in Table~\ref{Cone_Sym_Table}.  One Pauli operator from each of these sets is not independent, indicating that the structure of the line symmetry subgroup in each case is of the form $\mathbb{Z}_2^{2n-1}$.

We now wish to construct a cluster phase about the $(3,4,6,4)$ graph state fixed point.  The objective is to determine which symmetries give rise to locally acting symmetric unitaries that leave invariant the correspondence between the physical symmetries and underlying QCA structure.  As discussed in Sec.~\ref{SPTOreview}, this boils down to determining which products of $Z$ operators commute with all the symmetries in question.

Let us first attempt to construct a cluster phase protected by the line symmetries of the $(3,4,6,4)$ graph state.  The generators of this symmetry group along with the corresponding edge operators, up to vertical translation of their support, are shown in Fig.~\ref{3464lines}.  The simplest local product of $Z$ operators commuting with all symmetries is a product of two $Z$ operators supported on opposite corners of any four sided tile and also $\prod_{j\in\mathcal{N}(v)} Z_j$ for any vertex $v$.  We stress that the former is not stabilizer equivalent to some product of $X$ operators.  Furthermore, if this term is included in the Pauli expansion of the symmetric constant-depth unitary in Eq.~(\ref{Udef}), the local tensors $\mathcal{B}_\phi$ will not commute with the local action of the symmetry (recall the condition in Eq.~(\ref{Bsym})).  The local correspondence between QCA evolution and subsystem symmetries is lost and thus the resulting SPTO phase defined by the line symmetries is not a cluster phase. On the other hand, the commutant of the line symmetries of the $(4^4)$ and $(3^6)$ lattices consists of $Z$ operators of the form $\prod_{j\in\mathcal{N}(v)} Z_j$ and a pair of non-local $Z$ operators separated half way around the torus from each other. The latter operators, referred to as \emph{two-local} operators in Ref.~\cite{Raussendorf2018}, are omitted successfully from $U_\phi$ by an extra consideration that global operators cannot be implemented by $U_\phi$.  Notice that the key point in this argument is that the line symmetry group of the $(3,4,6,4)$ lattice is simply too small to allow one to define a cluster phase.

The line symmetries also restrict the available gate set from being universal in the case of the $(3,4,6,4)$ lattice.  This is apparent upon computing the tensors $C(x,y)$ defined by the line symmetries.  From Fig.~\ref{3464lines} it can be seen that the line symmetries have disjoint support.  This means that for any qubit at some site $(x,y)$ in the $n\times p\tau$ sized block of qubits, $C(x,y)$ will be an operator that anticommutes with only one operator from the set of gliders $\Gamma$ as defined in Eq.~(\ref{Gamma}).  Thus the available gate set cannot generate entanglement between encoded qubits at the edge in different blocks of size $\Delta=4$.  Again, we emphasize that for the $(4^4)$ and $(3^6)$ lattices, the available gate set defined by the line symmetries of each lattice can indeed be used to construct a universal gate set on a restricted subset of encoded qubits at the edge.  Namely, on the even or odd qubits.

Understanding that the line symmetries of the $(3,4,6,4)$ lattice fail to give a universal cluster phase, we now consider the full group of cone symmetries.  The generators of $\mathcal{S}_{\textrm{cone}}$ are depicted in Fig.~\ref{3464}.  Each symmetry operator has support on even number of neighbors of any vertex in the lattice.  The only place where this may not be true is near the boundary of the region depicted. However, evolving the edge operators through the QCA gives a new edge operator, which is some product of Pauli operators.  Thus, near the edge of the region shown the symmetry simply looks like a product of several generators.  If an operator commutes with all symmetries in the vicinity of the center of the region shown, it is guaranteed to commute with the symmetry everywhere.  The local operator commuting with all these symmetries is $\prod_{j\in\mathcal{N}(v)} Z_j$ which by the stabilizer relation is equivalent to $X_v$.  Therefore, we can construct a cluster-like SPT phase around the $(3,4,6,4)$ graph state defined by the cone symmetries.

We finally show that every point in the cluster phase constructed about the (3,4,6,4) lattice is universal for MBQC.  Since the quasi-1D MPS tensors are formed from blocks of size $n \times p\tau$ (i.e. a whole QCA cycle), we can perform identity gates and implement a segment of oblivious wire by measuring all qubits in the $X$ basis (Note that $p{\tau} = \frac{3n}{2}$ for the $(3,4,6,4)$ lattice).  Furthermore, preparation and readout can be performed by measuring the first column of qubits in a block in the $Z$ basis and measuring the remaining qubits in the $X$ basis.  Finally, we determine the relevant tensors for implementing a universal gate set to be,
\begin{eqnarray}
C(1,l) &=& Z_l \label{3464_Z}\\
C(p {\tau},l) &=& X_l \label{3464_X}\\
C(2,4l+1) &=& Z_{4l}X_{4l+1} Z_{4l+2} \label{3464_ZXZ}\\
C(p {\tau}-1,4l+2) &=& Y_{4l+2} Y_{4l+3} X_{4(l+1)}. \label{3464_YYX}
\end{eqnarray}
These tensor components were determined using the subsystem symmetry generators shown in Fig.~\ref{3464}.  The fact that each symmetry is diagonal in the $X$-basis and that each symmetry can be pushed through to the virtual level gives a set of commutation relations of each $C(x,y)$ with the edge representation of each symmetry (i.e. each single site Pauli operator on the virtual degrees of freedom).  This uniquely determines $C(x,y)$.  For more information we point the reader to Appendix~\ref{MBQCSSPT_1} where Eq.~(\ref{3464_ZXZ}) is derived explicitly as an example (see also Fig.~\ref{3464_Derive}).


To achieve universality we must fix the $4l+1^{\textrm{th}}$ and $4l+3^{\textrm{th}}$ qubits to be in the +1 eigenstates of $X$ and $Y$ respectively.  This procedure of fixing qubits to be in certain Pauli eigenstates can be done deterministically.  For more details we point the reader to Appendix~\ref{MBQCSSPT_5}.  The accessible universal gate set is given in Table~\ref{Cone_Sym_Table}. Similar results for the $(4^4)$ and $(3^6)$ lattices are worked out in detail in Appendix~\ref{Proofofphase}.

\subsection{Lattices with fractal symmetries}\label{Fractal_Symmetries}

\begin{table*}
\includegraphics[width=\linewidth]{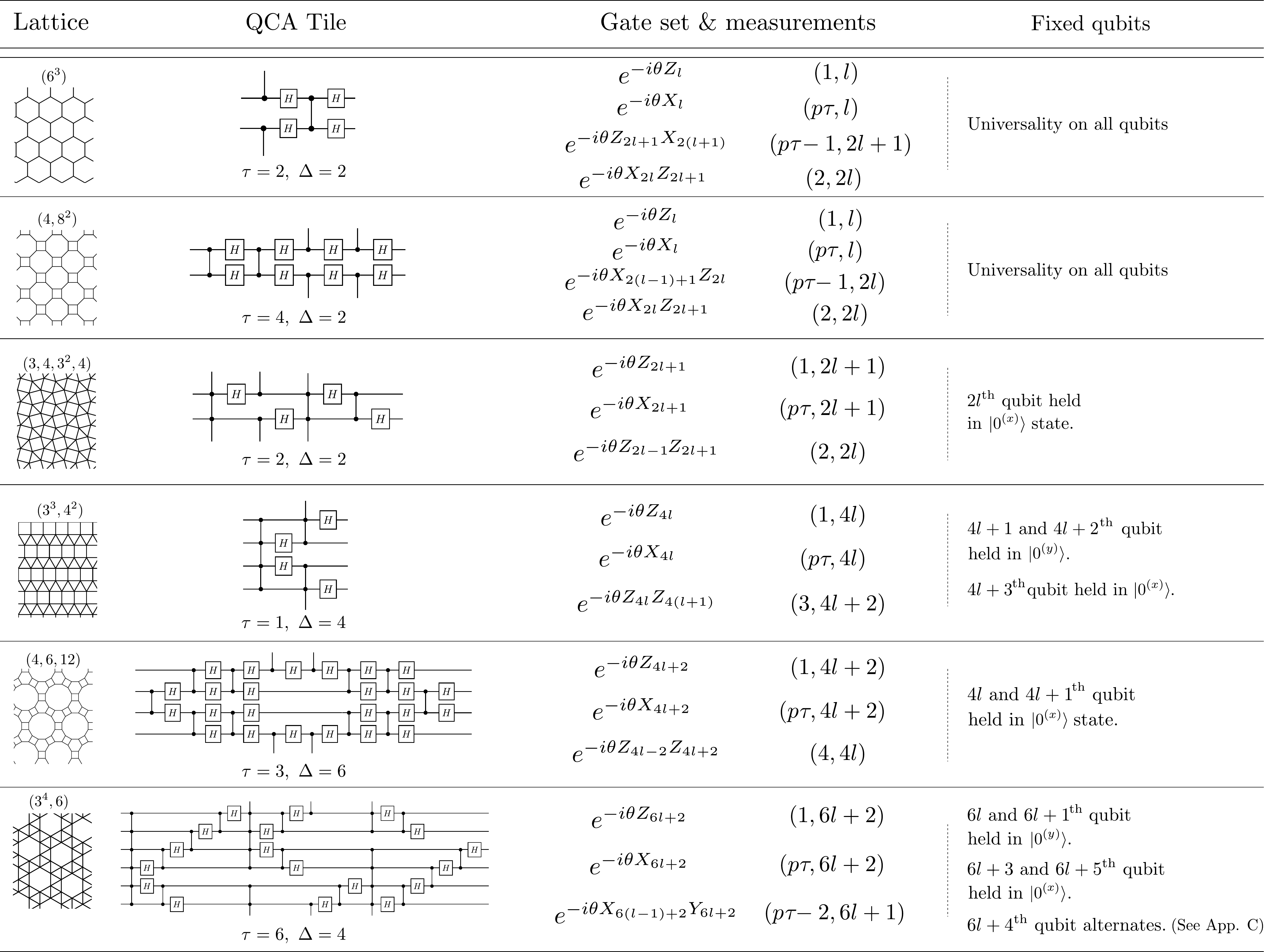}
\caption{The six lattices supporting fractal QCA are shown.  By tiling space with each lattice's QCA tile, the QCA for that lattice is obtained.  Also listed are the gate sets used to prove computational universality of each lattice.  To implement each gate, every qubit in a $n\times p {\tau}$ sized block must be measured in the $X$ basis except for one qubit located at site $(x,y)$, which is measured in the $X_{d\theta}$ basis. Furthermore, to achieve the gate set we must restrict some qubits to be always fixed in a specific state to get the appropriate two body interaction desired.}
\label{Fractal_Table}
\end{table*}

In this section, we study Archimedean lattices supporting fractal subsystem symmetries and underlying fractal QCA.  Six of the eleven Archimedean lattices have this property.  We confirm that in each case a computationally universal cluster phase protected by fractal subsystem symmetries can be constructed.  As an example, we will study the $(4,8^2)$ lattice in detail. Apart from being a new example of a lattice supporting a cluster phase protected by fractal subsystem symmetries, it has the added benefit of achieving universality on all $n$ qubits encoded at the edge.  We remark that the $(6^3)$ lattice also shares this property (c.f. the two site construction of \cite{Devakul2018, Stephen2019}).  Details for the other five lattices are worked out in Appendix~\ref{Proofofphase} and are listed in Table~\ref{Fractal_Table}.

\begin{figure}
\includegraphics[width= \linewidth]{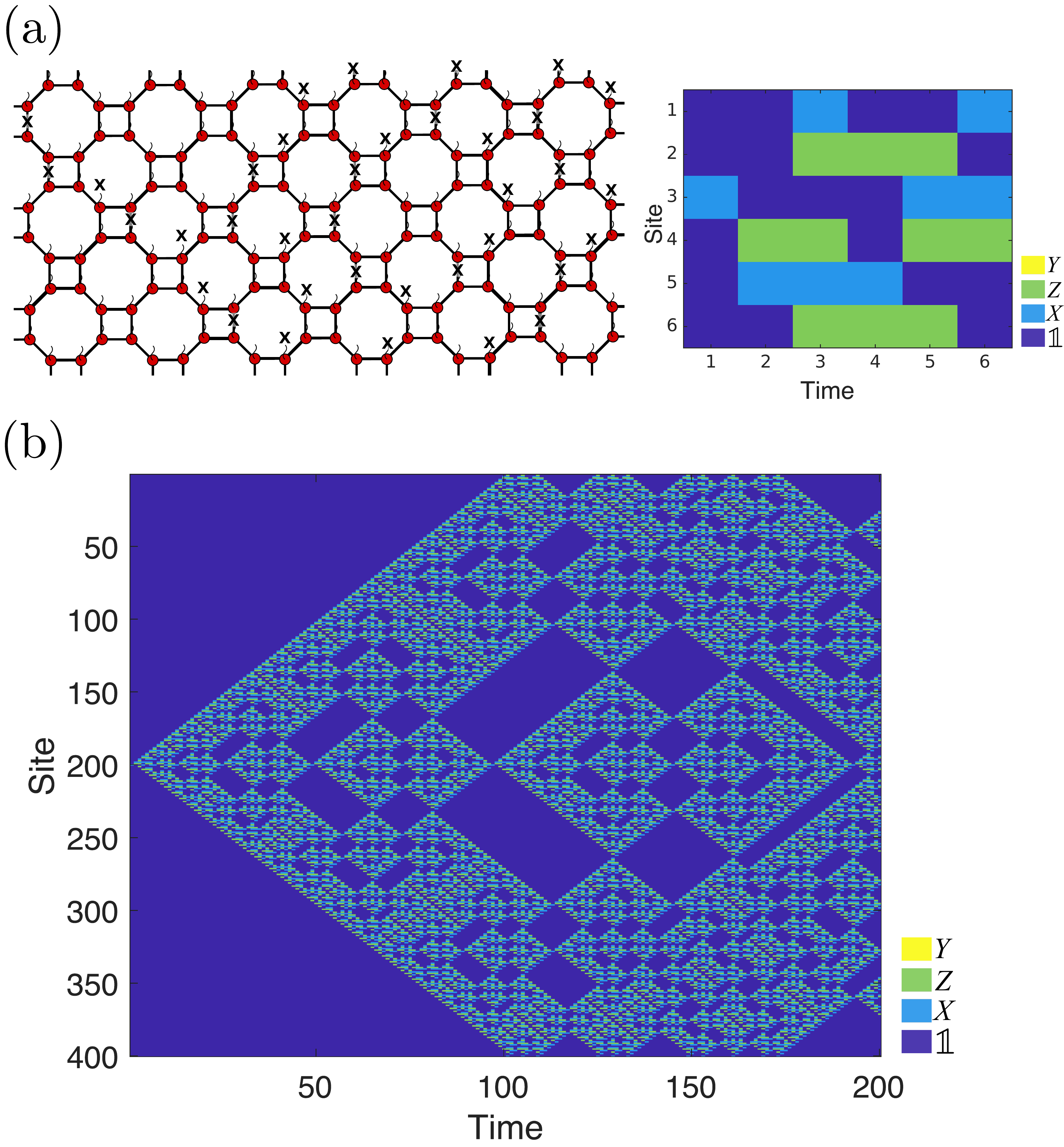}
\caption{Finite size fractal symmetry and corresponding QCA evolution for the $(4,8^2)$ lattice.  The self similar structure of the fractal QCA becomes apparent for large $n$, the circumference of the cylinder.}
\label{488cone_QCA}
\end{figure}

To study the $(4,8^2)$ lattice in detail, we must first determine the underlying QCA structure.  The translational invariance parameters for the $(4,8^2)$ lattice are $\Delta =2$ and $\tau = 4$.  We then use this information to construct a translationally invariant block of tensors for the tensor network description of the $(4,8^2)$ graph state.  The resulting Clifford circuit defining the QCA can easily be obtained from this and is given in Table~\ref{Fractal_Table}.  Simulating the evolution of Pauli operators through the circuit, we get fractal subsystem symmetries as depicted in Fig.~\ref{488cone_QCA}.  For the same reason discussed before, these symmetries again form a representation of $\mathbb{Z}_2^{2n}$.

\begin{figure}
\includegraphics[width=\linewidth]{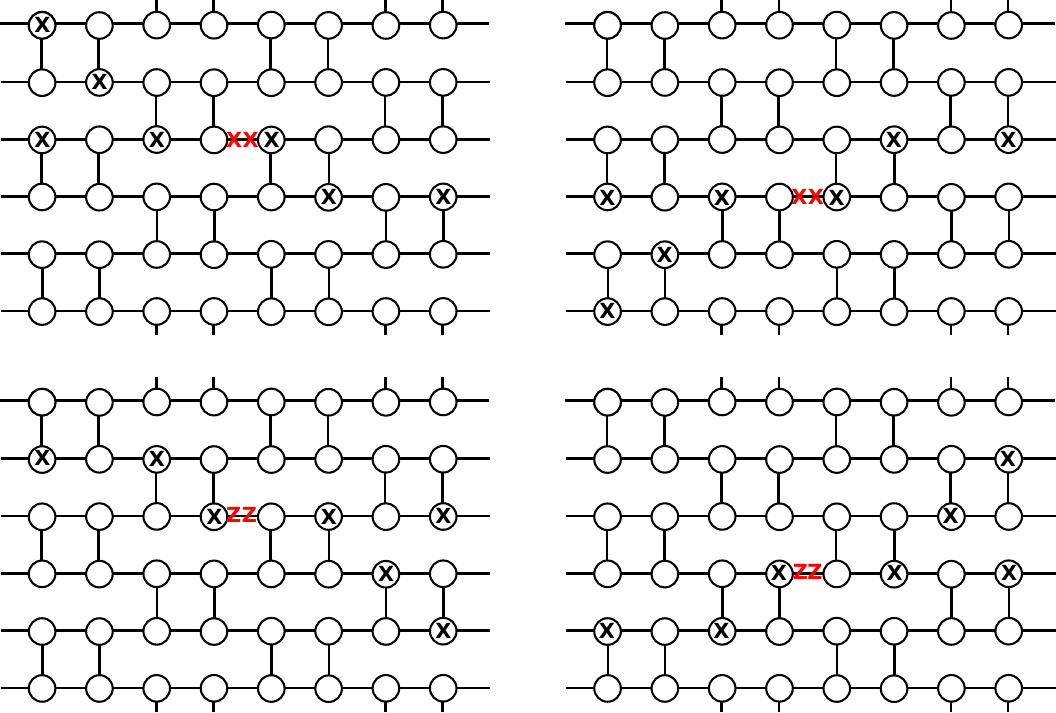}
\caption{Symmetry generators for the group of fractal symmetries of the $(4,8^2)$ lattice.  All conventions are the same as in Fig.~\ref{3464}.}
\label{488_Symmetry_Generators}
\end{figure}

The fractal symmetries of the $(4,8^2)$ lattice are capable of protecting a cluster phase.  Plotting the generators of the symmetry up to vertical translation in Fig.~\ref{488_Symmetry_Generators}, we see that each generator has support on an even number of sites in the neighborhood of any vertex.  Thus, the only product of $Z$ operators that commutes with all the subsystem symmetries is of the form $\prod_{j\in \mathcal{N}(v)}Z_j$ for any vertex $v$.  This meets the condition described in Sec.~\ref{SPTOreview} so we can define a cluster phase protected by the fractal subsystem symmetries.

Finally, the cluster-like phase defined by the fractal symmetries is universal for MBQC.  To determine the gate set available to us, we analyze Fig.~\ref{488_Symmetry_Generators} and employ the argument made in Sec.~\ref{Computational_Universality} to obtain the following set of relevant tensors.
\begin{eqnarray}
C(1,l)&=&Z_l\\
C(p {\tau},l)&=&X_l\\
C(2,2l)&=&X_{2l}Z_{2l+1}\\
C(p {\tau}-1,2l)&=&Z_{2l}X_{2(l-1)+1}
\end{eqnarray}
Measuring the corresponding qubits in the usual rotated basis we can exponentiate these operators to obtain the universal gate set shown in Table~\ref{Fractal_Table}.  Therefore the cluster phase constructed around the $(4,8^2)$ graph state is universal for MBQC on all $n$ qubits at the edge.

\subsection{Decorated Archimedean lattices and periodic QCA structure}\label{Periodic}

All Archimedean lattices possessing a QCA structure have given either a glider or fractal Clifford QCA.  Due to the incompatibility of the Hadamard and $CZ$ gates, one can never obtain a periodic QCA from a vertex translative lattice. To achieve a periodic QCA structure, it suffices to add an additional qubit along each edge that constituting a segment of wire in the QCA construction. This is analogous to a gauging procedure, and referred to as partially decorating the lattice.  This causes all Hadamard gates in the underlying QCA to cancel leaving behind a Clifford circuit consisting only of $CZ$ gates.  The resulting QCA has a period that is some constant dependent on the lattice geometry.

\begin{figure}
\includegraphics[width=\linewidth]{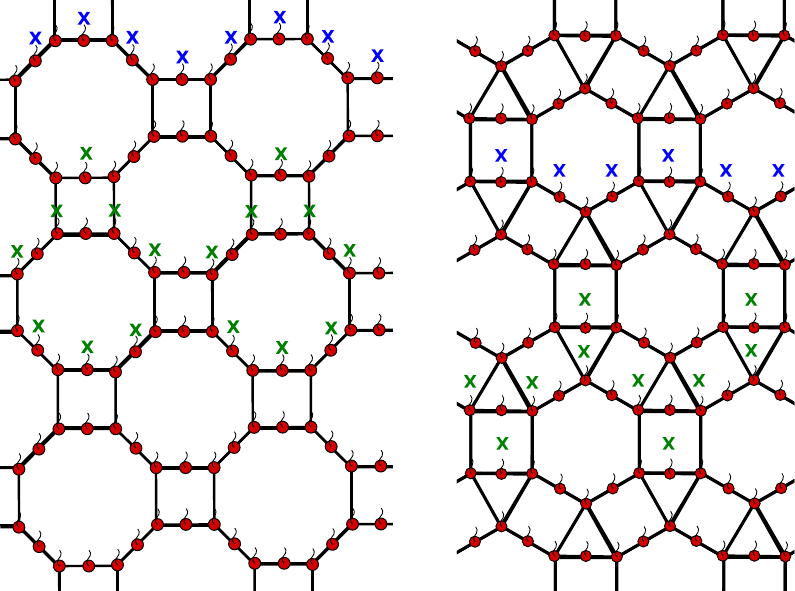}
\caption{Partially decorated Archimedean lattices and their ribbon symmetries.  Left: The dressed $(4,8^2)$ lattice and its line-like symmetries. Right: The dressed $(3,4,6,4)$ lattice and its line-like symmetries. The blue and green symmetries correspond to QCA evolution of $Z$ and $X$ operators at the edge respectively.}
\label{Dressed}
\end{figure}

Partially decorating each of the nine Archimedean lattices discussed previously, the resulting subsystem symmetries are ribbon symmetries with $2n$ generators resulting in a group structure isomorphic to $\mathbb{Z}_2^{2n}$.  We call these ribbon symmetries because the generators have bounded support in the spatial direction of the underlying $(1+1)$ dimensional circuit.  The generators again correspond to the evolution of generators of the Pauli group though the underlying QCA. 
We depict in Fig.~\ref{Dressed} the partially decorated lattices and resulting ribbon symmetries for the $(4,8^2)$ lattice, whose original symmetry is fractal, and the $(3,4,6,4)$ lattice, whose original symmetry is cone.  

The ribbon symmetries of each partially decorated lattice can protect a cluster phase in which every point is universal for MBQC.  This was discussed previously in Ref.~\cite{Stephen2019} for the $(4^4)$, or square, lattice.  There it was stated that since the QCA period is constant, they enjoy a quadratic reduction in the number of qubits to be measured in each quasi-1D segment of wire.  Due to this fact, partially decorated lattices are argued to be efficient for doing MBQC with this scheme.

\section{Archimedean lattices with 1-form symmetries}\label{1form}

\begin{figure}
\centering
\includegraphics[width=0.65\linewidth]{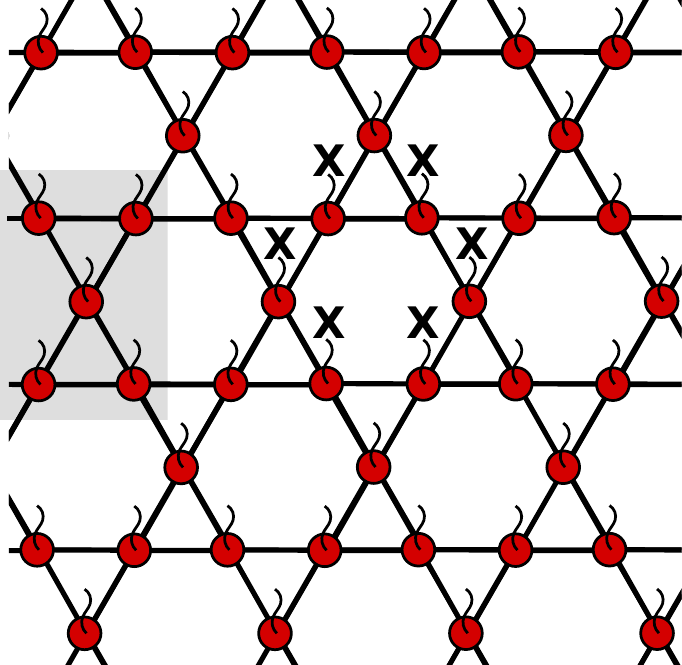}
\caption{1-form symmetries of the (3,6,3,6) lattice act around each hexagonal plaquette.  A bowtie subgraph is shaded in grey.}
\label{3636sym}
\end{figure}

In this section, we discuss the two Archimedean lattices for which there is no underlying Clifford QCA structure, i.e., the $(3,6,3,6)$ and $(3,12^2)$ lattices. By appropriate multiplication of cluster-state stabilizers, one can construct new operators that consist of a ring of $X$ operators around a single 6 or 12 side plaquette, such as that shown in Fig.~\ref{3636sym}. These are referred to as \emph{one-form symmetries}. In contrast to the other three classes of subsystem symmetries, one-form symmetries are deformable in the sense that multiplying a pair of loops of $X$ operators yields a larger loop. That is why they generate the group of products of $X$ operators that lie on homologically trivial loop configurations over the torus. The existence of one-form symmetries is an indicator of robustness to errors~\cite{raussendorf2006fault,roberts2017symmetry,roberts2018symmetry, kubica2018ungauging}. Below we shall see that the existence on such symmetries both precludes the construction of a cluster phase, and enables quantum teleportation on an encoded qubit protected by an error correction code.

It is the presence of these one-form symmetries that prevents these lattices from supporting a cluster phase.  The simplest product of $Z$ operators that commutes with all the 1-form symmetries is a product of three $Z$ operators acting on the vertices about any triangular tile on the lattice.  Such an operator cannot be recast as a product of $X$ operators by the stabilizer relations, and hence, by following analogous reasoning as in Sec.~\ref{Cone_Symmetries}, the resulting SPTO phase is not a cluster phase.  

\begin{figure}
\centering
\includegraphics[width=0.35\linewidth]{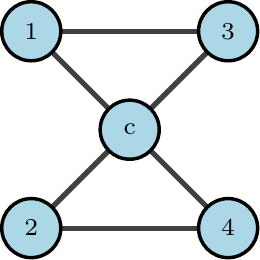}
\caption{Labeling of qubits for the bowtie subgraph of the $(3,6,3,6)$ lattice.}
\label{Bowtie_Subgraph}
\end{figure}
Despite their failure to support a cluster phase, both the $(3,6,3,6)$ and $(3,12^2)$ lattices have the feature that they are equivalent to a foliated classical repetition code capable of teleporting a single encoded qubit. Here we will focus on the $(3,6,3,6)$ lattice, treating the  $(3,12^2)$ lattice in Appendix~\ref{31212}. 

Consider the bowtie subgraph, shown in Fig.~\ref{3636sym}.  Labeling the vertices as shown in Fig.~\ref{Bowtie_Subgraph}, suppose qubits 1 and 2 encode logical inputs. We can write their corresponding logical operators as
\begin{eqnarray}
X_1^L &=& X_1 Z_c Z_3 \\
X_2^L &=& X_2 Z_c Z_4 \\
Z_1^L &=& Z_1  \\
Z_2^L &=& Z_2
\end{eqnarray}
and the graph state stabilizers of Eq.~(\ref{Stabilizer_Relation}).
Notice that,

\begin{equation}
X_1 X_2 X_c = S_c X_1^L Z_1^L X_2^L Z_2^L \equiv -Y_1^LY_2^L.
\end{equation}

\noindent Thus, measuring the first, second, and center qubits in the $X$ basis performs a logical measurement of $-Y_1^LY_2^L$, thereby projecting the input into the $(-1)^{m_1+m_2+m_3+1}$ eigenspace. 

\begin{figure}
\includegraphics[width=\linewidth]{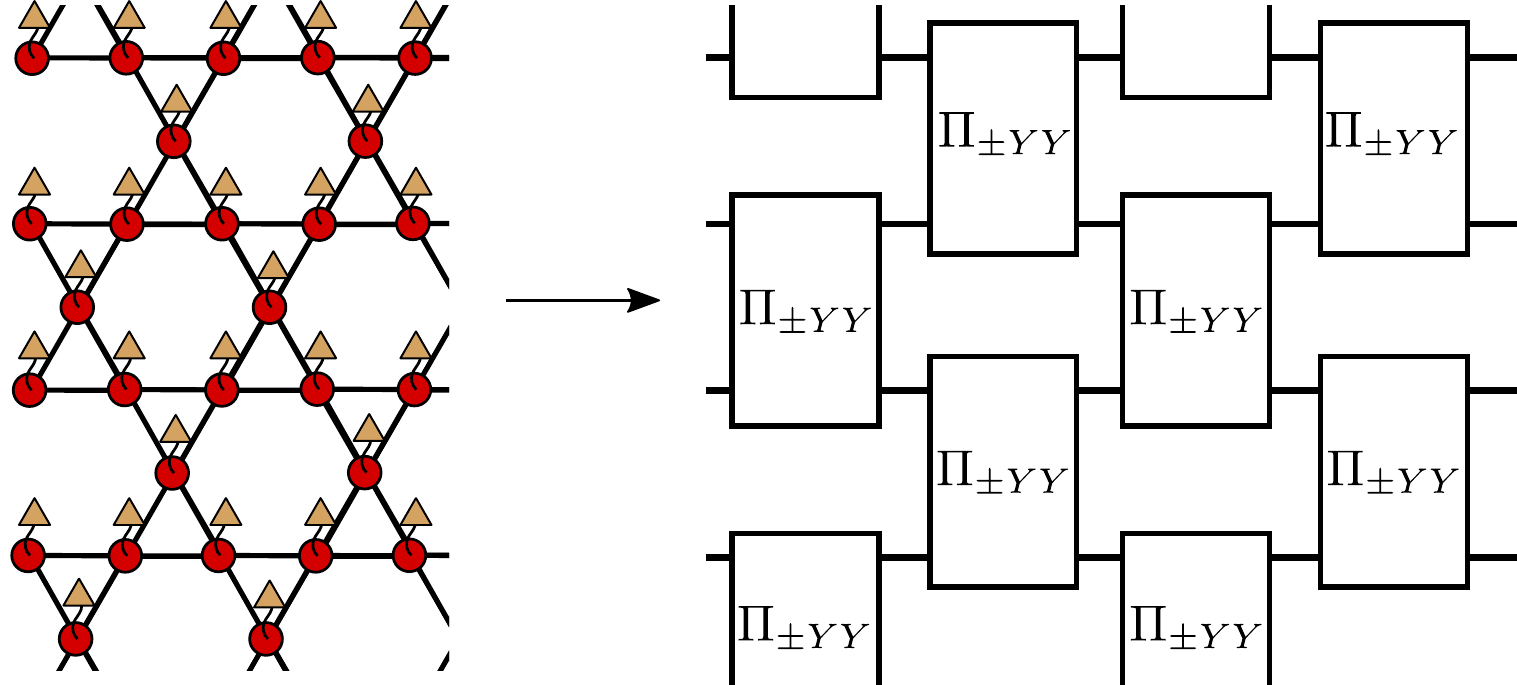}
\caption{The $(3,6,3,6)$ lattice as a foliated repetition code.  Measuring all qubits in the $X$-basis yields a circuit of outcome dependent projections onto an eigenspace of $YY$.  With classical post-processing single qubit $Z$ errors can be detected.}
\label{kagcircuitqca}
\end{figure}

Returning to the $(3,6,3,6)$ lattice, performing $X$ measurements on each qubit along each column implements a circuit consisting of $Y^L Y^L$ parity measurement on each pair of neighboring qubits as shown in Fig.~\ref{kagcircuitqca}. By the second time-step, information is automatically projected onto the single qubit code space (or error space) of the stabilizer code with stabilizer group equivalent to
\begin{align} 
\left\langle  \left\{Y^{L}_j Y^{L}_{j+1} \right\}_{\forall j}\right\rangle.
\end{align} 

\section{Changing foliation of time slices: effects of global topology}\label{Foliation}

The description of computational models in Secs.~\ref{Cone_Symmetries}, \ref{Fractal_Symmetries}, and \ref{Periodic} made use of a specific choice of the set of input qubits, the set of output qubits, and the way in which the lattice is embedded on to a cylinder/torus. In this section, we investigate the effects of global topology, such as specifying the direction of the periodic boundary conditions, on MBQC. 

First we consider varying the shape of the Cauchy surface (slice of constant time), which corresponds to the different ways that a torus can be cut open into a cylinder. This choice specifies the location of the inputs and outputs, corresponding to nodes on either end of the cylinder, respectively. It also defines which qubits lie within a common time slice. Though this choice has no effect on the physical symmetries of the state, it can affect the structure of each QCA block by simply changing the ordering gates in the circuit. 

\begin{figure}
\centering
\includegraphics[width=\linewidth]{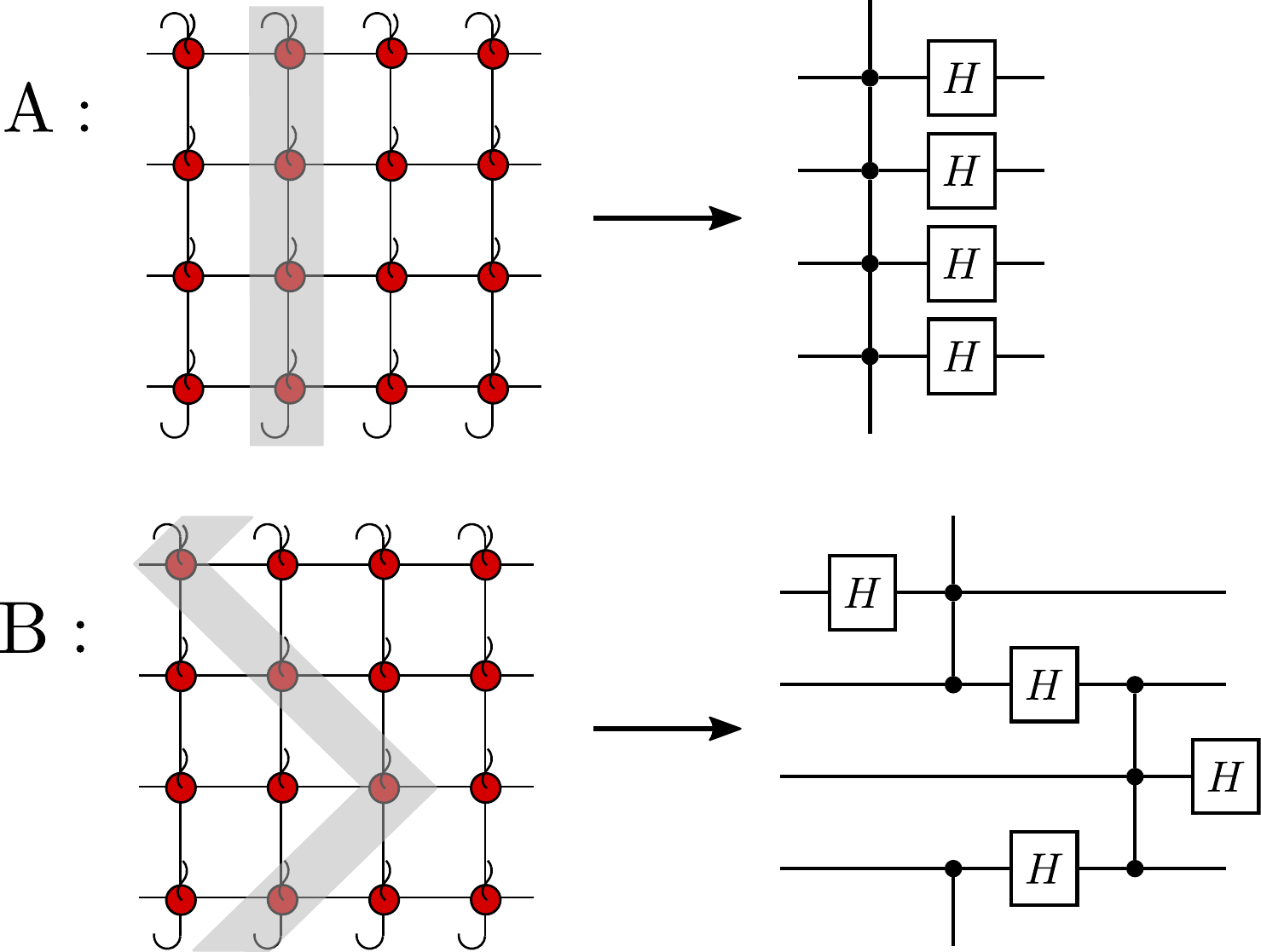}
\caption{Changing the definition of time slices, new QCA blocks are obtained that correspond to local translations of the temporal structure of the old blocks. Above: The original foliation (cut A) of the $(4^4)$ graph state. Below: A new non-translationally invariant QCA block obtained from deforming the original foliation (cut B).}
\label{inputs}
\end{figure}

To see this, consider the two distinct time-slices $A$ and $B$ of the $(4^4)$ lattice graph state as shown in Fig.~\ref{inputs}.  The choice of time slice affects the arrangements of the gates in the Clifford QCA structure. For a given time-slice cut, the QCA may not be translationally invariant, and thus, gates native to that case could be extremely nonlocal.  This could be advantageous for entangling many encoded qubits at once. 

As before, let the lattice be invariant under $\tau$ translations in the simulated time direction, and consider two distinct cuts $A$ and $B$. Let $T_{A}=T_{A\tau} \dotsm, T_{A1}$ and $T_{B}=T_{B\tau} \dotsm T_{B1}$ be the transfer matrix corresponding to each cut. Note that it is possible to transform the cylinder with cut $A$ into one with cut $B$ by performing $X$ measurements on a subset of the input nodes. Let $V_{A\rightarrow B}$ denote the Clifford circuit implemented by changing the time slice from $A$ to $B$ in this way. In a similar way, we can perform additional measurements to return from time-slice $B$ to $A$, implementing the Clifford circuit $V_{B\rightarrow A}$. Note that 
\begin{align}
T_{A}^{k+d}=V_{B\rightarrow A} T_{B}^{k} V_{A\rightarrow B}
\end{align}
and
\begin{align}
T_{B}^{k+d}=V_{A\rightarrow B} T_{A}^{k} V_{B\rightarrow A}
\end{align}
for all integers $k$, and where $d$ is a non-negative integer fixed by the choice of $A$ and $B$ (in particular, $d=2$ for the example shown in Fig.~\ref{inputs}). Modifying the time-slice cut preserves the trace of the transfer matrix, since  
\begin{align}
\text{tr}\left[ T_{A} \right] & = \text{tr}\left[ V_{A\rightarrow B}^{-1} T_{B}^{d+1} V_{B\rightarrow A}^{-1} \right] \\
& = \text{tr}\left[  T_{B}^{d+1} (V_{A\rightarrow B}V_{B\rightarrow A})^{-1} \right] \\
& = \text{tr}\left[  T_{B}^{d+1} (T_{B}^{d})^{-1} \right] \\
& = \text{tr}\left[  T_{B} \right].
\end{align}
Consequently, for any cuts $A$ and $B$, the transfer matrices $T_{A}$ and $T_{B}$ correspond to the same QCA class.

\begin{figure}
\centering
\includegraphics[width=\linewidth]{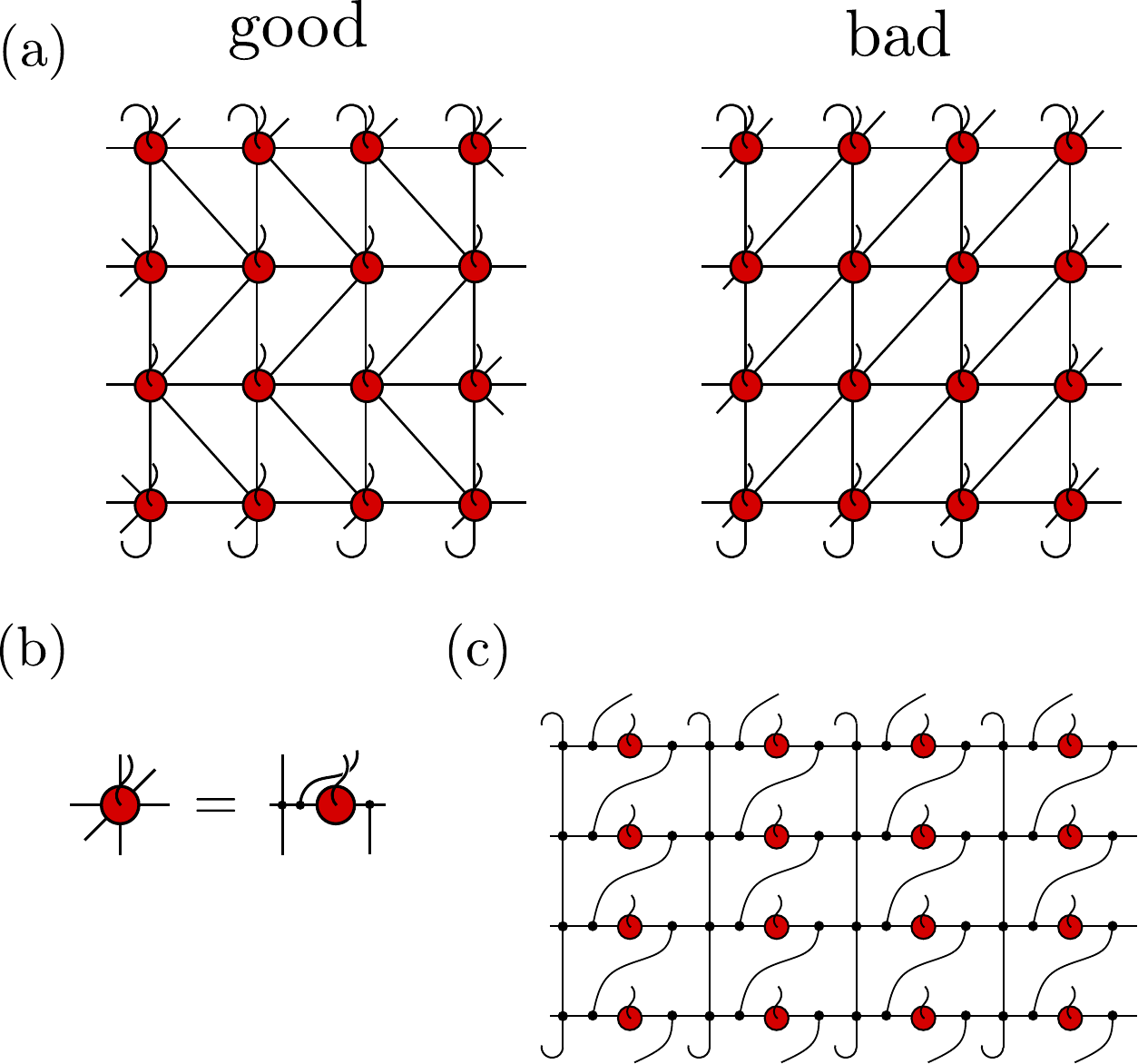}
\caption{Changing the foliation of the $(3^6)$ lattice by modifying the boundary conditions.  (a) On the left is the good foliation of the $(3^6)$ lattice that has underlying unitary CQCA structure described in Table~\ref{Cone_Sym_Table}.  By rotating the $(3^6)$ lattice and changing the boundary conditions the resulting foliation is bad in the sense that the virtual space has a nonunitary structure.  (b) Using push through properties each tensor can be expressed in terms of the 1D graph state MPS tensor with additional $CZ$ gates on the virtual legs. (c) Expressing the tensor network in this way, we see each collumn of tensors consists of an acausal cycle of $CZ$ gates.}
\label{MPS_Construction}
\end{figure}

The second, more nontrivial degree of freedom to vary is the choice of how the lattice becomes embedded onto the torus, i.e., the identification of edges with periodic boundary conditions. We focus our analysis to the $(3^6)$, on which the computational capability is highly dependent on how the lattice gets embedded. 

Recall that the $(3^6)$ lattice embedded as shown in Table~\ref{Cone_Sym_Table} has an underlying glider QCA structure. This was made by using the ``good'' embedding in Fig.~\ref{MPS_Construction} (a). However, an alternative ``bad'' embedding of Fig.~\ref{MPS_Construction}~(a), where the time slice is parallel to a line-like symmetry, results in a circuit with an invalid causal ordering (see Fig.~\ref{MPS_Construction}~(c)). When such circuits arise from MBQC, they can be interpreted as combination of unitary evolution and projective measurement~\cite{da2011closed}. Therefore, the $(3^6)$ lattice with the altered boundary conditions has a dramatically different computational capability.

\begin{figure}
\centering
\includegraphics[width=\linewidth]{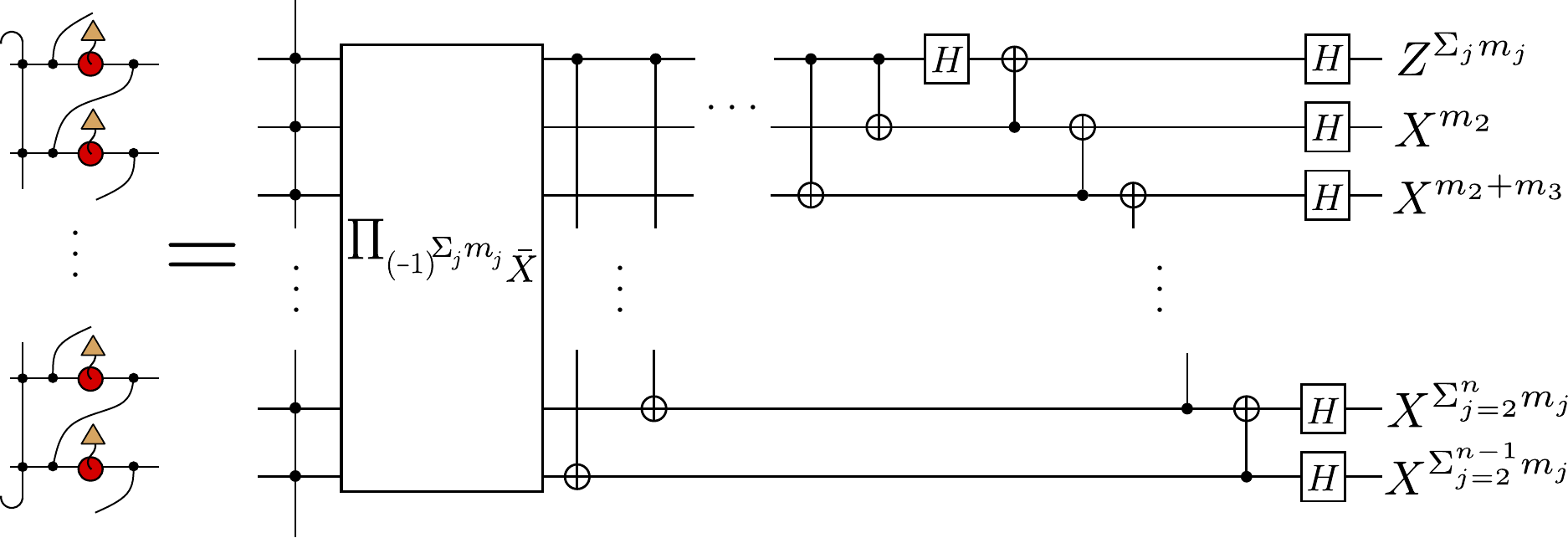}
\caption{The quantum circuit obtained from measurement of a ring of sites around the $(3^6)$ lattice with bad foliation.  The circuit consists of a projector and a unitary part that commute.}
\label{Code_QCA_Circuit}
\end{figure}

In Appendix~\ref{Nonunitary_Circuit}, we use the ZX-calculus~\cite{coecke2011interacting, jeandel2018complete} to explicitly compute the total non-unitary evolution operator, which is equivalent to the circuit shown in Fig.~\ref{Code_QCA_Circuit}. This circuit consists of a projection onto either of the $\pm1$ eigenspaces of the operator $\bar{X} = \prod_{j=1}^n X_j$, followed by unitary gates.

Intuitively, the non-unitary nature of this result can be understood by considering the original line-like symmetries of the triangular lattice. With this embedding, one of the three line-symmetry directions has become parallel the the time-slice of the input states, and thus, $X$ measurements made to teleport the inputs horizontally implement a projection onto the stabilizer code with the single generator
\begin{align}
\left\langle \prod_{j=1}^n X_j \right\rangle
\end{align}
Though very simple, this code can be used to detect a single $Z$ error. The logical operators for this code are
\begin{eqnarray}
X_k^L &=& X_{k+1} , \\
Z_k^L &=& Z_1 Z_{k+1}.
\end{eqnarray}
We also note that the unitary part of Fig.~\ref{Code_QCA_Circuit} preserves the code space, having the following action on the logical operators 
\begin{align}
&X_1^L \mapsto Z_1^L X_2^L ... X_{n-1}^L  \nonumber \\
&X_k^L \mapsto X_{k-1}^L Z_{k-1}^L X_{k}^L Z_{k}^L;\textrm{for } 2\leq k \leq n-1 \label{Logical_Map} \\
&Z_{k}^L \mapsto \prod_{l=k}^{n-1}X_{l}^L;~\forall k. \nonumber
\end{align}
 
To do MBQC, one must perform $\emph{encoded}$ logical operations. Note, however, that in MBQC we are restricted to local (single-site) measurements on the physical qubits. At the graph state fixed point, such measurements on edge qubits apply $\textrm{exp}(i\theta Z_k)$ on the corresponding virtual degree of freedom, which does not preserve the code space.  A code-space-preserving map such as a rotation by $Z_k^L$ requires an entangling (multiple-site) measurement, which is prohibited in MBQC. Therefore, universal MBQC is not possible for the $(3^6)$ lattice with the bad embedding on the torus, though one can use any state of this phase to teleport (i.e., perform the identity gate) $n-1$ logical qubits encoded in an error detection code down the length of the lattice.

\section{Conclusion}\label{Conclusion}

Our main results are summarized in the theorem at the beginning of Sec.~\ref{Main_Sec} and Table~\ref{SStable}. Using 2D vertex-translative Archimedean lattices, we showed that nine of these eleven lattices supported universal cluster phases, where three have glider QCA structures and six have fractal QCA structures. Moreover, the lack of universality in the two other cases can be attributed to the presence of one-form symmetries. Our {\em systematic} analysis on 2D lattice geometry led to several new insights specific to particular QCA classes. For glider QCA, we found that the line symmetries were---in some cases---insufficient for construct universal phases. For this reason, we emphasize the importance of cone symmetries in defining SPTO phases that are also cluster phases.  
Previous work with fractal cluster phases~\cite{Stephen2019,Devakul2018} required sparse usage of qubits by pairing sites in order to prove universality. We improve on this result by showing that, in some cases, the cluster phases afford more efficient usage, where MBQC is universal on all inputs.  
Our results on partially decorated lattices generalize the work of Ref.~\cite{Stephen2019} by showing that any lattice can be partially decorated, resulting in a change in the QCA structure from fractal or glider to a periodic QCA. 

As an outlook, there seem to remain interesting research directions whenever MBQC does not match a conventional picture of quantum computation, i.e., the quantum-circuit model. The lattices supporting one-form symmetries are interesting, since they precluded a unitary QCA and universal cluster phase, and yet, could also be imbued with certain protection from a foliated error correction code. Though in the present 2D and vertex-translative cases, it was not possible to support a non-trivial QCA structure at the logical level, it remains an open problem whether one can construct a cluster phase, in particular in 3D, that supports a foliated quantum error correcting code such that there is also a non-trivial QCA structure acting within the logical code space that enables universal quantum computation. Our investigation on the effects of modifying the temporal ordering of measurements and global boundary conditions on the torus is relevant as well. While we showed that the former cannot change the QCA class, the latter can result in dramatically different QCA structure. These considerations seem to be timely, given the recent interests in combing single-shot quantum error correction with measurement-based routes to universal quantum computation, such as Refs.~\cite{nickerson2018measurement, bombin20182d}.

 \begin{acknowledgments}
This work is supported by National Science Foundation grants PHY-1620651 and PHY-1915011. R. N. A. is supported by National Science Foundation grant PHY-1630114.
\end{acknowledgments}

\printbibliography

\appendix

\section{Tensor network notation}\label{TNNappendix}
Here we introduce the tensor network notation used throughout this article. For a more pedagogical introduction to tensor networks see Ref.~\cite{bridgeman2017hand}. All our tensors can be decomposed into one, two, and three index tensors that correspond to $X$ measurements, a Hadamard gate, and copying the value of an index (known as a copy tensor), respectively.  They are the following,

\begin{gather}
\includegraphics[scale=0.75]{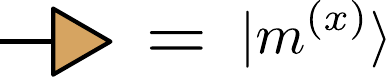}\label{measurement_tensor}\\
\includegraphics[scale=0.75]{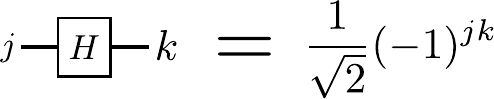} \\
\includegraphics[scale=0.75]{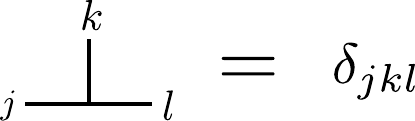}
\end{gather}
where $\delta_{jkl}=\delta_{jk}\delta_{kl}$.
Here $|m^{(x)}\rangle$ is defined by $X|m^{(x)}\rangle = (-1)^{m}|m^{(x)}\rangle$.  These satisfy the following relations,

\begin{gather}
\includegraphics[scale=0.75]{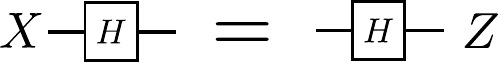} \\
\includegraphics[scale=0.75]{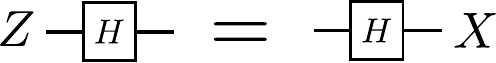} \\
\includegraphics[scale=0.75]{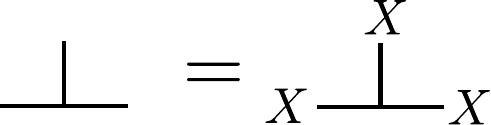}\\
\includegraphics[scale=0.75]{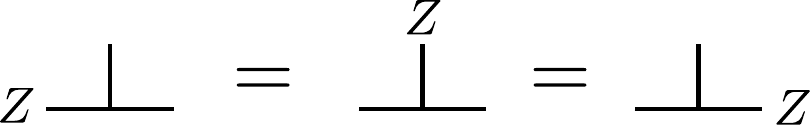} \\
\includegraphics[scale=0.75]{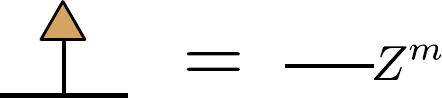}
\end{gather}

We can use these to construct two more objects that will show up frequently throughout this article.   The $CZ$ gate can be represented as,

\begin{equation}
\includegraphics[scale=0.75]{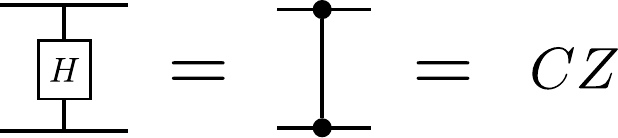}.
\end{equation}

\noindent Furthermore the matrix product state (MPS) tensor for the 1D graph state can be constructed as,

\begin{equation}
\includegraphics[scale=0.75]{MPS_Tensor_From_Tensors}.
\label{MPS_Tensor_From_Tensors}
\end{equation}

\noindent We will refer to the vertical wavy index as the physical index and the horizontal indices as the left and right virtual indices.  Notice then that this MPS tensor has the following symmetries:

\begin{gather}
\includegraphics[scale=0.75]{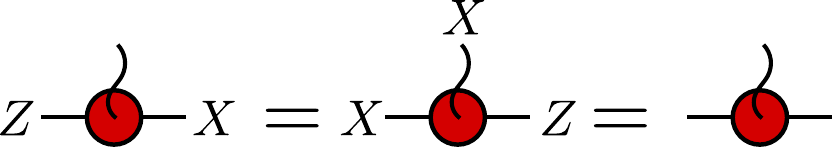} \label{1D_MPS_Symmetries}\\
\includegraphics[scale=0.75]{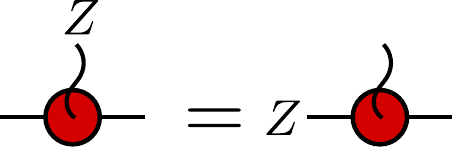}\label{1D_MPS_Z_Symmetries}.
\end{gather}

Comparing this to (\ref{MPS_Tensor_From_Tensors}), one sees that any operator diagonal in the $Z$ basis can be moved from the physical index to the left virtual index.  One such relation that will be useful to us is

\begin{equation}
\includegraphics[scale=0.75]{1D_MPS_CZ_Symmetry}, \label{czsym}
\end{equation}
which involves moving one end of a $CZ$ operator from the physical to the left virtual index.

\begin{figure}
\includegraphics[width=\linewidth]{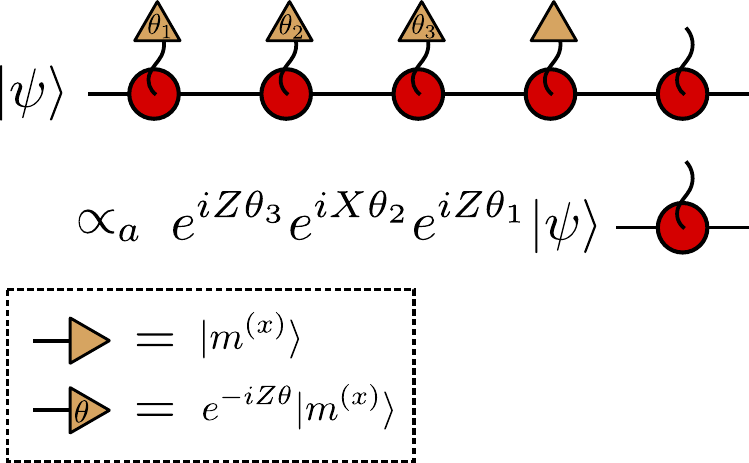}
\caption{The measurement pattern to achieve an arbitrary $SU(2)$ operation.  1-index yellow tensors represent $X$ measurments.  $\propto_a$ is equivalence after measuring adaptively and up to Pauli byproduct operators.}
\label{MBQCin1D}
\end{figure}

One benefit of this notation is that it makes single-qubit MBQC transparent, as shown in Fig.~\ref{MBQCin1D}.

\section{MBQC with quasi-1D SSPT phases}\label{MBQCSSPT}

In this section we review the fundamentals of performing MBQC with SPTO phases.  This section is a review of the results of Refs.~\cite{Raussendorf2017, Stephen2017} that are necessary for this work.

To use a subsystem SPTO phase for MBQC, the notion of locality that arises from the 2D lattice must be replaced by a quasi-1D notion of locality as follows. All Archimedean lattices can be deformed such that each vertex lies on a square grid.  We then embed the resulting lattice on a torus of dimension $n\times N$ where $n=m_1\Delta$  {and} $N=m_2 p {\tau}$  where $m_1,m_2,\in\mathbbm{N}$, $p$ is the period of the QCA, and the lattice has $\Delta~(\tau)$-site space (time) translational invariance. Next, we coarse grain the torus into a quasi-1D wire made up of $n\times p {\tau}$ sized cylindrically-shaped blocks. This quasi-1D structure is equivalent to a generalized 1D quantum wire with an MPS description that has physical Hilbert space dimension $2^{n p {\tau}}$ and bond dimension $2^{n}$.

The tensors used in this MPS description are determined by the symmetries of the system. Each MPS tensor has the so-called \emph{Clifford property}, by which $X$-type subsystem symmetries on the physical degrees of freedom map to Pauli operators on a connected virtual edge. In fact, the $2n$ generators of the $n$ qubit Pauli group are in one to one correspondence of the $2n$ generators of the subsystem symmetries. More precisely, $n$-qubit Pauli operators acting on the virtual degrees of freedom form a non-trivial projective representation of the symmetry group, which corresponds to a particular cohomology class that defines a 1D SPTO phase.  Since the physical lattice is actually 2D, such an SPTO phase is referred to as a quasi-1D SPTO phase.  This SPTO phase is protected by subsystem symmetry, which in the 1D picture acts on the coarse grained blocks in an onsite manner.

\subsection{Determining fixed point tensors}\label{MBQCSSPT_1}

The subsystem symmetries on the physical level form a reducible representation of $\mathbbm{Z}_2^{2n}$ that consists of tensor products of $X$ operators.  Thus, the symmetry group is a direct sum of 1D irreducible representations when written in the basis $\{|0^{(x)}\rangle,|1^{(x)}\rangle\}^{\otimes np {\tau}}$.  At the fixed point, the quasi-1D MPS tensors, denoted as $\mathcal{C}$, can be written in this basis as,
\begin{equation}
\mathcal{C} = \sum_{\mathbf{j}\in\mathbbm{Z}_2^{np {\tau}}} C\left(\mathbf{j}\right) \otimes |\mathbf{j}\rangle
\end{equation}
where, $\mathbf{j}\in\mathbbm{Z}_2^{np\tau}$ is a binary vector and the state $|\mathbf{j}\rangle$ represents the configuration where the $k^\textrm{th}$ physical qubit is in the $(-1)^{j_k}$ eigenstate of $X_k$. 

The action of the symmetry $u(g)$ on the physical degrees of freedom gives a phase $\chi_\mathbf{j}(g)$ to each component $\mathbf{j}$ since by definition $u(g)|\mathbf{j}\rangle = \chi_{\mathbf{j}}(g)|\mathbf{j}\rangle$.  However, we may also push the symmetry through to the virtual level where it acts via the projective representation $V(g)$.  Equating these two pictures we have,
\begin{equation}
\sum_{\mathbf{j}} \chi_\mathbf{j}(g)C(\mathbf{j})\otimes|\mathbf{j}\rangle =  \sum_{\mathbf{j}} V(g)C(\mathbf{j})V(g)^\dagger \otimes|\mathbf{j}\rangle.
\end{equation}
Equating each component we find,
\begin{equation}
V(g) C(\mathbf{j}) = \chi_{\mathbf{j}}(g) C(\mathbf{j}) V(g).
\label{TensorEqn}
\end{equation}


Eq.~(\ref{TensorEqn}) is of fundamental importance in determining the structure of the quasi-1D MPS tensors for a given 2D lattice embedded on a torus.  To determine the tensor $C(x,y)=C(\mathbf{e}_{(x,y)})$, we note that $\chi_{\mathbf{e}_{(x,y)}}(g) = \chi_{(x,y)}(g)$ will be $-1$ for any symmetry $g$ that has support on the site $(x,y)$ and +1 for all other symmetries. As a consequence of Eq.~(\ref{TensorEqn}),  $C(x,y)$ should anti-commute with the edge representation $V(g)$ of the symmetry operators supported at site $(x,y)$ and commute with all others.  Since the edge representations of the symmetry generators are exactly the generators of the $n$-qubit Pauli group, this uniquely specifies $C(x,y)$ as a product of Pauli operators.


For clarity, let us derive Eq. (\ref{3464_ZXZ}).  That is, we wish to determine $C(2,4l+1)$ for the $(3,4,6,4)$ lattice. First, for each model we study, $V(g)$ is always some string of Pauli operators.  Furthermore, since all symmetries consist purely of $X$ operators we have, $|\mathbf{j}\rangle \in \{|0^{(x)}\rangle, |1^{(x)}\rangle\}^{\otimes np\tau}$ and so $\chi_{\mathbf{j}}(g) = \pm1$ depending on $\mathbf{j}$ and $g$.  Thus, $C(\mathbf{j})$ will simply be some string of Pauli operators determined by the commutation relations (\ref{TensorEqn}).

\begin{figure*}
\centering
\includegraphics[width=\linewidth]{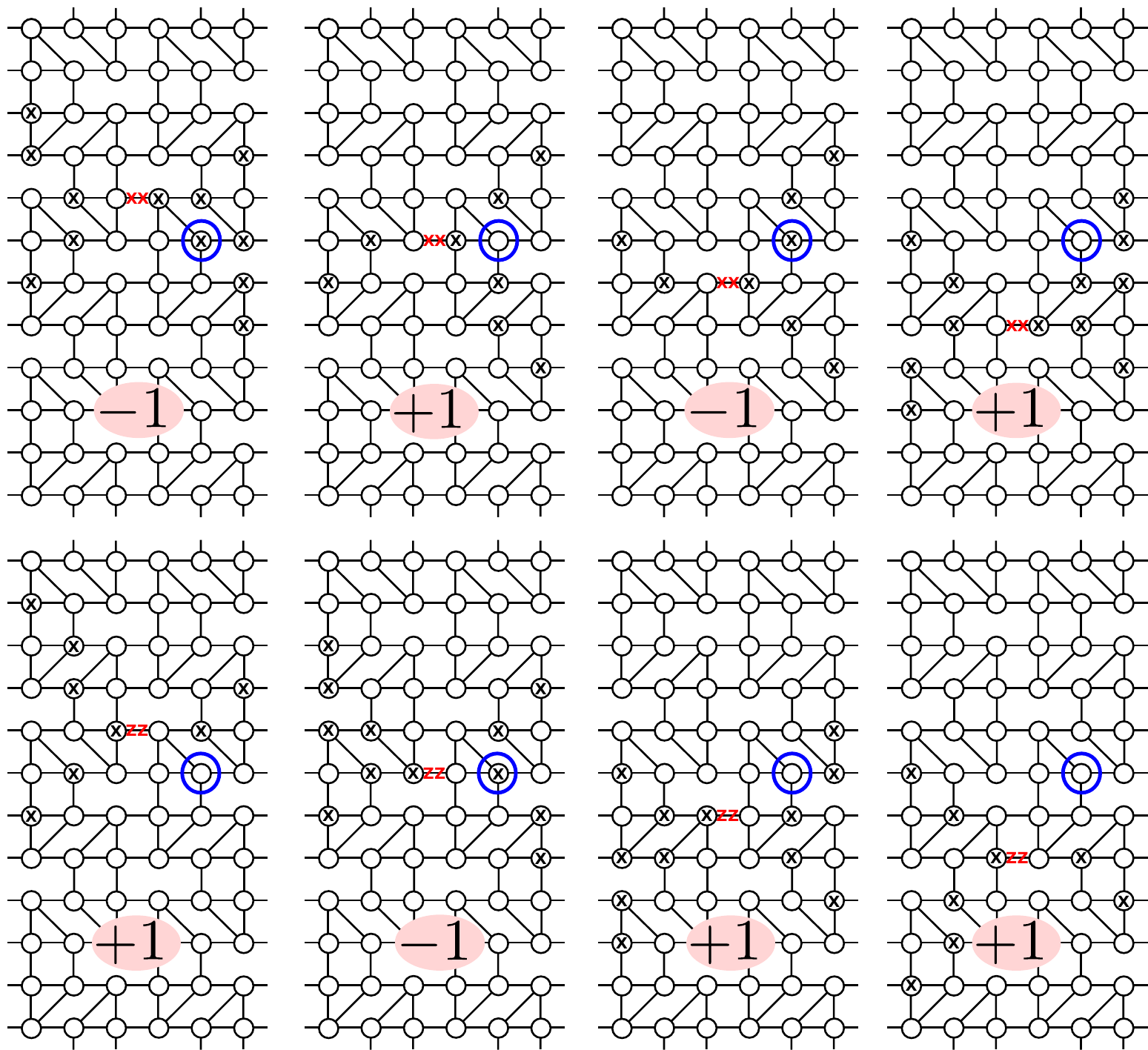}
\caption{Details of the derivation of Eq.~(\ref{3464_ZXZ}).  Here we have included Fig.~\ref{3464} with additional details to aid in following the derivation.  The symmetry generators of the $(3,4,6,4)$ lattice are displayed along with their respective projective representation at the edge written twice in red.  The site $(2,4l+1)$ is circled in blue.  For each symmetry generator, we have specified the eigenphase $\chi_{(2,4l+1)}(g) = \pm1$ obtained from the action of each symmetry generator $u(g)$ on the state $|\mathbf{e}_{(2,4l+1)}\rangle$ in the pink shaded bubbles.  The tensor $C(2,4l+1)$ is then the $n$-qubit Pauli operator that either commutes or anti-commutes with the virtual representation of each symmetry for $\chi_{(2,4l+1)}(g) = +1$ or $-1$, respectively.  A simple analysis leads to the conclusion that $C(2,4l+1) = Z_{4l} X_{4l+1} Z_{4l+2}$.}
\label{3464_Derive}
\end{figure*}

To determine $C(2,4l+1)$, we must analyze the symmetry generators in Fig.~\ref{3464}.  For completeness we have included it here (Fig.~\ref{3464_Derive}) with some additional details.  $C(2,4l+1)$ is the component corresponding to the physical state $|\mathbf{e}_{(2,4l+1)}\rangle$. This is a product state that has each physical qubit in the state $|0^{(x)}\rangle$ except for the qubit at site $(2,4l+1)$, which is in the state $|1^{(x)}\rangle$.  This site is circled in blue in Fig.~\ref{3464_Derive}.

The respective eigenphase, $\chi_{(2,4l+1)}(g) = \pm1$, obtained from the action of each symmetry generator $u(g)$ on the state $|\mathbf{e}_{(2,4l+1)}\rangle$ is denoted for each symmetry generator in Fig.~\ref{3464_Derive} in the pink shaded bubbles.  Notice eigenphase is $-1$ whenever the symmetry has support on the site $(2,4l+1)$ and is $+1$ otherwise.  We find that $C(2,4l+1)$ should have the following commutation/ anti-commutation relations with each virtual representation of the symmetry, denoted in red in Fig.~\ref{3464_Derive},
\begin{align}
X_{4l}C(2,4l+1) &= - C(2,4l+1) X_{4l} \\
Z_{4l+1}C(2,4l+1) &= - C(2,4l+1) Z_{4l+1} \\
X_{4l+2}C(2,4l+1) &= - C(2,4l+1) X_{4l+2} \\
P C(2,4l+1) &= C(2,4l+1) P~~\textrm{otherwise}.
\end{align}
Where $P$ is some single site Pauli operator.
Therefore, we find that,
\begin{equation}
C(2,4l+1) = Z_{4l} X_{4l+1} Z_{4l+2}.
\end{equation}

Once the tensors $\mathcal{C}$ have been determined, we can devise a method for implementing MBQC.  This is broken down into three key ingredients~\cite{Stephen2017}; oblivious wire, unitary gates, and preparation and readout. Below we review each of these in detail in the context of quasi-1D SPTO.

\subsection{Oblivious Wire}

At an arbitrary point in a quasi-1D SPT phase the MPS tensors take the form

\begin{equation}
\mathcal{A}_\phi = \sum_{\mathbf{j}\in\mathbbm{Z}_2^{np {\tau}}}C(\mathbf{j}) \otimes B(\mathbf{j}) |\mathbf{j} \rangle
\end{equation}

\noindent where the $B(\mathbf{j})$ are the so-called \emph{junk tensors} that depend on the location of the state within the phase and the $C(\mathbf{j})$ are the \emph{logical tensors} that are the same at every point in the phase. The support of these tensors partition the  virtual Hilbert space, and are referred to as the logical and junk subspaces/subsystems. The logical subspace is used to house the encoded input for MBQC. At generic points in the phase, measuring in a  basis rotated away from $X$ couples the logical and junk subsystems, resulting in unwanted leakage of the encoded information into the junk subspace.   

This effect can be mitigated by a procedure that uncouples the logical and junk subsystems between computational steps. This is known as implementing an \emph{oblivious wire} segment. It involves measuring $L$-many $n\times p {\tau}$ sized blocks in the symmetry protected basis $\{|\mathbf{j}\rangle\}$, where $L$ is assumed to be a large number.  Suppose the outcome of each measurement is given by $\{|\mathbf{j}_k\rangle | k=1,...,L\}$ for each block.  Representing the edge state as $\rho$, this measurement implements the map,
\begin{equation}
\prod_{k=1}^L C(\mathbf{j}_k) \otimes B(\mathbf{j}_k) \rho C(\mathbf{j}_k)^\dagger \otimes B(\mathbf{j}_k)^\dagger
\end{equation}
\noindent on the virtual edge state.  Correcting the logical part of the byproduct operator and ``forgetting" the measurement outcome---effectively averaging over all possible outcomes---implements a Krauss map, $\mathcal{L}$, for each $n\times p {\tau}$ sized block that is measured.  We may write one iteration of this map as,
 \begin{equation}
 \mathcal{L}(\rho) = \sum_{\mathbf{j}} \mathbbm{1}\otimes B(\mathbf{j}) ~ \rho ~ \mathbbm{1}\otimes B(\mathbf{j})^\dagger.
 \end{equation}
  Notice this map acts trivially on the logical space.  When the quasi-1D MPS tensors are injective and put in canonical form, this map has $\mathbbm{1}\otimes \rho_{\textrm{fix}}$ as a unique fixed point with eigenvalue one.  All other fixed points have eigenvalue less than one. Thus as $L\rightarrow\infty$ this procedure decouples the junk and logical subsystems by driving the junk subsystem to $\rho_{\textrm{fix}}$. The state in the logical subsystem is no longer entangled with the junk subsystem, and we denote the resulting state by $\sigma$.  Hence,
 
 \begin{equation}
 \lim_{L\rightarrow\infty} \mathcal{L}^L(\rho) = \sigma\otimes\rho_{\textrm{fix}}.
 \end{equation}
 
 This procedure can be used repeatedly to suppress information leakage from the logical subsystem to the junk subsystem.

\subsection{Unitary gates}
On the graph state, universal MBQC is possible via measuring some qubits in the rotated basis $X_{\theta}$, for some $\theta$ that is not an integer multiple of $\pi/2$. As described above, doing this at generic points in the SPTO phase is problematic, since it results in non-unitary evolution on the logical subspace. However, when the measured basis is only rotated by a small angle $d\theta$ away from the $X$ basis, then the non-unitary component of the evolution can be suppressed to second order in $d\theta$ by immediately implementing a section of oblivious wire. Iterating this procedure for sufficiently small $d\theta$ yields an arbitrarily good approximation to universal MBQC. Below, we review this procedure in more detail. 
 
After implementing a segment of oblivious wire, the state of the system can be written as 
 \begin{equation}
 \sum_{\mathbf{j},\mathbf{k}} A[\mathbf{j}] \left(\sigma\otimes\rho_{\textrm{fix}}\right) A[\mathbf{k}]^\dagger \otimes |\mathbf{j}\rangle\langle\mathbf{k}|
 \end{equation}
\noindent where $A[\mathbf{j}]=C(\mathbf{j})\otimes B(\mathbf{j})$.  

For the $n\times p {\tau}$ qubits in the next measurement round, let the standard basis vector $\mathbf{e}_\mathbf{j}\in\{0, 1\}^{n p {\tau}}$ label the $j^{\text{th}}$ qubit. We will assume that the $s^{\text{th}}$ qubit is measured in the basis
\begin{equation}
\begin{cases}
|0'\rangle = |0^{(x)}\rangle + e^{i\gamma}d\theta |1^{(x)}\rangle \\
|1'\rangle = |1^{(x)}\rangle - e^{-i\gamma}d\theta |0^{(x)}\rangle
\end{cases}
\end{equation}
and all others are measured in the $X$ basis.

Let $\mathbf{m}\in\{0, 1\}^{n p {\tau}}$ be the binary vector of outcomes of this measurement, and assume that the $s^{\text{th}}$ qubit outcome was $0^{\prime}$. Conditioned on these outcomes, the virtual space evolves via 
\begin{equation}
A_{0'} = A[\mathbf{m}] + e^{i\gamma}d\theta A[\mathbf{m}'],
\end{equation}
where $\mathbf{m}'=\mathbf{m}+\mathbf{e}_\mathbf{s}$. In terms of the logical and junk tensors we may write this as,
\begin{equation}
A_{0'} = C(\mathbf{m}) \left( \mathbbm{1}\otimes B(\mathbf{m}) + e^{i\gamma}d\theta C\otimes B(\mathbf{m}')\right),
\end{equation}  
where $C=C(\mathbf{m})^{-1}C(\mathbf{m}')$. Moreover, we can undo the $C(\mathbf{m})$ part by applying the byproduct operator:
\begin{align}
A_{0'}\mapsto C(\mathbf{m})^{-1}A_{0'}
\end{align}
For all cluster phases studied in this paper, $C(\mathbf{m})$ is a product of Pauli operators and $C = C(\mathbf{e}_{\mathbf{s}})$. Up to order $d\theta$, the measurement implements the map
\begin{widetext}

\begin{align}
 A_{0'}\left( \sigma\otimes\rho_{\textrm{fix}}\right) A_{0'}^\dagger = & \sigma\otimes B(\mathbf{m})\rho_{\text{fix}}B(\mathbf{m})^\dagger + d\theta \big[ e^{i\gamma}C\sigma\otimes B(\mathbf{m}')\rho_{\text{fix}}B(\mathbf{m})^\dagger e^{-i\gamma}\sigma C^\dagger \otimes B(\mathbf{m})\rho_{\textrm{fix}}B(\mathbf{m}')^\dagger  \big]. 
\end{align}
Next, we define
\begin{equation}
\lim_{L\rightarrow\infty}\mathcal{L}^L(B(\mathbf{j})\rho_{\textrm{fix}}B(\mathbf{k})^{\dagger}) = \nu_{\mathbf{j},\mathbf{k}} \rho_{\textrm{fix}}.
\end{equation}
Notice that since $\mathcal{L}(O)^\dagger =\mathcal{L}(O^\dagger)$ and $\rho_{\textrm{fix}}^\dagger = \rho_{\textrm{fix}}$ we have that $\nu_{\mathbf{j},\mathbf{k}} = \nu_{\mathbf{k},\mathbf{j}}^{*}$.  Also since $\sum_{\mathbf{j}} B(\mathbf{j})\rho_{\textrm{fix}}B(\mathbf{j})^\dagger=\rho_{\textrm{fix}}$ we have $\sum_{\mathbf{j}} \nu_{\mathbf{j},\mathbf{j}}=1$.
The state after implementing such a length $L$ segment of oblivious wire can be written to first order as 
\begin{align}
A_{0'}&\left(\sigma\otimes\rho_{\textrm{fix}}\right)A_{0'}^\dagger
= \big( \nu_{\mathbf{m},\mathbf{m}}\sigma + d\theta \big[ e^{i\gamma} \nu_{\mathbf{m}',\mathbf{m}} C\sigma +  e^{-i\gamma}\nu_{\mathbf{m},\mathbf{m}'}\sigma C^\dagger \big]\big)\otimes \rho_{\textrm{fix}} \\
&= \nu_{\mathbf{m},\mathbf{m}} \sigma\otimes\rho_{\textrm{fix}} + 
 \frac{d\theta}{2}\left( [e^{i\gamma}\nu_{\mathbf{m}',\mathbf{m}} C-e^{-i\gamma}\nu_{\mathbf{m}',\mathbf{m}}^*C^{\dagger},\sigma] + \{ e^{i\gamma}\nu_{\mathbf{m}',\mathbf{m}}C+e^{-i\gamma}\nu_{\mathbf{m}',\mathbf{m}}^{*}C^\dagger, \sigma \}\right)\otimes \rho_{\textrm{fix}}.\label{eq:ev0}
\end{align}

By an analogous calculation, the output conditioned on outcome $1'$ on qubit $s$ is
\begin{align}
A_{1'}\left(\sigma\otimes\rho_{\textrm{fix}}\right)A_{1'}^\dagger =& \nu_{\mathbf{m}',\mathbf{m}'} \sigma\otimes\rho_{\textrm{fix}} + \cdot\cdot\cdot \\
&\frac{d\theta}{2} \left( [e^{i\gamma}\nu_{\mathbf{m}',\mathbf{m}} C-e^{-i\gamma}\nu_{\mathbf{m}',\mathbf{m}}^*C^{\dagger},\sigma] - \{ e^{i\gamma}\nu_{\mathbf{m}',\mathbf{m}}C+e^{-i\gamma}\nu_{\mathbf{m}',\mathbf{m}}^{*}C^\dagger, \sigma \}\right)\otimes \rho_{\textrm{fix}}.\nonumber\label{eq:ev1}
\end{align}
If the measurement record is discarded, then the resulting evolution is the average of Eqs.~(\ref{eq:ev0}) and (\ref{eq:ev1}), resulting in 
\begin{eqnarray}
\sigma\otimes\rho_{\textrm{fix}} &\mapsto& \sum_{\mathbf{m}}\big(  (\nu_{\mathbf{m},\mathbf{m}}+ \nu_{\mathbf{m}',\mathbf{m}'})\sigma +  d\theta [ e^{i\gamma}\nu_{\mathbf{m}',\mathbf{m}} C -  e^{-i\gamma}\nu_{\mathbf{m}',\mathbf{m}}^*C^\dagger,\sigma]\big)\otimes \rho_{\textrm{fix}}. 
\end{eqnarray}

\end{widetext}

\noindent Let us write $\sum_{\mathbf{m}}  \nu_{\mathbf{m}',\mathbf{m}} = \nu_{\mathbf{s}}$.  In the case of the cluster-like phase (i.e. whenever $C = C(\mathbf{e_s})$ are Pauli operators) this procedure implements the unitary,

\begin{equation}
U(d\theta) = \textrm{exp}\left( id\theta \frac{e^{i\gamma}\nu_{\mathbf{s}}-e^{-i\gamma}\nu_{\mathbf{s}}^* }{i} C(\mathbf{e_s}) \right).
\end{equation}
We may simplify this to
\begin{equation}
U(d\theta,\gamma) = \textrm{exp} \left( -i 2d\theta |\nu_{\mathbf{s}}| \sin(\gamma+\delta) C(\mathbf{e_s}) \right)
\end{equation}
by writing $\nu_{\mathbf{s}} = |\nu_{\mathbf{s}}|e^{-i\delta}$. If we have determined which point of the phase we are at by first measuring the $\nu_{\mathbf{s}}$, we can choose $\gamma$ such that $\gamma+\delta= \frac{\pi}{2}$ to obtain the unitary,

\begin{equation}
U(d\theta) =  \textrm{exp} \left( -i 2d\theta |\nu_{\mathbf{s}}|C(\mathbf{e_s}) \right).
\end{equation}

Iterating this procedure many times, these small rotations compound to give a large rotation by angle $\theta$. Hence, we can implement the unitary

\begin{equation}
U(\theta) =  \textrm{exp} \left( -i \theta C(\mathbf{e_s}) \right)
\end{equation}

\noindent up to some $\epsilon$ of error.  Notice all errors come from the fact that if we were to carry out the calculation up to $\mathcal{O}(d\theta^2)$ we see the map implemented is no longer unitary.  To minimize error, we must make $d\theta$ as small as possible.

\subsection{Measurement}

Here we shall see that we may perform weak measurements of an observable corresponding any fixed point tensor component $C(\mathbf{e}_\mathbf{s})$.  These weak measurements may be performed many times to give a strong, approximately projective measurement of $C(\mathbf{e}_{\mathbf{s}})$.  For each lattice with an underlying QCA structure studied in this paper we find that $C(1,l) = Z_l$ for each $l=1,...,n$, which implies we can always do single qubit $Z$ measurements on any edge qubit.

To perform such a measurement, we must completely break the symmetry and measure the qubit at site $\mathbf{s}$ in the $Z$-basis.  All other qubits in the $n\times p\tau$ sized block are measured in the $X$-basis. This is immediately followed by a segment of oblivious wire, and the results of all $X$ measurements are discarded after applying the relevant correction/ byproduct operators. The resulting map acts trivially on the junk subsystem (it gets driven to the fixed point), and thus, we are free to ignore the $\otimes \rho_{\textrm{fix}}$ factor present on the output state.  The overall map implemented on the logical part of the edge state $\sigma$ is,
\begin{align}
&\Lambda_0\left( \sigma \right)= \sigma + \nu_{\textbf{s}} C(\mathbf{e_s})\sigma + \nu_{\textbf{s}}^*\sigma C(\mathbf{e_s})^\dagger+  C(\mathbf{e_s})\sigma C(\mathbf{e_s})^\dagger\\
&\Lambda_1\left( \sigma \right)= \sigma - \nu_{\textbf{s}} C(\mathbf{e_s})\sigma - \nu_{\textbf{s}}^*\sigma C(\mathbf{e_s})^\dagger +  C(\mathbf{e_s})\sigma C(\mathbf{e_s})^\dagger.
\end{align}

Using the eigenbasis, 
\begin{equation}
C(\mathbf{e_s}) |\phi_j\rangle= e^{i\phi_j}|\phi_j\rangle
\end{equation}
we can expand the density matrix of the edge state as
\begin{equation}
\sigma = \sum_{j,j'}\langle\phi_j|\sigma|\phi_j'\rangle \otimes |\phi_j\rangle\langle\phi_{j'}|
\end{equation}
and similarly for the conditional output states $\Lambda_{0(1)}\left( \sigma \right)$, 
\begin{align}
\Lambda_{0(1)}\left( \sigma \right)= \sum_{j,j'}f_{0(1)}(\phi_j,\phi_{j'})\langle\phi_j|\sigma |\phi_{j'}\rangle \otimes |\phi_{j}\rangle\langle\phi_{j'}|
\end{align}
where,
\begin{equation}
f_{0(1)}(\phi_j,\phi_{j'}) =  1 \pm \nu_{\textbf{s}} e^{i\phi_j} \pm \nu_{\textbf{s}}^*e^{-i\phi_{j'}} + e^{i(\phi_j-\phi_{j'})}
\end{equation}
is the so-called filtering function. 

Repeating this procedure $N$ times, obtaining $N_0$ outcomes of 0 and $N_1$ outcomes of 1, the map induced on the edge state is
\begin{align}
&\sigma \mapsto \sum_{j,j'} f_{0}(\phi_j,\phi_{j'})^{N_{0}}f_{1}(\phi_j,\phi_{j'})^{N_{1}}|\phi_{j}\rangle\langle\phi_j|\sigma |\phi_{j'}\rangle\langle\phi_{j'}|
\end{align}
By considering the diagonal elements ($j=j'$) and maximizing $ f_{0}(\phi,\phi)^{N_{0}}f_{1}(\phi,\phi)^{N_{1}}$ with respect to $\phi$, one sees that this procedure implements a strong, approximately projective measurement. The maximum can be found by solving
\begin{equation}
\phi = \frac{f_{0}(\phi,\phi)}{f_{1}(\phi,\phi)} = \frac{N_0}{N_1}
\end{equation}
for $\phi$ and finding the eigenphase $\phi_j$ closest to $\phi$.  The measurement outcome is the eigenstate corresponding to this eigenphase.

\subsection{Initialization}\label{MBQCSSPT_5}
The measurement scheme can also be used to initialize all edge qubits for computation.  For many of the computationally universal cluster phases on different Archimedean lattices, it is necessary to fix some edge qubits to be in the Pauli eigenstates $|0^{(x)}\rangle$ or $|0^{(y)}\rangle$ to achieve the necessary two qubit interactions to build a universal gate set.  We now describe how such states can be initialized deterministically.

For each lattice with an underlying QCA structure, the tensor $C(p\tau, y)$ is a Pauli operator that is diagonal in the $X$-basis for each $y =1,...,n$.  These come in two varieties. The first type---seen in $(4^4)$, $(3,4,6,4)$, $(6^3)$, $(4,8^2)$, and $(4,6,12)$---are single site $X_y$ operators.  For example, for the (3,4,6,4) lattice,
\begin{align}
C (p\tau,4l) &= X_{4l} \\ 
C (p\tau,4l+1) &= X_{4l+1} \\
C (p\tau,4l+2) &= X_{4l+2} \\
C (p\tau,4l+3) &=  X_{4l+3}.
\end{align}
This structure allows us to prepare all edge qubits to be individually in some Pauli $X$ eigenstate.
In all other cases---$(3^6)$, $(3^4,6)$, $(3^3,4^2)$, and $(3,4,3^2,4)$---these are strings of sequential Pauli operators build up from a single site $Z$-operator recursively.  For example, for the $(3^4,6)$ lattice,
\begin{align}
C(p\tau, 6l) &= X_{6l} X_{6l+1} \\
C(p\tau, 6l+1) &= X_{6l+1} \\
C(p\tau, 6l+2) &= X_{6l+2} \\
C(p\tau, 6l+3) &= X_{6l+2} X_{6l+3} \\
C(p\tau, 6l+4) &= X_{6l+2} X_{6l+3} X_{6l+4} \\
C(p\tau, 6l+5) &= X_{6l+2} X_{6l+3} X_{6l+4} X_{6l+5}.
\end{align}
This structure allows us to also prepare all edge qubits to be individually in some $X$-eigenstate.  However, we must now prepare them sequentially, starting from the edge qubits corresponding to the tensors that are single site Pauli operators and working our way out.

Finally, in all lattices, the tensor $C(1,y)$ is always $Z_y$ for each $y=1,...,n$, which gives the capability to perform the single qubit rotation $\exp{(i\theta Z_y)}$ for each edge qubit $y$.  Therefore, after initializing each edge qubit to some $X$-eigenstate, which we can determine from the measurement record, we can utilize this $Z$-rotation to rotate each edge qubit to the state $|0^{(x)}\rangle$ or $|0^{(y)}\rangle$.

\section{Proofs of phase and universality}\label{Proofofphase}

In this section we prove that each lattice discussed in Sec.~\ref{Cone_Symmetries} and Sec.~\ref{Fractal_Symmetries} constitutes a computationally universal cluster phase, as defined in Sec.~\ref{SPTOreview}.  We analyze each lattice case by case, first proving the existence of the cluster phase and then use the symmetries to determine the universal gate set native to each lattice.  Recall from Sec.~\ref{SPTOreview} that determining whether or not a graph state can be used to construct a cluster phase requires us to identify the products of $Z$ operators that commute with all the generators of the symmetry group.  If these turn out to be stabilizer equivalent to a product of $X$ operators, the lattice can be used to construct a cluster phase.

The symmetry generators are products of $X$ operators on the physical degrees of freedom that arise from propagating single site Pauli operators through the virtual degrees of freedom of the tensor network representation.  Due to the inherent translational invariance of the lattices studied, we need only consider the symmetry generators up to translation by $\Delta$ sites in the spatial direction.  Therefore, we need only check commutation relations with $2\Delta$ symmetry generators generated within the space translationally invariant block.

To show each cluster phase is universal for MBQC, we must determine the $C(x,y)$ tensors defined in Sec.~\ref{Computational_Universality}.  Using the techniques described in Appendix~\ref{MBQCSSPT}, the $C(x,y)$ may be exponentiated to implement gates of the form $\textrm{exp}\left( -i\theta C(x,y) \right)$.  In many cases it will be necessary to fix some qubits to be in certain Pauli eigenstates to achieve the 2-body interactions necessary for proving computational universality.  Below we prove that each lattice supports a computationally universal cluster phase.

\subsection{$(4^4)$ cluster phase}

The line symmetries of the $(4^4)$ lattice were proven in Ref.~\cite{Raussendorf2018} to protect a cluster phase.  As discussed in Sec.~\ref{Cone_Symmetries}, the same cluster phase arises from using either line symmetries, or the larger group of cone symmetries as the symmetry group.  Here we focus on a construction based on the cone symmetries.

\begin{figure}
\centering
\includegraphics[width= 0.7 \linewidth]{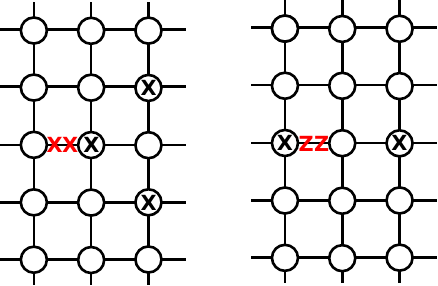}
\caption{Symmetry generators for the group of cone symmetries of the $(4^4)$ lattice.  All conventions are the same as in Fig.~\ref{3464}.}
\label{44syms}
\end{figure}

The generators of the cone symmetry group are depicted in Fig.~\ref{44syms} up to vertical translation.  One can see that for each vertex, each symmetry generator has support on an even number of its neighbors.  Thus, the simplest product of $Z$ operators that commutes with all the symmetry generators is of the form $\prod_{j\in\mathcal{N}(v)}Z_j$ for each vertex $v$.  Such a product of $Z$ operators can be visually seen to commute with all symmetries as depicted in Fig.~\ref{Commuting_Z}.  This operator is stabilizer equivalent to $X_v$ and hence the resulting phase defined by the subsystem symmetries is a cluster phase.

Furthermore, making use of Eq.~(\ref{TensorEqn}), one can determine the logical part of the MPS tensors for a $n\times p {\tau}$ sized quasi-1D block $C(x,y)$.  The relevant tensors for MBQC are,

\begin{eqnarray}
C(1,l) &=& Z_{l} \\
C(p {\tau},l) &=& X_{l} \\
C(2,l) &=& Z_{l-1}X_{l}Z_{l+1}
\end{eqnarray}
By measuring the corresponding qubits at site $(x,y)$ in the $X_\theta$ basis we can implement the gates,
\begin{eqnarray}
U_{1,l}(\theta) &=& e^{-i\theta Z_{l}} \\
U_{p {\tau},l}(\theta) &=& e^{-i\theta X_{l}} \\
U_{2,l}(\theta) &=& e^{-i\theta Z_{l-1}X_{l}Z_{l+1}}
\end{eqnarray}
By fixing every even qubit, indexed by 2$l$, to be in the $|0^{(x)}\rangle$ state this becomes a universal gate set on $\frac{n}{2}$ qubits:
\begin{eqnarray}
U_{1,2l+1}(\theta) &=& e^{-i\theta Z_{2l+1}} \\
U_{p {\tau},2l+1}(\theta) &=& e^{-i\theta X_{2l+1}} \\
U_{2,2l}(\theta) &=& e^{-i\theta Z_{2l-1}Z_{2l+1}}.
\end{eqnarray}

\subsection{$(3^6)$ cluster phase}

\begin{figure}
\centering
\includegraphics[width = 0.7 \linewidth]{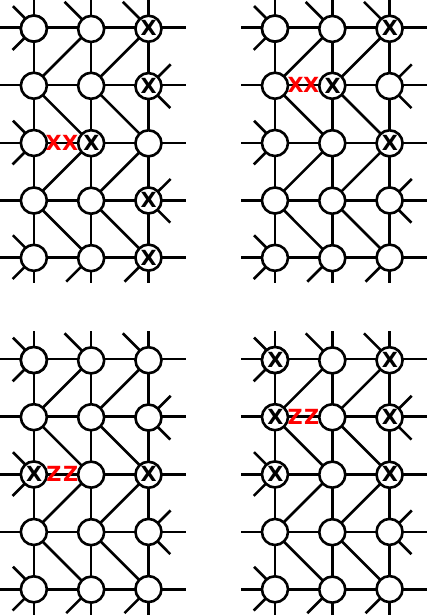}
\caption{Symmetry generators for the group of cone symmetries of the $(3^6)$ lattice.  All conventions are the same as in Fig.~\ref{3464}.}
\label{36syms}
\end{figure}

The generators of the cone symmetry group are depicted in Fig.~\ref{36syms} up to vertical translation.  One can see that for each vertex, each symmetry generator has support on an even number of its neighbors.  Thus, the simplest product of $Z$ operators that commutes with all the symmetry generators is of the form $\prod_{j\in\mathcal{N}(v)}Z_j$ for each vertex $v$.  This operator is stabilizer equivalent to $X_v$ and hence the resulting phase defined by the subsystem symmetries is a cluster phase.

Furthermore, making use of Eq.~(\ref{TensorEqn}), one can determine the logical part of the MPS tensors for a $n\times p {\tau}$ sized quasi-1D block $C(x,y)$.  The relevant tensors for MBQC are,

\begin{eqnarray}\label{36Tensors}
C(1,2l)&=&Z_{2l}\\
C(p {\tau},2l)&=&X_{2l-1}X_{2l}X_{2l+1}\\
C(2,2l+1)&=&Z_{2l}X_{2l+1}Z_{2l+2}
\end{eqnarray}
By measuring the corresponding qubits at site $(x,y)$ in the $X_\theta$ basis we can implement the gates,
\begin{eqnarray}
U_{1,2l}(\theta) &=& e^{-i\theta Z_{2l}} \\
U_{p {\tau},2l}(\theta) &=& e^{-i\theta X_{2l-1}X_{2l}X_{2l+1}} \\
U_{2,l}(\theta) &=& e^{-i\theta Z_{2l}X_{2l+1}Z_{2(l+1)}}
\end{eqnarray}
By fixing every odd qubit, indexed by $2l+1$, to be in the $|0^{(x)}\rangle$ state this becomes a universal gate set on $\frac{n}{2}$ qubits:
\begin{eqnarray}
U_{1,2l}(\theta) &=& e^{-i\theta Z_{2l}} \\
U_{p {\tau},2l}(\theta) &=& e^{-i\theta X_{2l}} \\
U_{2,2l}(\theta) &=& e^{-i\theta Z_{2l}Z_{2(l+1)}}.
\end{eqnarray}

\subsection{$(3,4,6,4)$ cluster phase}

See Sec.~\ref{Cone_Symmetries}.

\subsection{$(6^3)$ cluster phase}

\begin{figure}
\includegraphics[width = \linewidth]{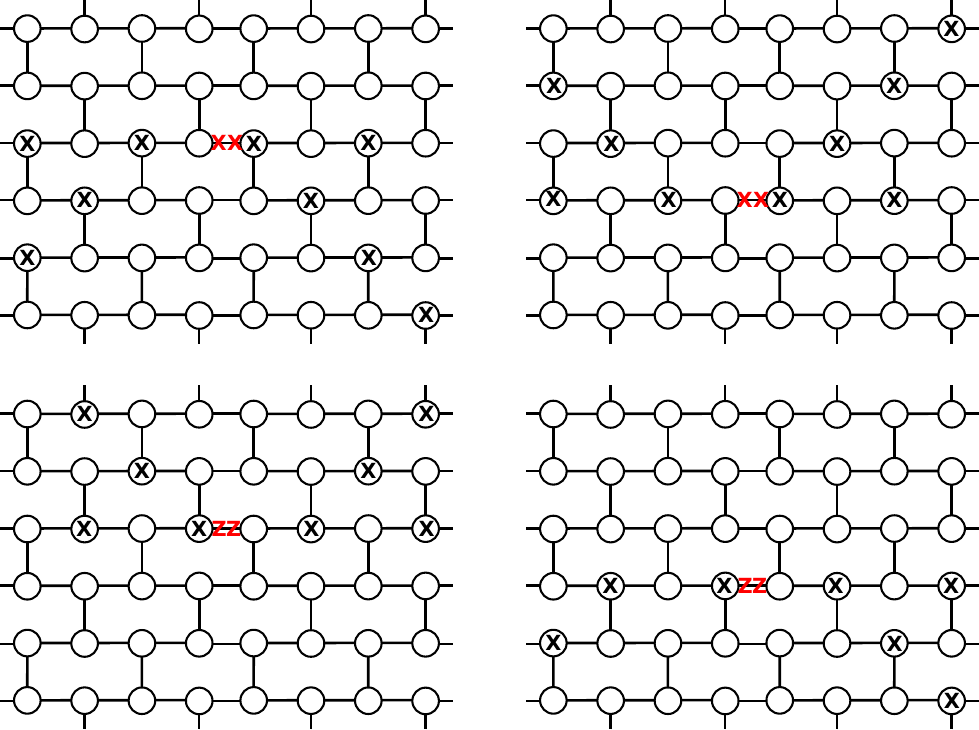}
\caption{Symmetry generators for the group of fractal symmetries of the $(6^3)$ lattice.  All conventions are the same as in Fig.~\ref{3464}.}
\label{63syms}
\end{figure}

The generators of the fractal symmetry group are depicted in Fig.~\ref{63syms} up to vertical translation.  One can see that for each vertex, each symmetry generator has support on an even number of its neighbors.  Thus, the simplest product of $Z$ operators that commutes with all the symmetry generators is of the form $\prod_{j\in\mathcal{N}(v)}Z_j$ for each vertex $v$.  This operator is stabilizer equivalent to $X_v$ and hence the resulting phase defined by the subsystem symmetries is a cluster phase.

Furthermore, making use of Eq.~(\ref{TensorEqn}), one can determine the logical part of the MPS tensors for a $n\times p {\tau}$ sized quasi-1D block $C(x,y)$.  The relevant tensors for MBQC are,

\begin{eqnarray}\label{63Tensors}
C(1,l)&=&Z_{l}\\
C(p {\tau},l)&=&X_{l}\\
C(2,2l)&=&X_{2l}Z_{2l+1}\\
C(p {\tau}-1,2l+1)&=&Z_{2l+1}X_{2(l+1)}
\end{eqnarray}
By measuring the corresponding qubits at site $(x,y)$ in the $X_\theta$ basis we can implement the gates,
\begin{eqnarray}
U_{1,l}(\theta) &=& e^{-i\theta Z_{l}} \\
U_{p {\tau},l}(\theta) &=& e^{-i\theta X_{l}} \\
U_{2,2l}(\theta) &=& e^{-i\theta X_{2l}Z_{2l+1}} \\
U_{p {\tau}-1,2l+1}(\theta) &=& e^{-i\theta Z_{2l+1}X_{2(l+1)}}
\end{eqnarray}
Together these gates form a universal gate set on all $n$ qubits encoded at the edge.

\subsection{$(4,8^2)$ cluster phase}

See Sec.~\ref{Fractal_Symmetries}.

\subsection{$(4,6,12)$ cluster phase}

\begin{figure*}
\centering
\includegraphics[width=0.7\linewidth]{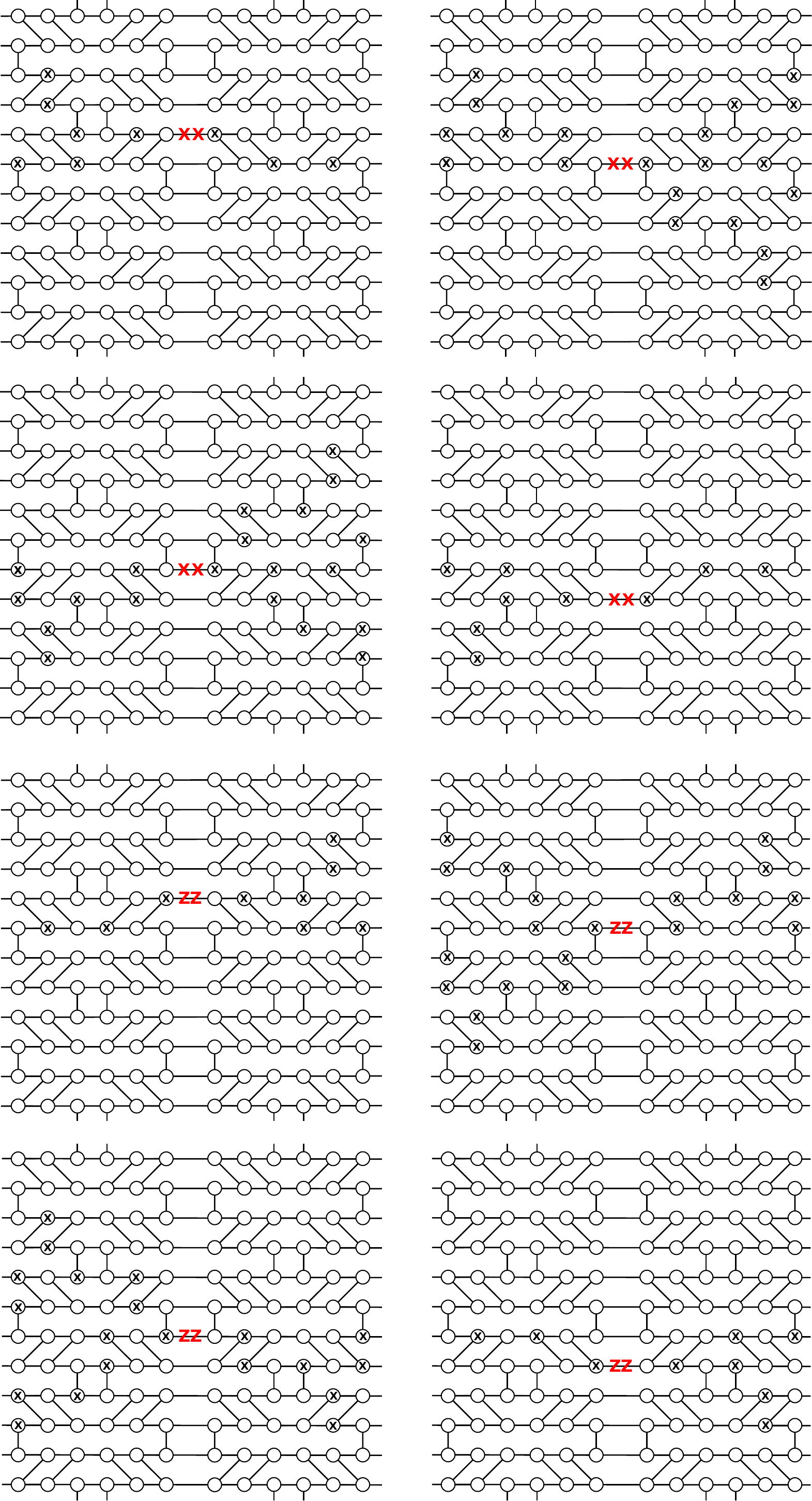}
\caption{Symmetry generators for the group of fractal symmetries of the $(4,6,12)$ lattice.  All conventions are the same as in Fig.~\ref{3464}.}
\label{4612syms}
\end{figure*}

The generators of the fractal symmetry group are depicted in Fig.~\ref{4612syms} up to vertical translation.  One can see that for each vertex, each symmetry generator has support on an even number of its neighbors.  Thus, the simplest product of $Z$ operators that commutes with all the symmetry generators is of the form $\prod_{j\in\mathcal{N}(v)}Z_j$ for each vertex $v$.  This operator is stabilizer equivalent to $X_v$ and hence the resulting phase defined by the subsystem symmetries is a cluster phase.

Furthermore, making use of Eq.~(\ref{TensorEqn}), one can determine the logical part of the MPS tensors for a $n\times p {\tau}$ sized quasi-1D block $C(x,y)$.  The relevant tensors for MBQC are,

\begin{eqnarray}\label{4612Tensors}
C(1,4l+2)&=&Z_{4l+2}\\
C(p {\tau},4l+2)&=&X_{4l+2}\\
C(2,4l)&=&Z_{4(l-1)+2}X_{4l}X_{4l+1}Z_{4l+2}.
\end{eqnarray}
By measuring the corresponding qubits at site $(x,y)$ in the $X_\theta$ basis we can implement the gates:
\begin{eqnarray}
U_{1,4l+2}(\theta) &=& e^{-i\theta Z_{4l+2}} \\
U_{p {\tau},4l+2}(\theta) &=& e^{-i\theta X_{4l+2}} \\
U_{2,4l}(\theta) &=& e^{-i\theta Z_{4(l-1)+2}X_{4l}X_{4l+1}Z_{4l+2}}.
\end{eqnarray}
By fixing every qubit indexed by $4l$ and $4l+1$ to be in the $|0^{(x)}\rangle$ state this becomes a universal gate set on $\frac{n}{4}$ qubits.
\begin{eqnarray}
U_{1,4l+2}(\theta) &=& e^{-i\theta Z_{4l+2}} \\
U_{p {\tau},4l+2}(\theta) &=& e^{-i\theta X_{4l+2}} \\
U_{2,4l}(\theta) &=& e^{-i\theta Z_{4(l-1)+2}Z_{4l+2}}.
\end{eqnarray}

\subsection{$(3^4,6)$ cluster phase}

\begin{figure*}[t]
\includegraphics[width=\linewidth]{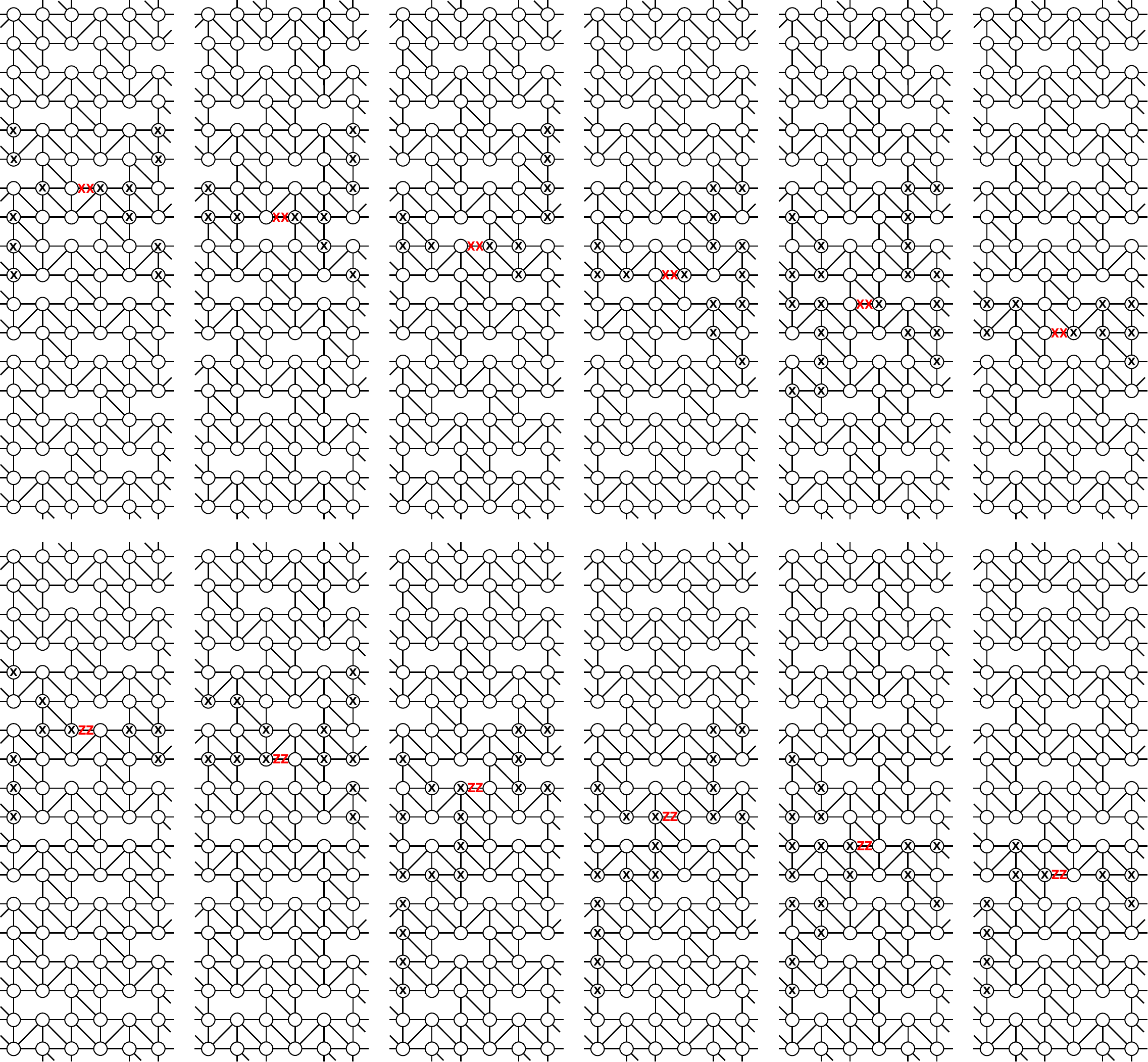}
\caption{Symmetry generators for the group of fractal symmetries of the $(3^4,6)$ lattice.  All conventions are the same as in Fig.~\ref{3464}.}
\label{33336syms}
\end{figure*}

The generators of the fractal symmetry group are depicted in Fig.~\ref{33336syms} up to vertical translation.  One can see that for each vertex, each symmetry generator has support on an even number of its neighbors.  Thus, the simplest product of $Z$ operators that commutes with all the symmetry generators is of the form $\prod_{j\in\mathcal{N}(v)}Z_j$ for each vertex $v$.  This operator is stabilizer equivalent to $X_v$ and hence the resulting phase defined by the subsystem symmetries is a cluster phase.

Furthermore, making use of Eq.~(\ref{TensorEqn}), one can determine the logical part of the MPS tensors for a $n\times p {\tau}$ sized quasi-1D block $C(x,y)$.  The relevant tensors for MBQC are,

\begin{eqnarray}\label{33336Tensors}
C(1,6l+2)&=&Z_{6l+2}\\
C(p {\tau},6l+2)&=&X_{6l+2}\\
C(p {\tau}-2,6l+1)&=&X_{6(l-1)+2}X_{6(l-1)+3} X_{6(l-1)+4}\nonumber\\
& & \times X_{6(l-1)+5}Y_{6l}Y_{6l+1}Y_{6l+2}Y_{6l+4} \nonumber \\
\end{eqnarray}
By measuring the corresponding qubits at site $(x,y)$ in the $X_\theta$ basis we can implement the gates,
\begin{eqnarray}
U_{1,6l+2}(\theta) &=& e^{-i\theta Z_{6l+2}} \\
U_{p {\tau},6l+2}(\theta) &=& e^{-i\theta X_{6l+2}} \\
U_{p {\tau}-2,6l+1}(\theta) &=& e^{-i\theta C(p-2,6l+1)}
\end{eqnarray}
The scheme for universal MBQC is slightly more complicated with this lattice.  First, fix every qubit indexed by $6l$ and $6l+1$ to be in the $|0^{(y)}\rangle$ state.  Also fix every qubit indexed by $6l+2$ and $6l+3$ to be in the $|0^{(x)}\rangle$ state.  Finally, notice that the $6(l-1)+4^{\textrm{th}}$ qubit should be fixed in the $|0^{(x)}\rangle$ state whereas the $6l+4^{\textrm{th}}$ qubit should be in the $|0^{(y)}\rangle$ state.  Luckily, there is a way to rotate between these two states so that we can perform entangling gates between any neighboring qubits at sites indexed by $6l+2$. 

To be consistent, the entangling gates must be broken up into two steps.  Let $l$ be an even integer and refer to the qubits at site $6l+2$ as even qubits and qubits at $6(l-1)+2$ as odd qubits.  To generate entanglement between any even qubit and the previous odd qubit fix each qubit indexed by $6(l-1)+4$ to be fixed in the $|0^{(x)}\rangle$ state and each indexed by  $6l+4$ to be in the $|0^{(y)}\rangle$ state.  Then, to generate entanglement between any even qubit and the next odd qubit first note that $C(1,6j+4)=Z_{6j+4}$ for any integer $j$.  Since this allows us to perform $Z$ rotations on the $6j+4^{\textrm{th}}$ qubits, the qubits at site $6(l-1)+4$ may be rotated to be in the ${|0^{(y)}\rangle}$ state and those at site $6l+4$ may be rotated to be in the $|0^{(x)}\rangle$ state.  We may then perform the same entangling gate before between any even qubit and the next odd qubit.

This scheme allows us to implement the universal gate set on $\frac{n}{6}$ qubits,
\begin{eqnarray}
U_{1,6l+2}(\theta) &=& e^{-i\theta Z_{6l+2}} \\
U_{p {\tau},6l+2}(\theta) &=& e^{-i\theta X_{6l+2}} \\
U_{p {\tau}-2,6l+1}(\theta) &=& e^{-i\theta X_{6(l-1)+2}Y_{6l+2}}.
\end{eqnarray}

\subsection{$(3^3,4^2)$ cluster phase}

\begin{figure}[h]
\includegraphics[width=\linewidth]{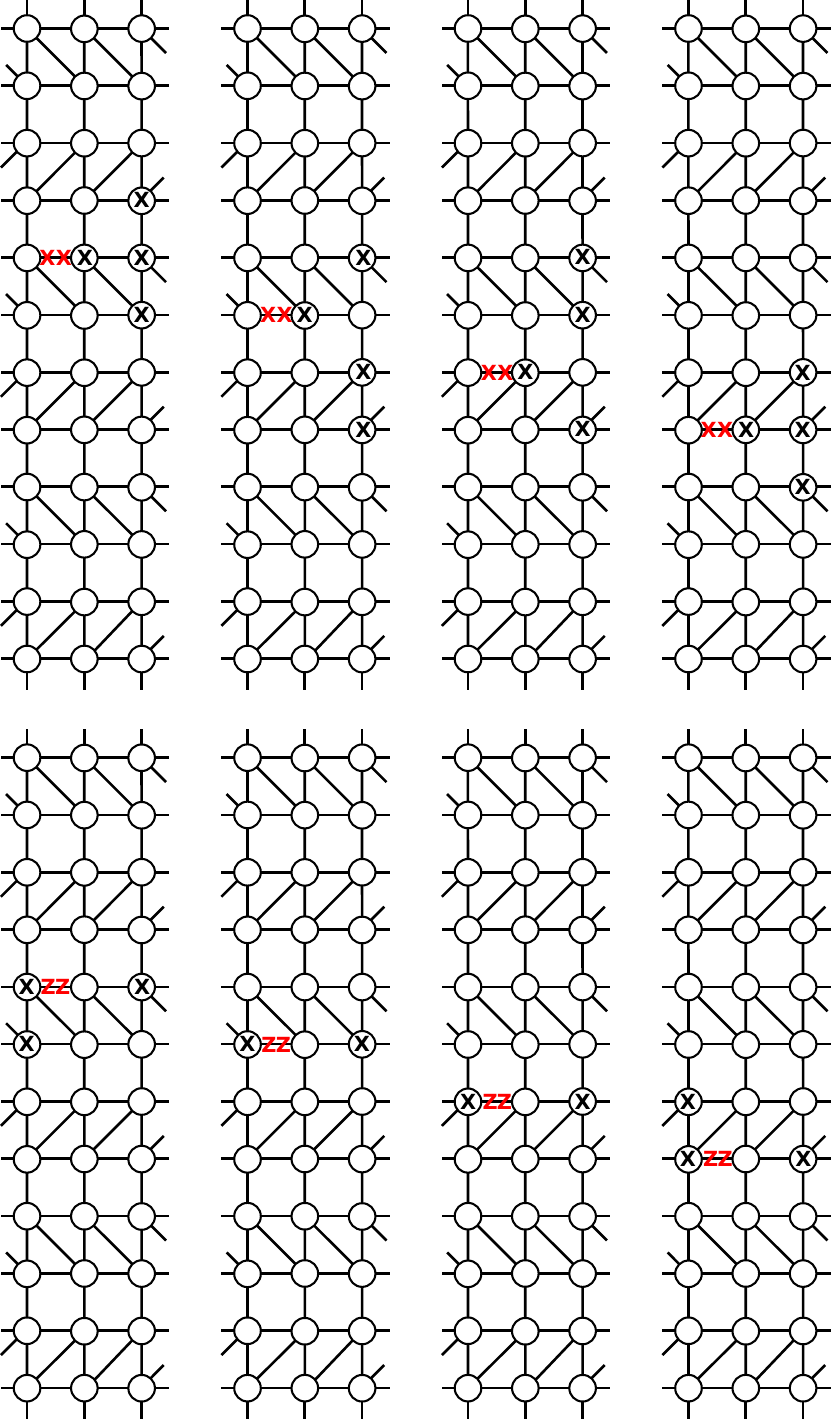}
\caption{Symmetry generators for the group of fractal symmetries of the $(3^3,4^2)$ lattice.  All conventions are the same as in Fig.~\ref{3464}.}
\label{33344syms}
\end{figure}

The generators of the fractal symmetry group are depicted in Fig.~\ref{33344syms} up to vertical translation.  One can see that for each vertex, each symmetry generator has support on an even number of its neighbors.  Thus, the simplest product of $Z$ operators that commutes with all the symmetry generators is of the form $\prod_{j\in\mathcal{N}(v)}Z_j$ for each vertex $v$.  This operator is stabilizer equivalent to $X_v$ and hence the resulting phase defined by the subsystem symmetries is a cluster phase.

Furthermore, making use of Eq.~(\ref{TensorEqn}), one can determine the logical part of the MPS tensors for a $n\times p {\tau}$ sized quasi-1D block $C(x,y)$.  The relevant tensors for MBQC are,

\begin{align}\label{36Tensors}
&C(1,4l)=Z_{4l}\\
&C(p {\tau},4l)=X_{4l}\\
&C(3,4l+2)=Z_{4l}Y_{4l+1}Y_{4l+2}X_{4l+3}Z_{4(l+1)}.
\end{align}
By measuring the corresponding qubits at site $(x,y)$ in the $X_\theta$ basis we can implement the gates,
\begin{eqnarray}
U_{1,4l}(\theta) &=& e^{-i\theta Z_{4l}} \\
U_{p {\tau},4l}(\theta) &=& e^{-i\theta X_{4l}} \\
U_{3,4l+2}(\theta) &=& e^{-i\theta Z_{4l}Y_{4l+1}Y_{4l+2}X_{4l+3}Z_{4(l+1)}}.
\end{eqnarray}
By fixing every qubit indexed by $4l+1$ and $4l+2$ to be in the $|0^{(y)}\rangle$ state and every qubit indexed by $4l+3$ to be in the $|0^{(x)}\rangle$ state, this becomes a universal gate set on $\frac{n}{4}$ qubits.
\begin{eqnarray}
U_{1,4l}(\theta) &=& e^{-i\theta Z_{4l}} \\
U_{p {\tau},4l}(\theta) &=& e^{-i\theta X_{4l}} \\
U_{3,4l+2}(\theta) &=& e^{-i\theta Z_{4l}Z_{4(l+1)}}.
\end{eqnarray}

\subsection{$(3,4,3^2,4)$ cluster phase}

\begin{figure}
\includegraphics[width=\linewidth]{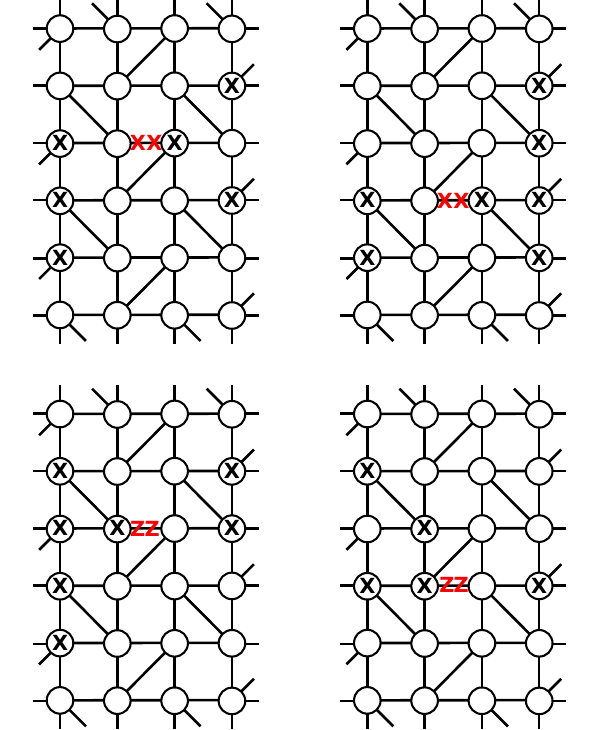}
\caption{Symmetry generators for the group of fractal symmetries of the $(3,4,3^2,4)$ lattice.  All conventions are the same as in Fig.~\ref{3464}.}
\label{33434syms}
\end{figure}

The generators of the fractal symmetry group are depicted in Fig.~\ref{33434syms} up to vertical translation.  One can see that for each vertex, each symmetry generator has support on an even number of its neighbors.  Thus, the simplest product of $Z$ operators that commutes with all the symmetry generators is of the form $\prod_{j\in\mathcal{N}(v)}Z_j$ for each vertex $v$.  This operator is stabilizer equivalent to $X_v$ and hence the resulting phase defined by the subsystem symmetries is a cluster phase.

Furthermore, making use of Eq.~(\ref{TensorEqn}), one can determine the logical part of the MPS tensors for a $n\times p {\tau}$ sized quasi-1D block $C(x,y)$.  The relevant tensors for MBQC are,

\begin{eqnarray}\label{33434Tensors}
C(1,2l+1)&=&Z_{2l+1}\\
C(p {\tau},2l+1)&=&X_{2l+1}\\
C(2,2l)&=&Z_{2l-1}X_{2l}Z_{2l+1}.
\end{eqnarray}
By measuring the corresponding qubits at site $(x,y)$ in the $X_\theta$ basis we can implement the gates,
\begin{eqnarray}
U_{1,2l+1}(\theta) &=& e^{-i\theta Z_{2l}} \\
U_{p {\tau},2l+1}(\theta) &=& e^{-i\theta X_{2l}} \\
U_{2,2l}(\theta) &=& e^{-i\theta Z_{2l-1}X_{2l}Z_{2l+1}}.
\end{eqnarray}
By fixing every even qubit, indexed by $2l$, to be in the $|0^{(x)}\rangle$ state this becomes a universal gate set on $\frac{n}{2}$ qubits:
\begin{eqnarray}
U_{1,2l+1}(\theta) &=& e^{-i\theta Z_{2l}} \\
U_{p {\tau},2l+1}(\theta) &=& e^{-i\theta X_{2l}} \\
U_{2,2l}(\theta) &=& e^{-i\theta Z_{2l-1}Z_{2l+1}}.
\end{eqnarray}

\section{1-form symmetry of the $(3,12^2)$ lattice}\label{31212}

\begin{figure}[h]
\centering
\includegraphics[width=0.6\linewidth]{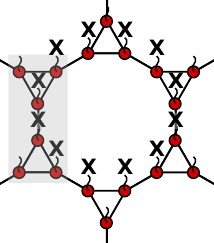}
\caption{1-form symmetry of the $(3,12^2)$ graph state.  There is one symmetry generator for each 12 sided shape in the lattice.  The bowtie subgraph of interest is shaded in grey.}
\label{31212_1form}
\end{figure}

In this section we describe implications of the 1-form symmetry of the graph state on the $(3,12^2)$ lattice.  As seen in Fig.~\ref{31212_1form}, the 1-form symmetry is generated by loops of $X$ operators around any 12 sided shape in the lattice.  To understand the computational properties of this state we analyze the capacity of a bowtie subgraph (shaded in grey in Fig.~\ref{31212_1form}) to teleport two qubits encoded at the left edge to the two at the right edge.

\begin{figure}
\centering
\includegraphics[width=0.35\linewidth]{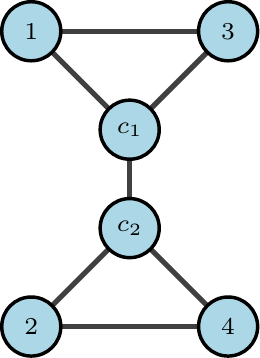}
\caption{Labeling of qubits for the bowtie subgraph of the $(3,12^2)$ lattice.}
\label{Bowtie_Subgraph_2}
\end{figure}

Let us follow the labeling of qubits shown in Fig.~\ref{Bowtie_Subgraph_2}.  For two qubits encoded on the left edge the logical operators are

\begin{eqnarray}
X_1^L &=& X_1 Z_{c_1} Z_3 \\
X_2^L &=& X_2 Z_{c_2} Z_4 \\
Z_1^L &=& Z_1 \\
Z_2^L &=& Z_2.
\end{eqnarray}
Also, there are the standard graph state stabilizers of Eq.~(\ref{Stabilizer_Relation}), $S_v$, centered at all other qubits.  Notice that,
\begin{equation}
X_1 X_2 X_{c_1} X_{c_2} = -S_{c_1}S_{c_2}Z_1^L X_1^L Z_2^L X_2^L \equiv Y_1^L Y_2^L.
\end{equation}
Thus, measuring the first, second, and two center qubits in the $X$ basis performs a logical measurement of $Y_{1}^L Y_{2}^L$, thereby projecting the input into the $(-1)^{m_1+m_2+m_{c_1}+m_{c_2}}$ eigenspace.

Returning to the $(3,12^2)$ lattice, we see that performing $X$ measurements on each qubit along each column implements a circuit consisting of $Y_{j}^L Y_{j+1}^L$ parity measurements.  Thus, this lattice acts as a foliated repetition code with stabilizer generators
\begin{align}
\langle \left\{Y^{L}_j Y^{L}_{j+1} \right\}_{\forall j}\rangle.
\end{align} 

\begin{figure*}
\centering
\includegraphics[width=0.7 \linewidth]{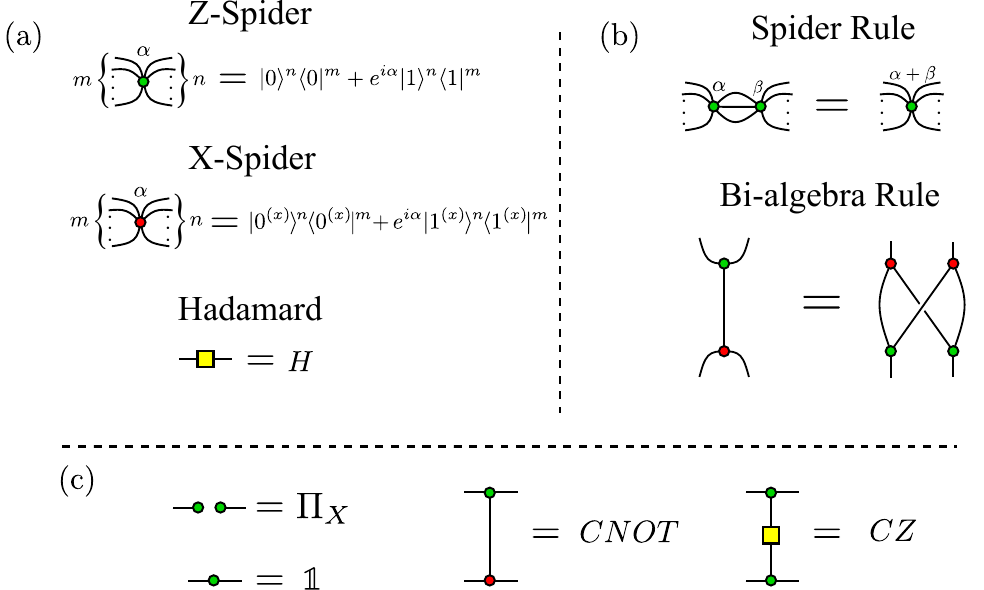}
\caption{The basics of the ZX-calculus. (a)The ZX-Calculus consists of three basic elements. Z-Spiders, X-Spiders, and the Hadamard.  Notice X and Z-spiders are related by applying a Hadamard to each leg. (b) The only two rewrite rules we will need are the spider and bi-algebra rules.  These can be easily derived using the definitions of the X and Z-spiders.  (c) Simple identities that will be useful for our purposes.  Notice for the $CNOT$ operation the target is the red spider.}
\label{ZX_Calculus}
\end{figure*}

\begin{figure*}
\centering
\includegraphics[width=.8\linewidth]{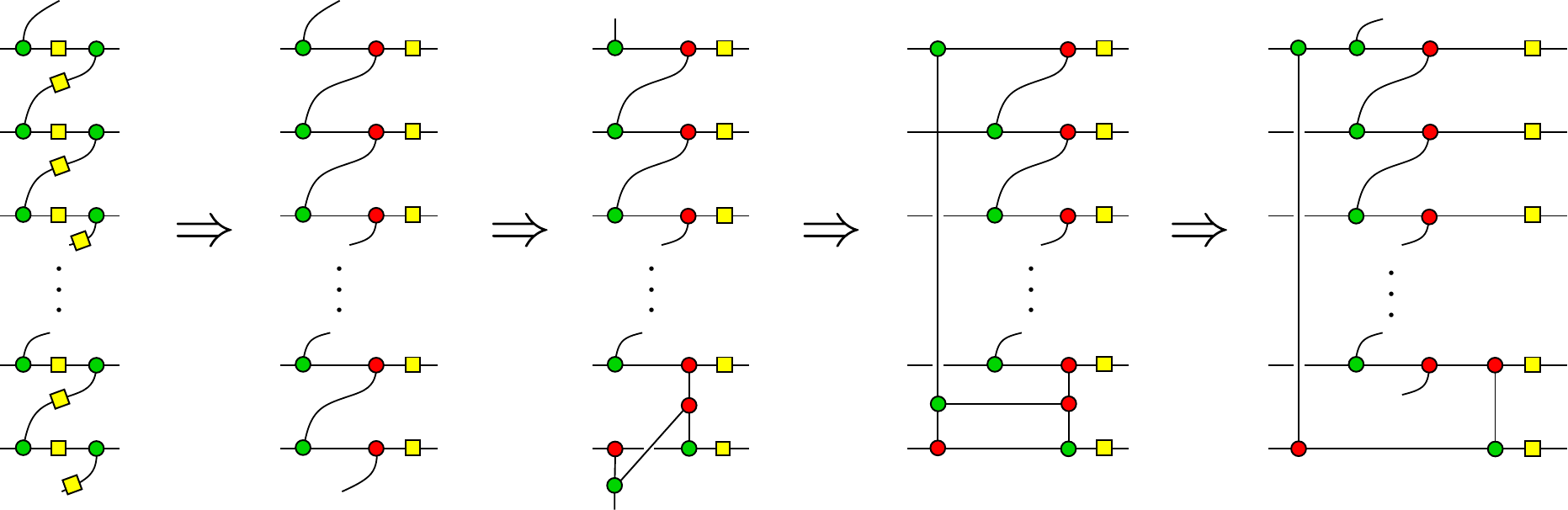}
\caption{The ZX-diagram for the a-causal circuit.  By pushing Hadamards to the end, applying the bi-algebra rule, and deforming the resulting diagram, the a-causal part of the circuit is reduced to $n-1$ wires.}
\label{ZX_Part_1}
\end{figure*}

\begin{figure*}
\centering
\includegraphics[width=.9\linewidth]{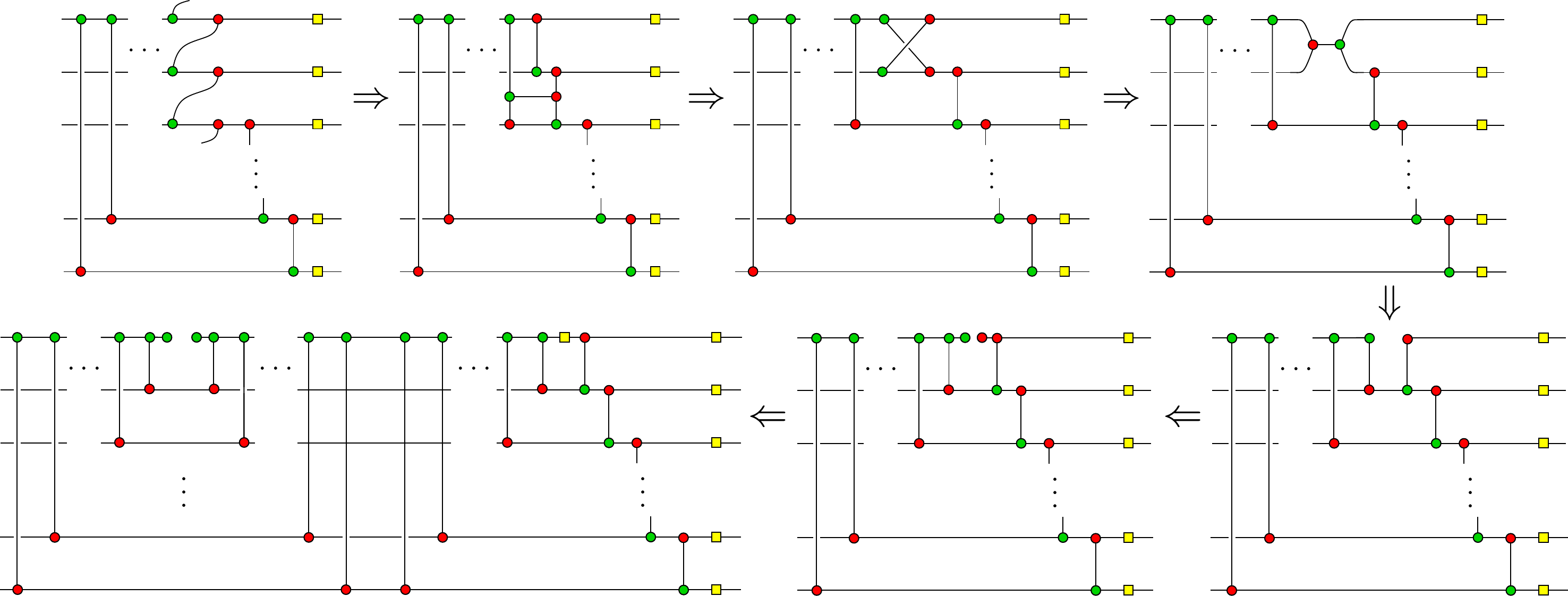}
\caption{Simplifying the ZX-diagram.  Applying the bi-algebra and spider rule iteratively, the a-causality of the diagram can be shown to have the structure of a circuit in which one qubit is measured.}
\label{Circuit_Part_2}
\end{figure*}

\section{Proof of QCA circuit for alternatively foliated $(3^6)$ lattice}\label{Nonunitary_Circuit}

In this section we derive the expression for the circuit shown in Fig.~\ref{Code_QCA_Circuit} corresponding to the ring tensors for the triangular lattice graph state with the alternative foliation, shown in Fig.~\ref{MPS_Construction}. First, contract each physical index around the ring with an arbitrary $X$ eigenstate (i.e. the measurement tensor in Eq.~(\ref{measurement_tensor})).  By Eq.~(\ref{1D_MPS_Z_Symmetries}) any $Z$ operator can be pulled through to the left virtual index of its respective site,
\begin{equation}
\includegraphics[width=0.5\linewidth]{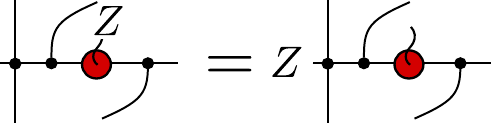}.
\end{equation}
Notice that $|m^{(x)}\rangle = Z^m|0^{(x)}\rangle$ so contracting a physical index with $|m^{(x)}\rangle$ is equivalent to first applying $Z^{m}$, and then contracting with $|0^{(x)}\rangle$. Decomposing each measurement tensor in this way and pushing the outcome dependent $Z$ operator to the virtual level, each tensor can be replaced with a Hadamard gate by Eq.~(\ref{MPS_Tensor_From_Tensors}),
\begin{equation}
\includegraphics[width=0.5\linewidth]{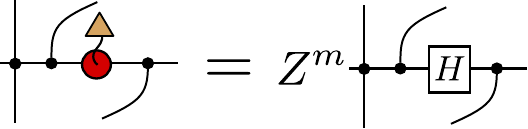}.
\end{equation}

Contracting the upper and lower indices of these tensors around a ring gives a quantum circuit that contains an acausal component.  To exorcise the acausality we appeal to the ZX-calculus~\cite{coecke2011interacting, jeandel2018complete}, a diagramatic language for simplifying and rewriting quantum operations.  The basic rules of this formalism needed for this calculation are summarized in Fig.~\ref{ZX_Calculus}.

Let us focus on the acausal part of the circuit.  This is recast as the ZX-diagram in Fig.~\ref{ZX_Part_1}.  By first moving all Hadamard gates to the right, applying the bi-algebra rule on the bottom wire untwists that wire out of the acausal loop.  Leveraging the spider rule, the resulting diagram contains an acausal loop on the remaining $n-1$ wires with the last wire separated out.

Iterating this procedure until the acausal part only acts on three wires, one more iteration leaves us with a diagram containing a structure that looks like the right hand side of the bi-algebra law on the top two wires.  Apply the bi-algebra rule on this structure and insert an identity as a Z and X-spider on the left and right, respectively.  By the spider rule we may then pull out a single node of an Z and X-spider from these identities.  Pulling out a Hadamard on the right hand side and inserting a resolution of the identity in the form of,
\begin{equation}
\mathbbm{1} = \left(\prod_{j=2}^n CNOT_{1,j}\right)^2,
\end{equation}
we end up with the diagram at the bottom of Fig.~\ref{Circuit_Part_2}.

Noting that,

\begin{equation}
\left(\prod_{j=2}^n CNOT_{1,j} \right) \Pi_X \left(\prod_{j=2}^n CNOT_{1,j} \right) = \Pi_{\bar{X}},
\end{equation}

\noindent the operation on the left hand side of the last diagram in Fig.~\ref{Circuit_Part_2} may be replaced by $\Pi_{\bar{X}}$.  The resulting circuit is then,

\begin{equation}
\includegraphics[width=\linewidth]{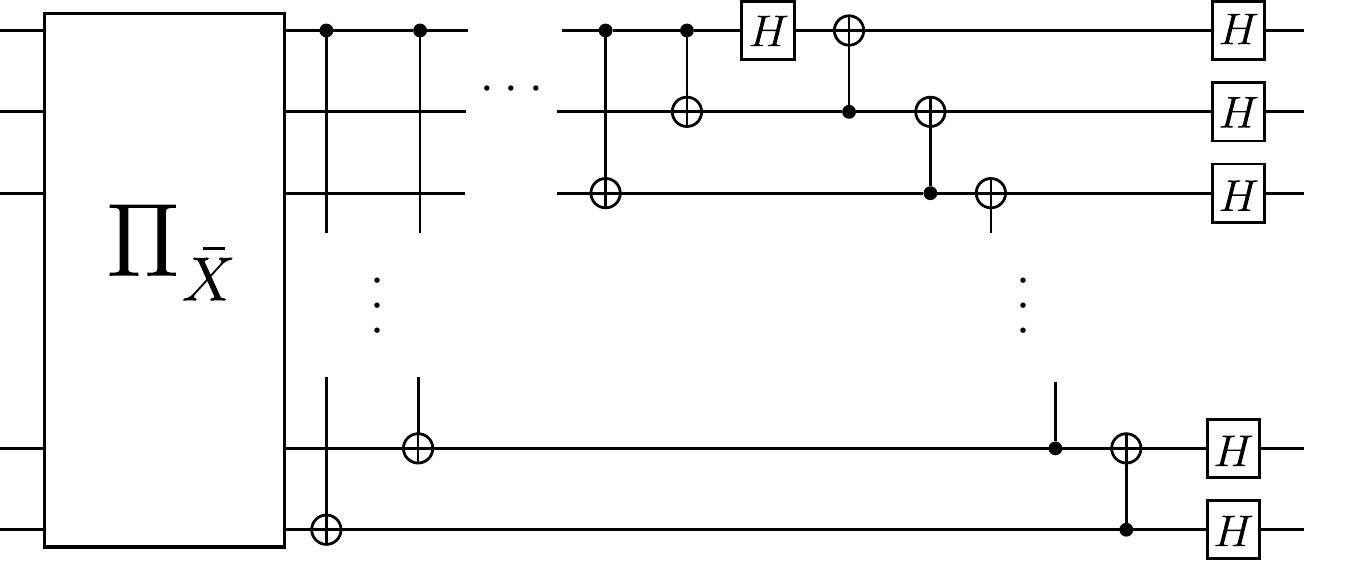}.
\end{equation}

Returning to the full circuit, we may Heisenberg evolve the outcome dependent $Z$ operators through the circuit to obtain an expression for the contracted ring tensor.  Therefore, preforming $X$ measurements on each site around a ring at the boundary of the state implements the following circuit,

\begin{equation}
\includegraphics[width=\linewidth]{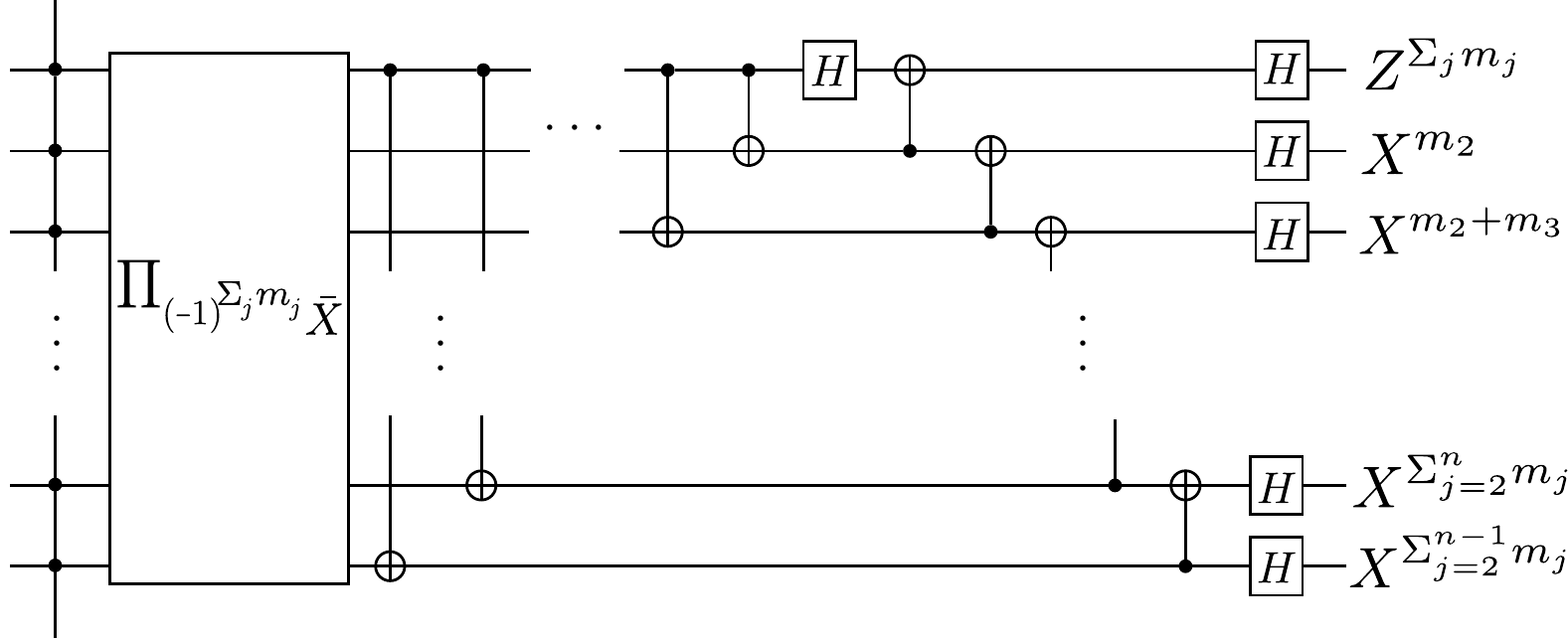}.
\end{equation}

\end{document}